\documentclass[PAPER, atlasdraft=false, texlive=2023, UKenglish,cernpreprint, texmf, orcidlogo]{atlasdoc}
 
\usepackage[subfigure]{atlaspackage}
\usepackage{atlasbiblatex}
 
\usepackage{atlasphysics,atlasjetetmiss,atlasprocess,atlasparticle}

\graphicspath{{logos/}{figures/}}
 
\usepackage{ANA-EGAM-2021-02-PAPER-defs}
 
\usepackage{adjustbox}

\AtlasTitle{Electron and photon energy calibration with the ATLAS detector using LHC Run~2 data}
 
\AtlasVersion{2.0}
 
\AtlasAbstract{
This paper presents the electron and photon energy calibration
obtained with the ATLAS detector using 140 fb$^{-1}$ of LHC
proton--proton collision data recorded at $\sqrt{s}=13$~TeV between 2015
and 2018. Methods for the measurement of electron and photon energies are outlined, along with the current knowledge of the passive material in front of the ATLAS electromagnetic calorimeter.
The energy calibration steps are discussed in detail, with emphasis on the improvements introduced in this paper. The absolute energy scale is set using a large sample of $Z$-boson
decays into electron--positron pairs, and its residual dependence on the electron energy is used for the first time to further constrain
systematic uncertainties. The achieved calibration
uncertainties are typically 0.05\% for electrons from resonant $Z$-boson decays, 0.4\% at $E_\text{T}\sim 10$~GeV, and 0.3\% at $E_\text{T}\sim 1$~TeV; for photons at $E_\text{T}\sim 60$~GeV, they are 0.2\% on average. This is more than twice as precise as the previous calibration. The new energy calibration is validated using $J/\psi \to ee$ and radiative $Z$-boson decays.
}

\AtlasRefCode{EGAM-2021-02}
 
\PreprintIdNumber{CERN-EP-2023-128}

\AtlasJournal{JINST}
\AtlasJournalRef{JINST 19 (2024) P02009}
\AtlasDOI{10.1088/1748-0221/19/02/P02009}

\hypersetup{pdftitle={ATLAS document},pdfauthor={The ATLAS Collaboration}}
 
\begin{document}
 
\maketitle
 
\clearpage
\tableofcontents

\clearpage
\section{Introduction}

 
During the 2015--2018 data-taking period (\RunTwo) of the Large Hadron Collider at CERN, the ATLAS experiment accumulated a large sample of proton--proton collisions at $\rts = 13$~\tev, corresponding to an integrated luminosity of 140~\ifb. Such a sample provides significant opportunities for improvements in detector performance and calibration precision, further exploration of the Standard Model and searches for new physics. Optimal energy reconstruction and calibration of the electromagnetic calorimeter is necessary for all analyses involving electrons and photons, and especially for precise measurements of the masses and properties of the Higgs, $W$ and $Z$ bosons. The present paper describes the calorimeter energy calibration using the \RunTwo data sample.
 
The calibration scheme comprises several steps: a simulation-based optimization of the energy measurement for electrons and photons, corrections for observed differences between data and simulation, calibration of the layers of the calorimeter and a final adjustment of the global energy scale using the abundant sample of electron--positron pairs from $Z$-boson decays. The resulting calibration corrections are validated using electrons from \jpsi decays and photons from radiative $Z$-boson decays.
 
The procedure applied to the full \RunTwo dataset is similar to the one in Refs.~\cite{PERF-2013-05,PERF-2017-03}.
Compared to the previous publication, the methodology has been updated in order to reduce the impact of the dominant sources of uncertainty: the offline reconstruction of electrons and photons in the calorimeter moved from a clustering algorithm that produced fixed-size clusters to one producing variable-size \enquote{superclusters}~\cite{EGAM-2018-01}; muons are now used for the presampler calibration, instead of electrons and photons; the layer intercalibration is now obtained by combining scales extracted using both electrons and muons; finally, dedicated data allowed further studies of the intercalibration of the high- and medium-gain electronics readouts. For the first time, the resulting calibration uncertainties are further constrained using a precise measurement of the energy dependence, or linearity, of the calorimeter response.
 
The calibration steps are discussed in the following, with special focus where improved methods have been utilized.  Section~\ref{sec:detector} briefly describes the ATLAS detector and  summarizes the electron and photon reconstruction algorithms  applied in this analysis. Section~\ref{sec:samples} describes the data and simulated event samples used for the studies, as well as the present knowledge of the passive material in front of the calorimeter. Section~\ref{sec:overview} gives an overview of the calibration procedure and details the changes relative to the previous procedure. Section~\ref{sec:unifstab} enumerates corrections applied to the data to balance geometrical inhomogeneities, and to account for residual non-linearity in the electronics readout. This section also details an improved analysis of the transition between the high- and medium-gain electronics readouts. Section~\ref{sec:layerInterCal} describes the calibration of the calorimeter layers, namely constraints on the presampler energy scale and the relative responses of the first and second compartments. Section~\ref{sec:insitu}  combines the results of these studies to extract the final adjustment of the global energy scale. Calibration corrections specific to photons are described in Section~\ref{sec:phSpec}. Calibration uncertainties at this stage are discussed in Section~\ref{sec:uncer}. The linearity of the electron energy scale is studied in Section~\ref{sec:linearity}. Finally, the validity of the energy calibration is established using independent samples of electron--positron pairs from \jpsi decays, and photons from radiative $Z$-boson decays, as discussed in Section~\ref{sec:xcheck}.


\section{Electron and photon reconstruction with the ATLAS detector}

 
\subsection{The ATLAS detector\label{sec:detector}}
 
The ATLAS experiment~\cite{PERF-2007-01} at the LHC is a multipurpose detector with cylindrical geometry\footnote{ATLAS uses a right-handed coordinate system with its origin at the nominal interaction point (IP) in the centre of the detector and the $z$-axis along the beam pipe. The $x$-axis points from the IP to the centre of the LHC ring, and the $y$-axis points upward. Cylindrical coordinates $(r,\phi)$ are used in the transverse plane, $\phi$ being the azimuthal angle around the $z$-axis. The pseudorapidity is defined in terms of the polar angle $\theta$ as $\eta=-\ln\tan(\theta/2)$. The transverse energy is defined as $\et = E/\cosh\eta$.} covering almost 4$\pi$ in solid angle. Closest to the collision point, ATLAS is instrumented with an inner tracking detector (ID) covering the pseudorapidity range of $|\eta |<2.5$ and consisting of a silicon pixel detector,
including the insertable B-layer installed as a new innermost layer before \RunTwo~\cite{ATLAS-TDR-19,PIX-2018-001},
followed by a silicon strip detector (SCT), and a transition radiation tracker (TRT) in the region $|\eta|<2.0$. The ID is surrounded by a superconducting solenoid producing an axial magnetic field of 2~T such that the ensemble enables efficient reconstruction of the tracks and momenta of charged particles, measurement of primary and secondary vertices, discrimination between electrons and pions, and reconstruction of photon conversions in ID material at radii up to 800~mm.
 
The solenoid surrounding the ID is encompassed by an electromagnetic (EM) calorimeter, which consists of lead absorbers folded in an accordion geometry and immersed in a liquid-argon (LAr) bath. The EM calorimeter is divided into three regions, each contained in a separate cryostat: the barrel section (EMB) covering the central pseudorapidity region $|\eta|<1.475$ and two endcaps (EMEC) covering the acceptance regions $1.375<|\eta|< 3.2$. The EMB and EMEC are segmented longitudinally into three (two) layers within $|\eta|<2.5$ ($2.5 <|\eta|<3.2$) with variable cell sizes, such that the direction of photon showers can be measured. The first layer (Layer~1) spans the regions $|\eta|<1.45$ and $1.5<|\eta|< 2.4$, and has a thickness between three and five radiation lengths ($X_{0}$), depending on $\eta$, and cells with a fine segmentation of $0.003\times0.1$ in $\Delta \eta \times \Delta \phi$ in the EMB, providing excellent discrimination between single photon showers and the showers of two nearly collinear photons from high-momentum pion decay. The second layer (Layer~2), with a cell granularity of $0.025 \times 0.025$ in $\Delta \eta \times \Delta \phi$, has a thickness between 17\,$X_{0}$ and 20\,$X_0$ and collects most of the energy deposited in the calorimeter by electron and photon showers. A third layer (Layer~3) with a thickness of 2--10\,$X_{0}$ and a coarser granularity of $0.05\times 0.025$ in $\Delta \eta \times \Delta \phi$ is used to collect the energy tails of very energetic showers. A thin presampler (PS) layer, placed in front of the accordion layers and covering the region $|\eta|<1.8$, is used to correct for energy losses upstream of the calorimeter. This detector consists of a 1~cm (0.5~cm) active LAr layer in the barrel (endcap) region with a coarse granularity of $0.025 \times 0.1$ in $\Delta \eta \times \Delta \phi$. The typical structure of the EM barrel calorimeter is illustrated in Figure~\ref{fig:archiLAr}. Scintillators are placed between the barrel and endcap cryostats ($1.37<|\eta|<1.52$) to improve the energy measurement in this region.
 
\begin{figure*}
\centering
\includegraphics[width=.7\textwidth]{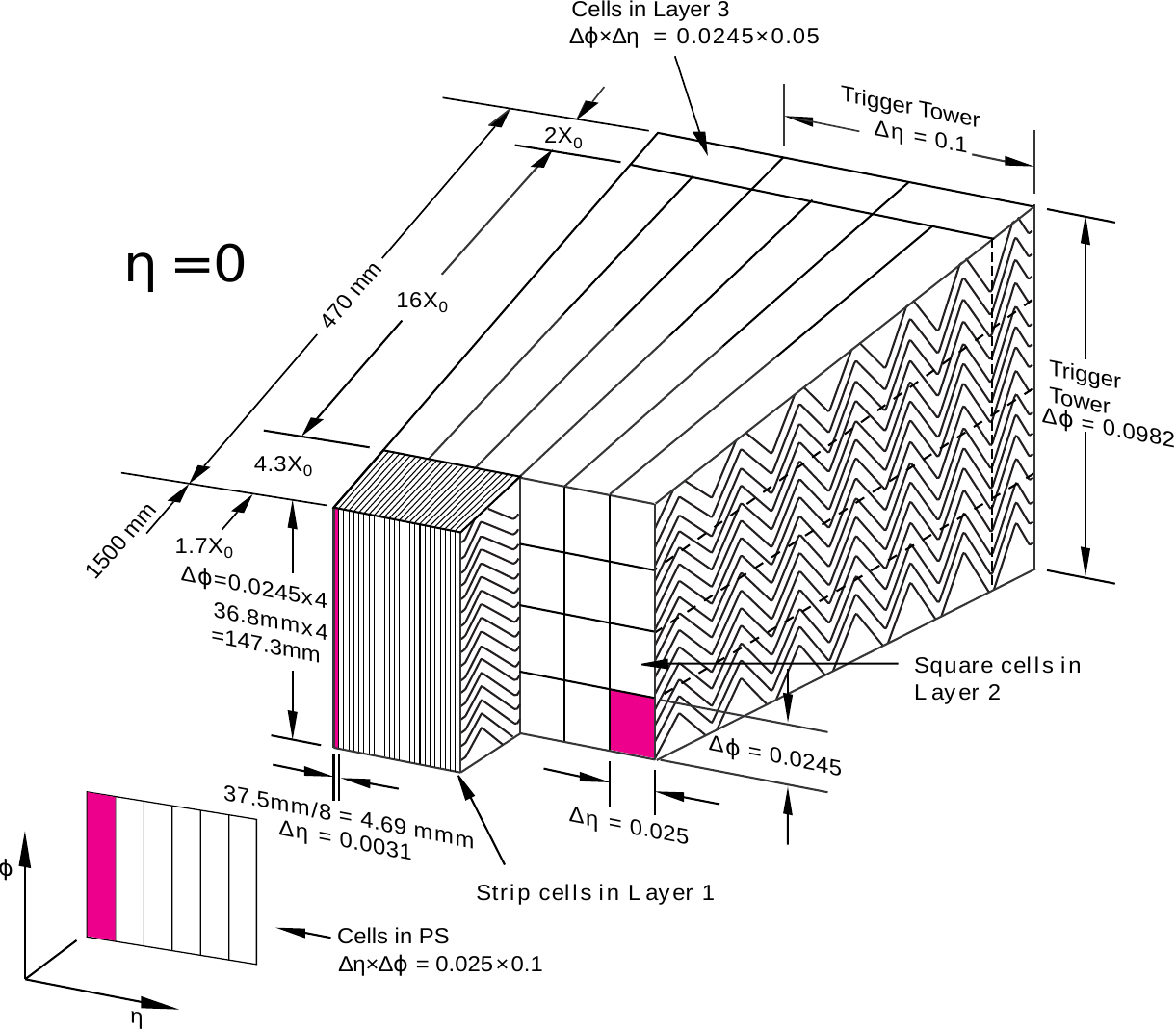}
\caption{Sketch of a barrel EM calorimeter module~\cite{PERF-2007-01}.\label{fig:archiLAr}}
\end{figure*}

The EM calorimeter is surrounded by an iron/scintillator hadron calorimeter in the region $|\eta|<1.7$. In the endcap regions, copper/LAr calorimeters are used up to $|\eta|=3.2$. Energy measurements at higher $|\eta|$, up to $|\eta|=4.9$, are made using a combination of forward copper/LAr and tungsten/LAr modules placed inside the endcap cryostats with the EMEC.
Muons are accurately measured and identified up to $|\eta|=2.7$ by a muon spectrometer located behind the calorimeters. It consists of three air-core superconducting toroids with eight coils each, precision tracking chambers, and fast chambers for triggering up to $|\eta|=2.4$.
 
A two-level trigger system is used to select events.
The first-level trigger is implemented in hardware and uses a subset of the
detector information to accept events at a rate below 100~kHz.
This is followed by a software-based high-level trigger that
reduces the accepted event rate to 1~kHz on average
depending on the data-taking conditions.

\subsection{Energy measurement, electron and photon reconstruction and identification}
 
The current generated in an EM calorimeter cell by ionizing particles is collected, amplified, and shaped to reduce the impact of out-of time showers, especially in high instantaneous luminosity conditions~\cite{Abreu:2010zzc}. The signal is sampled at 40~MHz and digitized by a 12-bit analogue-to-digital converter (ADC) in three different electronics readout gains, high, medium and low, in the front-end boards. The signal observed in the sample matching the trigger time defines which gain to use for the readout. Four digitized samples (two before the sample with highest energy, and one after) are sent to the back-end electronics. The energy deposited in the calorimeter cell is estimated through an optimal filtering procedure applied to the four samples after pedestal subtraction~\cite{Cleland:1994upr}, corrected by factors describing the conversion from ADC count to current and from current to energy. The pedestal, ADC-to-current conversion and signal shape of all calorimeter cells are derived from specific electronics calibration data. The \pileup dependence of the resulting energy value is corrected for using the measured instantaneous luminosity for the considered bunch crossings.
 
Electrons and photons are reconstructed from energy deposits in the cells, using a dynamic, variable cluster-size algorithm to form superclusters~\cite{PERF-2014-07,EGAM-2018-01}, allowing the recovery of energy from bremsstrahlung photons or from electrons from photon conversions. In this method an electron candidate is identified as a supercluster matching a track reconstructed in the ID. If a match is found, the track is re-fitted to account for bremsstrahlung. Superclusters without a matching track in the ID define unconverted-photon candidates. Converted-photon candidates are defined as a cluster matching a track that originates from a conversion vertex. The fraction of photons that convert varies from 20\% in the central region to 65\% in the endcaps.
 
Selection criteria are applied after reconstruction to identify genuine electrons while rejecting a large fraction of fake electrons. Depending on the desired background rejection factor and the specific needs of each analysis, four operating points, called Very Loose, Loose, Medium and Tight, are optimized for electrons in bins of $\eta$ and \et. The criteria were chosen by using a likelihood discriminant based on a number of track and cluster properties for which probability distributions are derived from electron and pion candidates in data. The identification efficiency of the selection is measured using electrons from $Z$-boson and \Jpsi decays. Descriptions of the methods, the used samples and the obtained results are given in Refs.~\cite{EGAM-2018-01,PERF-2017-01,EGAM-2021-01}.
 
Prompt photons are identified using two selection criteria, Loose and Tight, which are based on the EM shower shapes. The working points are defined in bins of $|\eta|$ and, in the case of the Tight selection, also in bins of \et.  The Loose criterion is independent of the conversion status of the photon. The Tight selection uses shower shape information from the first calorimeter layer and it is optimized separately for the two cases, accounting for the opening angle of the \epem pair in the magnetic field, which may impact the response for converted photons. The identification efficiency is measured using distinct data samples: inclusive photon production, photons from radiative $Z$-boson decays, and electrons from $Z$-boson decays after modifying their shower shapes to resemble those of photons. The corresponding analyses are described in detail in Refs.~\cite{PERF-2017-02,EGAM-2018-01,EGAM-2021-01}.


\section{Collision data and simulation}

\label{sec:samples}
 
\subsection{Dataset}
 
The analyses described in this paper use the full \pp collision dataset recorded by ATLAS between 2015 and 2018
with the LHC operating at a centre-of-mass energy of $\rts = 13$~\tev\ and 25~ns bunch spacing.
The sample corresponds to an integrated luminosity of 140~\ifb after quality cuts~\cite{DAPR-2018-01,DAPR-2021-01}; the typical mean number of interactions per bunch crossing, \avgmu,
was on average 13, 25 and 37 for the 2015, 2016 and 2017--2018 data, respectively.
Special samples, called \enquote{low-pile-up} samples in the following, were recorded in 2017 and 2018 at low instantaneous luminosity, with $\avgmu \sim 2$;
after applying data-quality requirements, the corresponding integrated luminosity amounts to 340~\ipb.

The measurements of the electromagnetic energy response
use a large sample of \Zee\ events selected with single-electron and dielectron triggers~\cite{TRIG-2018-05}. The dielectron
high-level triggers use a transverse energy threshold ranging from \SI{12}{\GeV} (2015) to 17 or \SI{24}{\GeV} (2016--2018)
and a Loose (2015) or Very Loose (2016--2018) identification criterion.
The single-electron high-level trigger has a transverse energy threshold ranging from \SI{24}{\GeV} in 2015 and most of 2016 to
\SI{26}{\GeV} at the end of 2016 and during 2017 and 2018; it applies Tight identification and loose track-based
isolation criteria~\cite{PERF-2017-01}. The offline selection for the energy calibration measurement requires two electrons
satisfying Medium identification, loose isolation and $\et > \SI{27}{\GeV}$, resulting in ${\sim}57$ million \Zee\ candidate events.
 
A sample of \jpsiee\ events with at least two electron candidates with $\et > 5~\gev$ and $|\eta| < 2.4$
is used to validate the electron energy scale at low \et. It was collected
using dedicated prescaled dielectron triggers with asymmetric \et thresholds ranging from 4 to \SI{14}{\GeV}.
The sample contains ${\sim}260\;\!000$ events.
 
Samples of $\Zll\gamma$ events ($\ell = e, \mu$), used to validate the photon energy scale, were selected with the same triggers as for the \Zee\ sample for the electron channel
and with single-muon or dimuon triggers~\cite{TRIG-2018-01} for the muon channel. The high-level dimuon (single-muon) trigger's transverse momentum threshold was 14~(26)~\gev; a loose track-based isolation criterion was applied in the high-level single-muon trigger.
The $\mu\mu\gamma$ ($ee\gamma$) samples, after requiring two muons (electrons) with Medium identification~\cite{PERF-2015-10},
transverse momentum $\pt > \SI{15}{\GeV}$ (\SI{18}{\GeV}) and one tightly identified and loosely isolated photon with $\et > \SI{15}{\GeV}$, contain ${\sim}210\;\!000$ (${\sim}100\;\!000$) events.
 
\subsection{Simulation samples}
 
Large Monte Carlo (MC) samples of \Zll and \Wln events
were simulated at next-to-leading order (NLO) in QCD using \POWHEG{}~\cite{Alioli:2008gx} and the \Pythia[8]~\cite{Sjostrand:2007gs}
parton shower model.
The \CT[10]~\cite{Lai:2010vv} parton distribution function (PDF) set was used in the matrix element calculation.
The \AZNLO set of tuned parameters~\cite{STDM-2012-23} and the \CTEQ[6L1]~\cite{Pumplin:2002vw} PDF set were used in the
modelling of non-perturbative effects.
\PHOTOSpp[3.52]~\cite{Davidson:2010ew} was used for QED emissions from electroweak vertices and charged leptons.
 
Both non-prompt (originating from $b$-hadron decays) and prompt (not originating from $b$-hadron decays) \jpsiee\ samples were generated
using \Pythia[8]. The A14 set of tuned parameters~\cite{ATL-PHYS-PUB-2014-021} was used together with the \CTEQ[6L1] PDF set,
and \EVTGEN[1.2.0]~\cite{Lange:2001uf} was used to model $b$- and $c$-hadron decays.
 
Samples of $Z \to \ell\ell\gamma$ events with photon $\et > \SI{10}{\GeV}$ were generated
with \Sherpa[2.2.4]~\cite{Bothmann:2019yzt} using QCD leading-order matrix elements with up to three additional
partons in the final state. The \NNPDF[3.0nnlo] PDF set was used in conjunction with the dedicated parton shower tuning
developed by the \SHERPA authors.
 
The energy resolution of the new reconstruction algorithm was optimized using samples of 40~million single-electron and single-photon events simulated without \pileup.
Their transverse energy distribution covers the range from 1~\gev\ to 3~\tev. Smaller samples with a flat \avgmu\ spectrum between 0 and 60 were also simulated to assess the performance as a function of \avgmu.
 
An extensive software suite~\cite{ATL-SOFT-PUB-2021-001} is used in data simulation, in the reconstruction and analysis of real and simulated data, in detector operations, and in the trigger and data acquisition systems of the experiment. The generated events were processed through the full ATLAS detector simulation~\cite{SOFT-2010-01} based on \GEANT~\cite{Agostinelli:2002hh}.
The MC events were simulated with additional interactions in the same or neighbouring bunch crossings to match the \pileup conditions
during LHC operations.
The overlaid \pp collisions were generated with the soft QCD processes of \Pythia[8]
using the A3 set of
tuned parameters~\cite{ATL-PHYS-PUB-2016-017} and the \NNPDF[2.3lo] PDF~\cite{Ball:2012cx}. Although this set of tuned parameters improves the modelling
of minimum-bias data relative to the A2 set~\cite{ATL-PHYS-PUB-2012-003} used previously, it overestimates
the hadronic activity measured using charged-particle tracks by roughly 3\%. Simulated events
were weighted to reproduce the distribution of the average number of interactions per bunch crossing in data, scaled
down by a factor 1.03.

\subsection{Passive material model\label{sec:passivemat}}

 
Measurements of electron and photon energies are affected by the passive material in front of the EM calorimeter. The simulation-based energy calibration accounts for this effect, but any differences between the simulated detector model and the actual detector produce discrepancies between the energy responses. For electrons of $\et~\approx~40~\gev$, such discrepancies are absorbed through the $Z$-based energy calibration, but biases remain that depend on the particle type (electron, unconverted or converted photon) and energy. The model used for the passive material and its associated uncertainties was derived from \RunOne and partial \RunTwo data~\cite{PERF-2013-05,PERF-2017-01}, and is summarized briefly below.
 
The model of the ID is based on measurements performed during its construction~\cite{PERF-2007-01}, leading to a 5\% uncertainty in the amount of material. In \RunTwo, a 10\% uncertainty is assigned to the material of the insertable B-layer, and 25\% to that of the inner-detector service patch panel~\cite{PERF-2015-07}, affecting the high-$|\eta|$ region.
 
The material between the ID and the LAr calorimeter was probed~\cite{PERF-2013-05} using the longitudinal development of electron and photon showers. The ratio of the energies deposited in the first and second accordion layers was found to be sensitive to the total amount of material traversed by these particles before entering the calorimeter. After calibration of the layers' energy response, this ratio was measured with a relative precision of about 1\% in the barrel, and about 2\% in the endcaps, leading to a determination of the passive material with a typical precision of 5\%$X_0$ for $|\eta|<1.3$, 10\%$X_0$ for $1.6<|\eta|<2.1$, and up to 20\%$X_0$ for $2.1 <|\eta|<2.4$. The impact of the passive material uncertainties on the energy measurement was parameterized using simulation as a function of the particle type, \et, and $|\eta|$. The corresponding calibration uncertainties mostly affect electrons at low \et and unconverted photons, and typically reach 1\textperthousand--2\textperthousand. No measurement was performed in the barrel--endcap transition regions, so these regions have larger uncertainties.
 
The resulting material model is presented in Figure~\ref{fig:passivea}, which summarizes the passive material between the interaction point and the calorimeter. The measured passive material is compared with the simulation and its uncertainties in Figure~\ref{fig:passiveb}.
 
\begin{figure*}
\centering
\subfigure[]{\includegraphics[width=.49\textwidth]{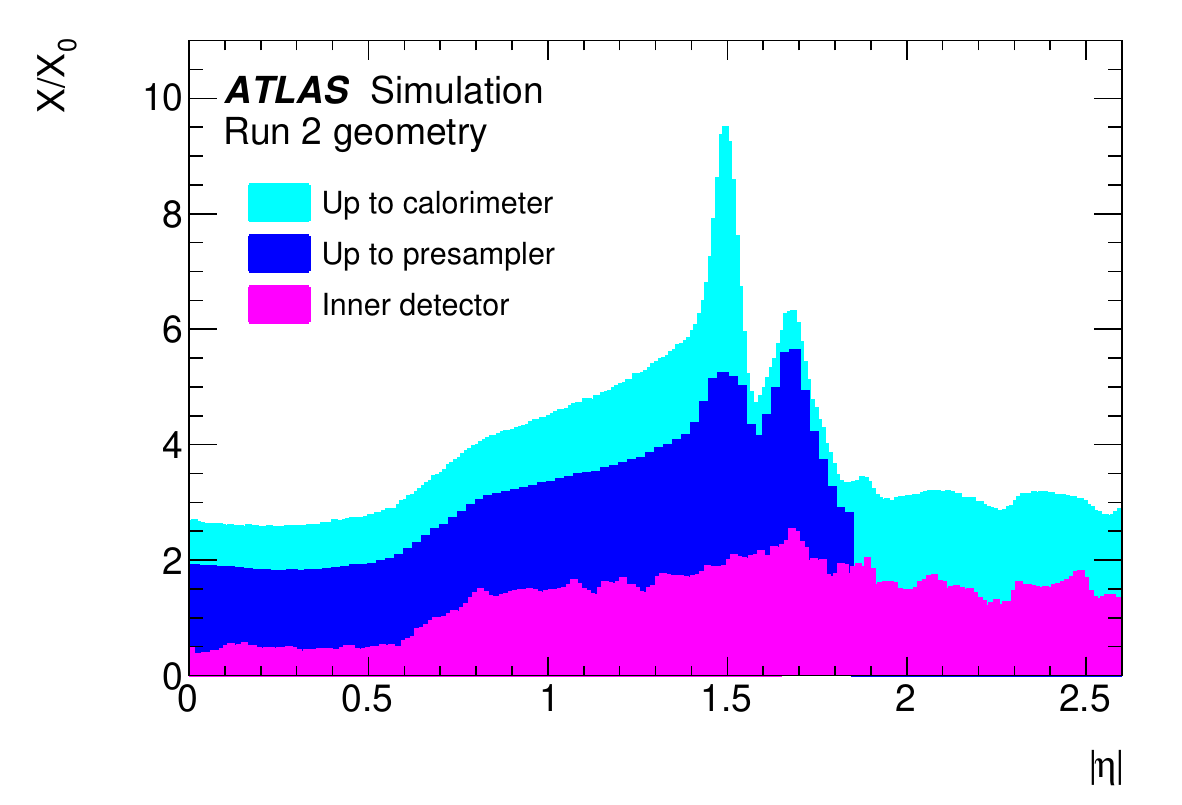}\label{fig:passivea}}
\subfigure[]{\includegraphics[width=.49\textwidth]{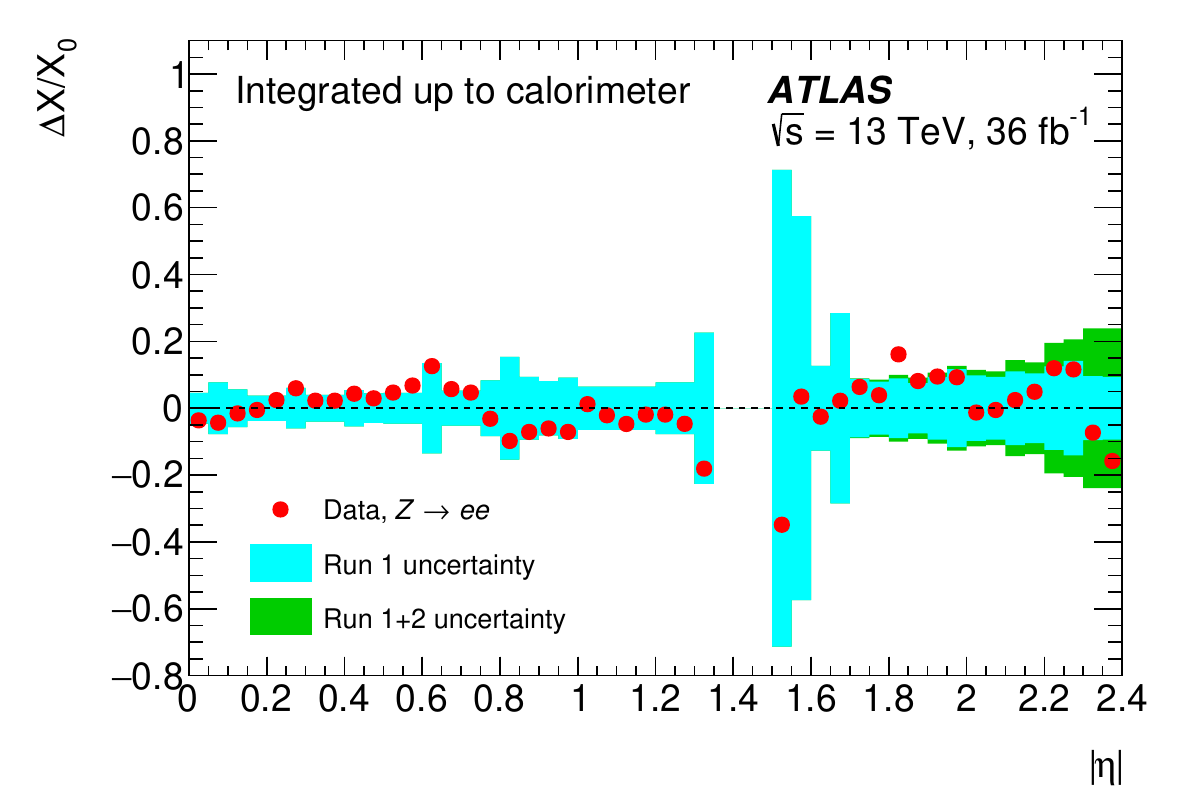}\label{fig:passiveb}}
\caption{(a) Amount of material traversed by a particle, in units of radiation lengths $X/X_0$, as a function of $|\eta|$ in the nominal simulation.
(b) Measured difference between the data and the nominal simulation of the detector material up to the first layer of the EM calorimeter. The data points are obtained using the longitudinal shower profile of electrons from $Z$-boson decays in partial \RunTwo data, and have negligible statistical uncertainties.
The blue band summarizes the material uncertainties, relative to the nominal simulation, as determined in \RunOne calibration studies.
The green band includes additional uncertainties related to the introduction of new material into the detector for \RunTwo, and covers potential mis-modelling of the insertable B-layer and a modified inner-detector patch panel (PP0). At high $|\eta|$, these uncertainties are dominated by a 50\% uncertainty in the simulated PP0 material.}
\end{figure*}



\section{Overview of the calibration procedure}

\label{sec:overview}
 
The different steps in the procedure used to calibrate the energy response for electrons and photons described in this paper are illustrated in Figure~\ref{fig:Flowchart}, and summarized below.
\begin{figure*}[h]
\centering
\includegraphics[trim={0 4cm 0 4cm},clip,width=\textwidth]{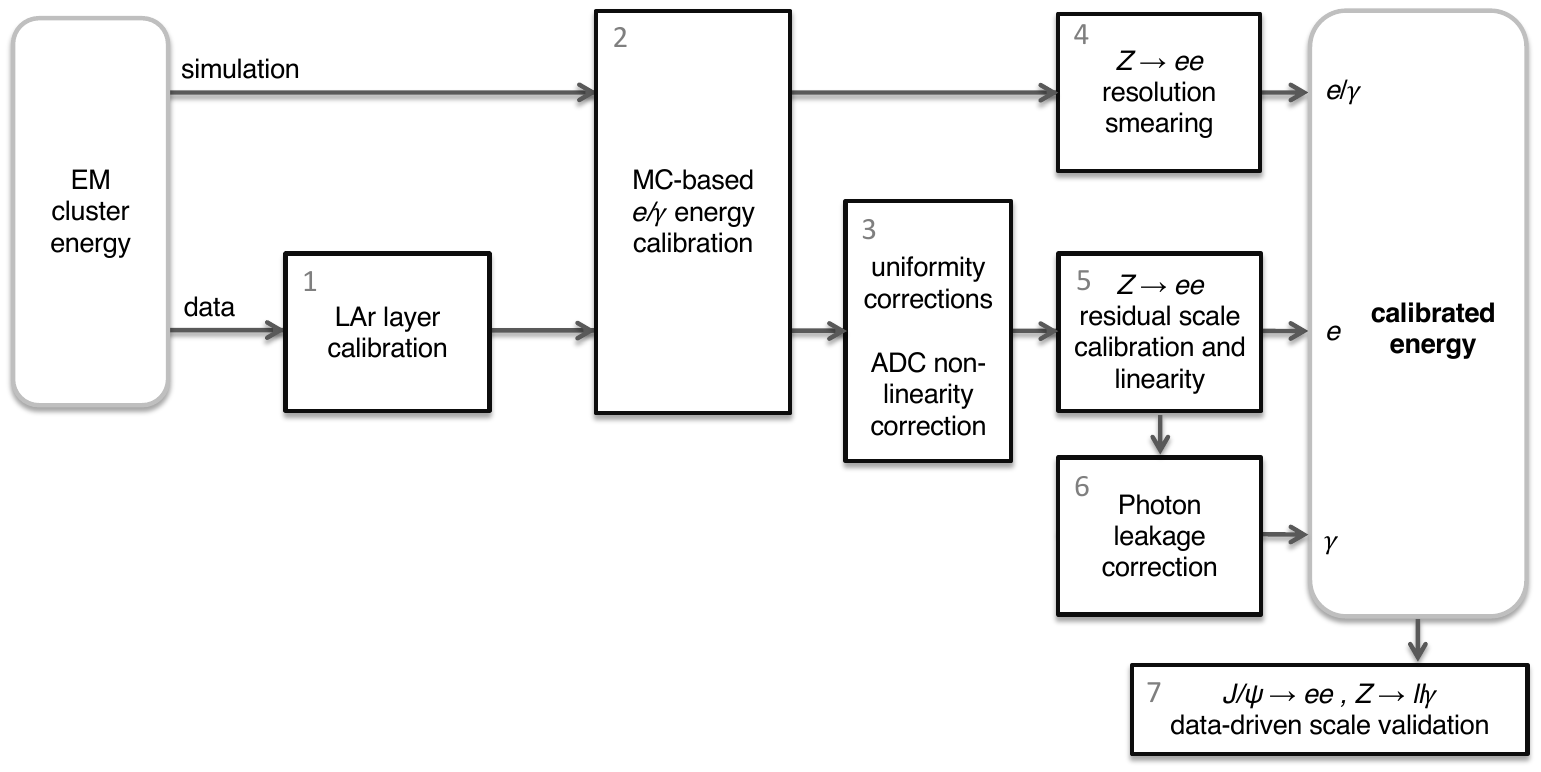}
\caption{Schematic overview of the electron and photon energy calibration procedure in ATLAS.}
\label{fig:Flowchart}
\end{figure*}
 
The energy of an electron or photon candidate is built from the energy of a cluster of cells in the electromagnetic
calorimeter. The measurement of electron and photon energies is optimized using a simulation-based boosted-decision-tree regression algorithm, combining energy deposits belonging to the reconstructed supercluster, in the presampler and in the calorimeter layers. The optimization is performed separately for electrons, converted photons and unconverted photons, taking into account the particle position. It is the same as the one used in Ref.~\cite{EGAM-2018-01} and the methodology is discussed in detail in Ref.~\cite{PERF-2017-03}.
 
Since the EM calorimeter is segmented in depth, the longitudinal layers should be calibrated separately to provide a correct description of the calorimeter response as a function of \et (step 1). After these corrections, the simulation-based calibration is applied identically to the cluster energies reconstructed from collision data and simulated event samples (step 2).
 
A set of additional corrections is applied to data to account for response variations not included in the simulation in specific detector regions, e.g.\ regions with non-optimal
high voltage, azimuthal non-uniformities, or biases associated with the liquid-argon calorimeter's electronics calibration (step 3). The stability of the calorimeter response as a function of azimuth, time and \pileup is also studied.
 
A final adjustment of the calorimeter response is derived from samples of \Zee events, so that the peak of the $Z$ resonance reconstructed in data coincides with that in the simulation. The response corrections are applied to the data. Using the same event samples, it is found that the resolution in data is slightly worse than that in simulation, and appropriate corrections are derived and applied to the simulation to match the data (steps 4, 5). The passive material model and the intercalibration corrections carry uncertainties that affect the energy calibration with a specific dependence on the particle \et. Measuring the residual energy dependence of the energy scale thus allows further adjustments of the calibration model, and provides additional constraints on the associated uncertainties. High-precision measurements,  such as those measuring the masses of the Higgs and $W$ bosons, will profit significantly from these uncertainty reductions.
 
The calibration factors extracted from \Zee\ events are assumed to reflect the intrinsic response of the calorimeter, and are thus applied identically to electrons and photons. Nevertheless, photon-specific corrections are needed to account for differences in the lateral development of electron and photon showers (step 6). Finally, the calibration chain is validated in data with low-\et electron candidates from \jpsiee\ decays, and with photon candidates from radiative $Z$-boson decays (step 7).


\section{Effects on the uniformity and stability of the energy response}

\label{sec:unifstab}
 
This section discusses the stability of the calorimeter energy response as a function of azimuthal angle, time, and \pileup. The dependence of the energy reconstruction on the readout electronics (ADC calibration and readout gain) is also discussed. Not all effects are modelled in the simulation. Corrections are defined for the ADC calibration and the azimuthal non-uniformity, while a systematic uncertainty is assigned for the observed dependence of the energy measurement on the readout gain.
 
\subsection{Uniformity}
 
Gravity-induced mechanical deformations of the calorimeter cause variations in the size of the liquid-argon gaps between the absorbers as a function of azimuthal angle. The resulting energy response variations are at the level of 0.1\%--0.2\% in the barrel, and up to 1\% in the endcaps. Energy corrections are derived from the modulations of the response as a function of $\phi$, separately in six intervals of absolute pseudorapidity (0--0.6, 0.6--1.0, 1.0--1.37, 1.37--1.55, 1.55--1.82, 1.82--2.47). In each $|\eta|$ interval, the relative response is defined from profiles of $\et/\avg{\et}$ in $Z$-boson decays, where \et is the electron transverse energy at a given $\phi$ value, and $\avg{\et}$ is the average over $\phi$.
 
The effect of this correction is illustrated in Figure~\ref{fig:uniformity1}. While the $\phi$-averaged energy response is unchanged by construction, the better uniformity is expected to yield a small improvement in the overall energy resolution. In practice, the resolution's constant term, discussed in Section~\ref{sec:insitu}, is reduced by 5\%--10\% in the endcaps, where the correction is most significant.
 
\begin{figure*}[h]
\centering
\includegraphics[width=.49\textwidth]{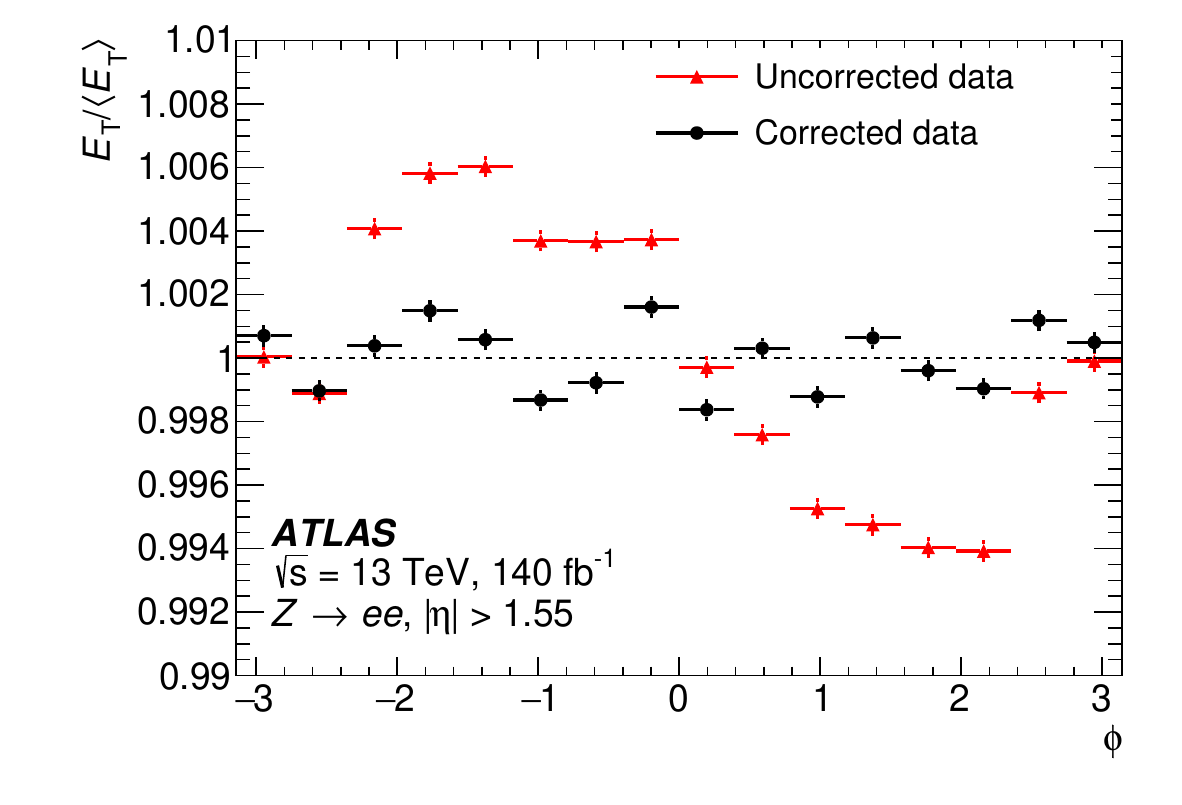}
\vspace{-0.3cm}
\caption{Electron energy response as a function of electron azimuth $\phi$, in the endcaps, before (red dots) and after (black dots) the $\phi$-uniformity energy correction. This correction was computed from 2015+2016 data and applied to the full \RunTwo sample. The observed non-zero residuals have no impact on the final result}
\label{fig:uniformity1}
\end{figure*}
 
\subsection{Stability}
 
The  stability of the calorimeter response is studied using the reconstructed peak position of the dielectron mass distribution, $m_{ee} / \langle m_{ee}\rangle$, in \Zee candidate events. This is illustrated in Figure~\ref{fig:uniformity2a} for the data taken between 2015 and 2018. Stability at the level of 0.05\% is observed over the full data-taking period.
 
Figure~\ref{fig:uniformity2b} shows $m_{ee} / \langle m_{ee}\rangle$ as a function of the average number
of interactions per bunch crossing for the data collected between 2015 and 2018.
The bipolar shaping of the calorimeter signals~\cite{Abreu:2010zzc} protects the energy measurement against \pileup fluctuations, and after correcting for bunch-to-bunch variations of the instantaneous luminosity, the residual dependence of the energy scale on \avgmu is below 0.1\%.
The small increase in energy observed in data is consistent with the MC expectation over most of the $\avg{\mu}$ range and is related to the new dynamical clustering used for the energy measurement~\cite{EGAM-2018-01}. The high-$\avg{\mu}$ region mostly reflects data taken in 2017, and the discrepancies observed in this region justify the extraction of dedicated energy-scale corrections for each data-taking year. More details are given in Section~\ref{sec:insitu}.
 
The stability of the response as a function of the number of reconstructed collision vertices (\nvtx)~\cite{PERF-2015-01}  is shown in Figure~\ref{fig:uniformity2c}. Classifying events according to \nvtx, related to the number of interactions in the specific bunch crossing, biases the \pileup activity of colliding bunches relative to the average.
In this case the compensation of the \pileup contributions to the reconstructed energy by the bipolar shaping becomes imperfect, giving rise to the observed slope.
The description of this effect in the simulation is accurate to 0.1\% for $\avgmu<25$, rising to 0.5\% at high \nvtx. The larger discrepancy at high \nvtx\ reflects the effect observed for \avg{\mu}.
\begin{figure*}[h]
\centering
\subfigure[]{\includegraphics[width=.49\textwidth]{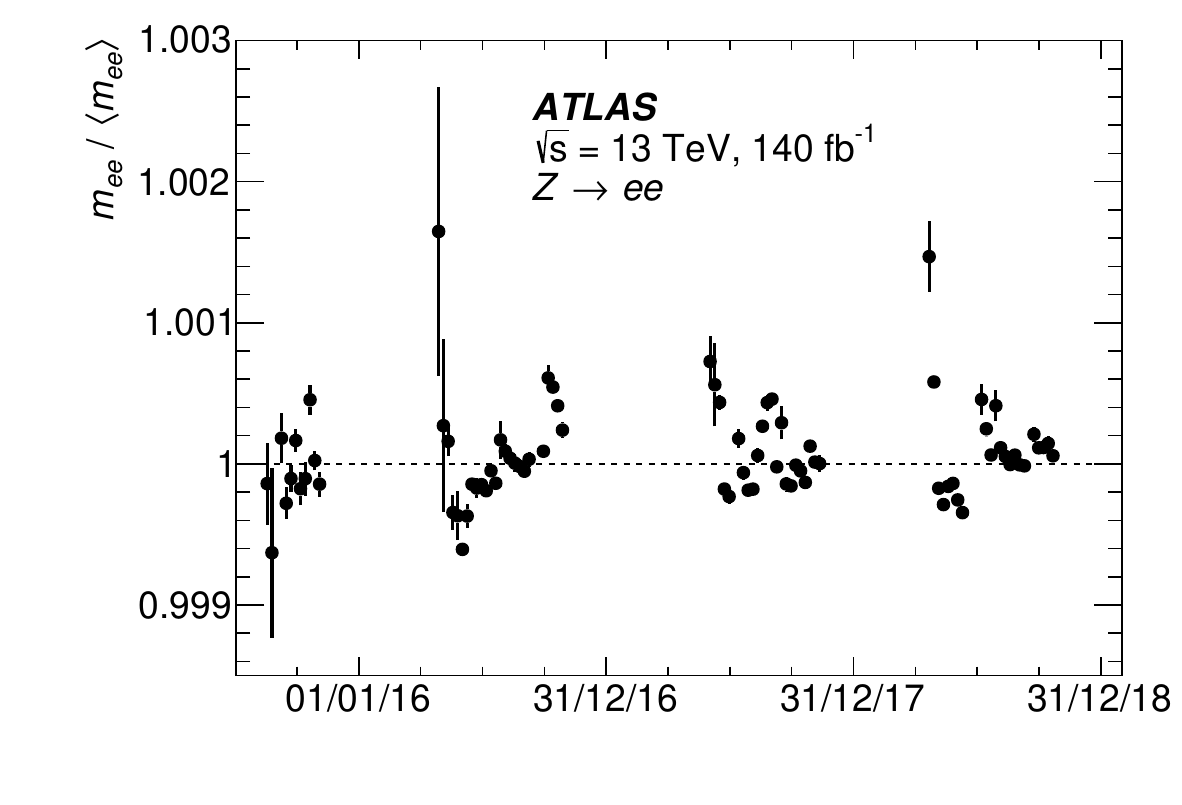}\label{fig:uniformity2a}} \\
\subfigure[]{\includegraphics[width=.49\textwidth]{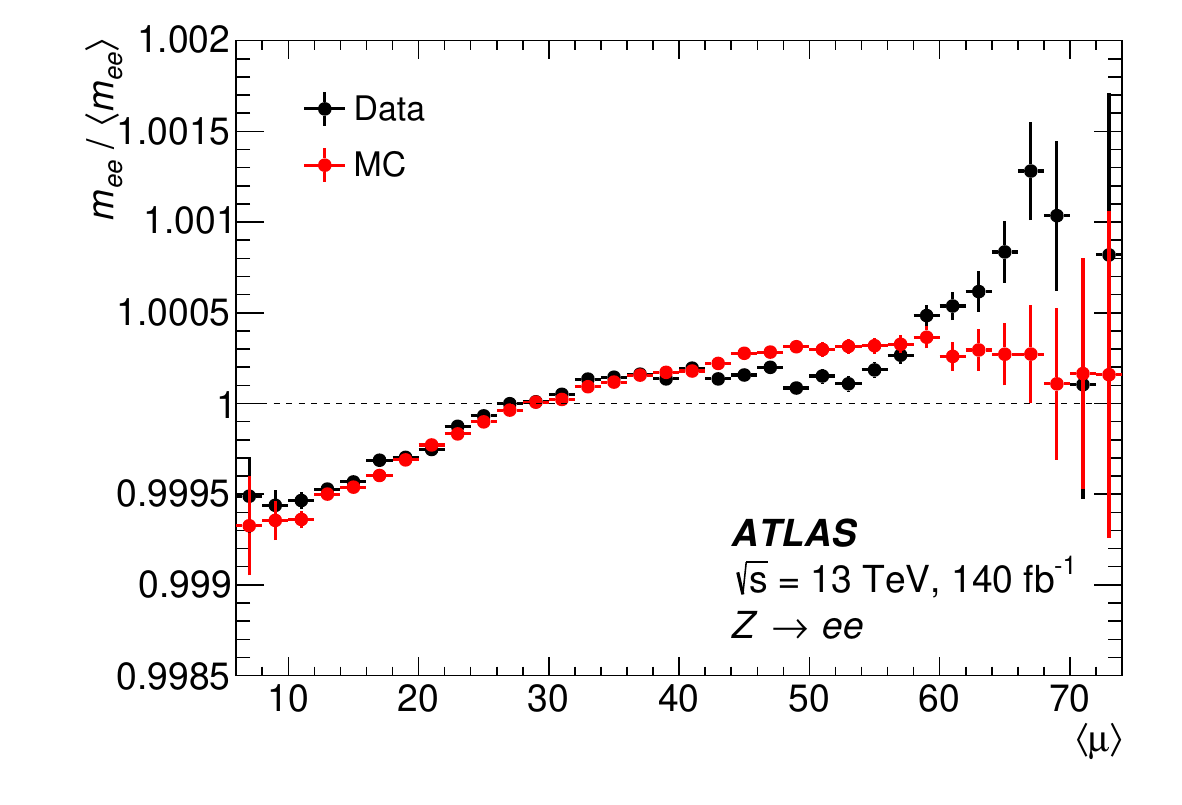}\label{fig:uniformity2b}}
\subfigure[]{\includegraphics[width=.49\textwidth]{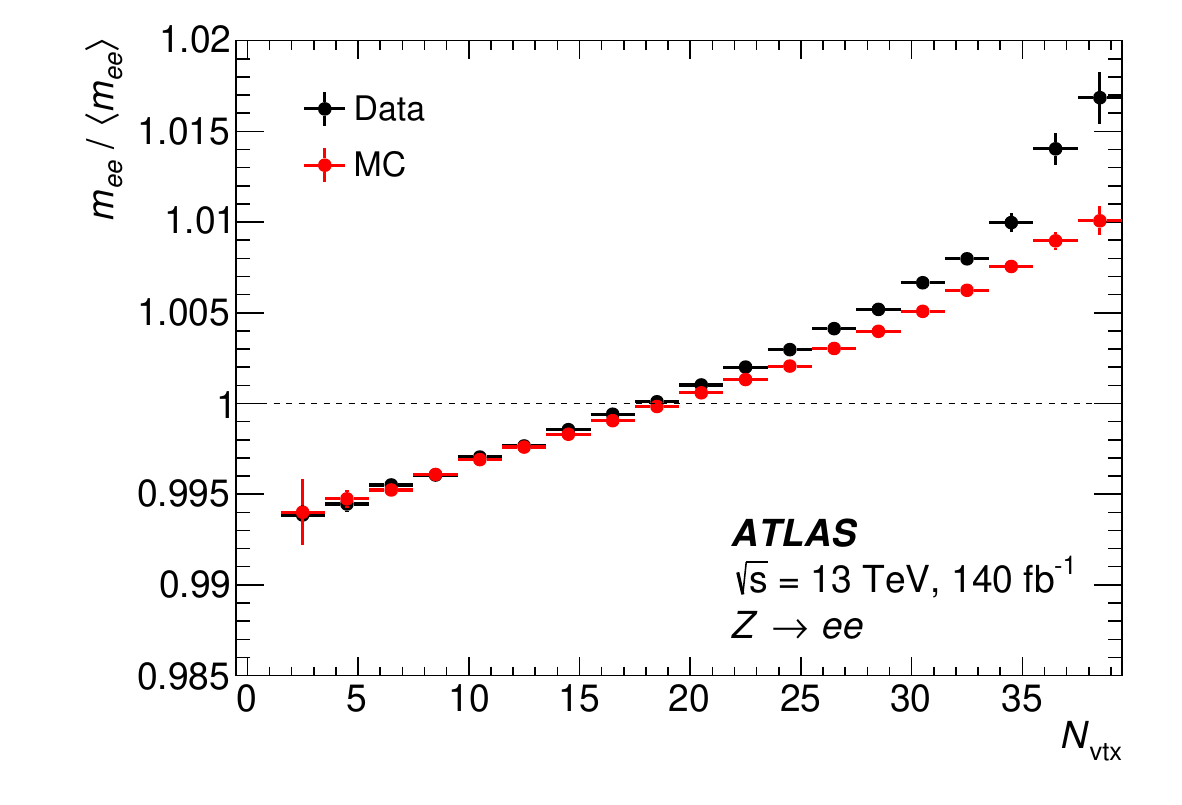}\label{fig:uniformity2c}}
\vspace{-0.2cm}
\caption{Relative variation of the peak position of the reconstructed dielectron mass distribution in \Zee events as a function of (a) time, (b) the average number of \pileup interactions, \avgmu, and (c) the number of reconstructed collision vertices, \nvtx.}
\end{figure*}

\subsection{ADC non-linearity correction}
\label{sec:adclin}
 
The energy reconstruction in a LAr calorimeter cell is based on a linear conversion from ADC counts to current, and an additional factor converting current into energy. In practice, the energy response of each cell is determined during dedicated electronics calibration runs, parameterizing the relation between the injected current and the measured ADC count using a linear function. The fits to the calibration data show non-zero residuals, caused by intrinsic non-linear behaviour of the electronics. This is illustrated in Figure~\ref{fig:adccorr} for an example cell and a calibration run performed in medium gain (MG). The residuals deviate from a linear fit by about 0.3\%; comparable deviations are observed in high-gain (HG) calibration runs.
 
The implications of this non-linearity are twofold. First, the residuals of the ADC-to-current correction contribute to the non-linearity of measurements of electron and photon energies, and must therefore be accounted for in view of the linearity analysis in Section~\ref{sec:linearity}. A correction is implemented, based on fits of a fifth-order polynomial to the residuals, separately for each cell of the calorimeter, as exemplified in Figure~\ref{fig:adccorr}. The parameterized residual is then added to the cell energy estimate. The impact on electron or photon cluster energies is estimated by repeating this procedure for all cells belonging to the cluster, and computing the modified cluster energy. The final correction is built such that it does not
modify the cluster energies of electrons at $\et = 40\ \gev$, where the global energy scale is set (see Section~\ref{sec:insitu}).
Results of this procedure for electrons, unconverted photons and converted photons are illustrated in Figure~\ref{fig:adccorrImp}. Cluster energies are increased by about 0.4\% at low \et, and decreased by about 0.2\% at high \et, with a moderate dependence on particle type and pseudorapidity. A relative uncertainty of 30\% is assigned to this correction.
 
\begin{figure*}[h]
\centering
\includegraphics[width=.49\textwidth]{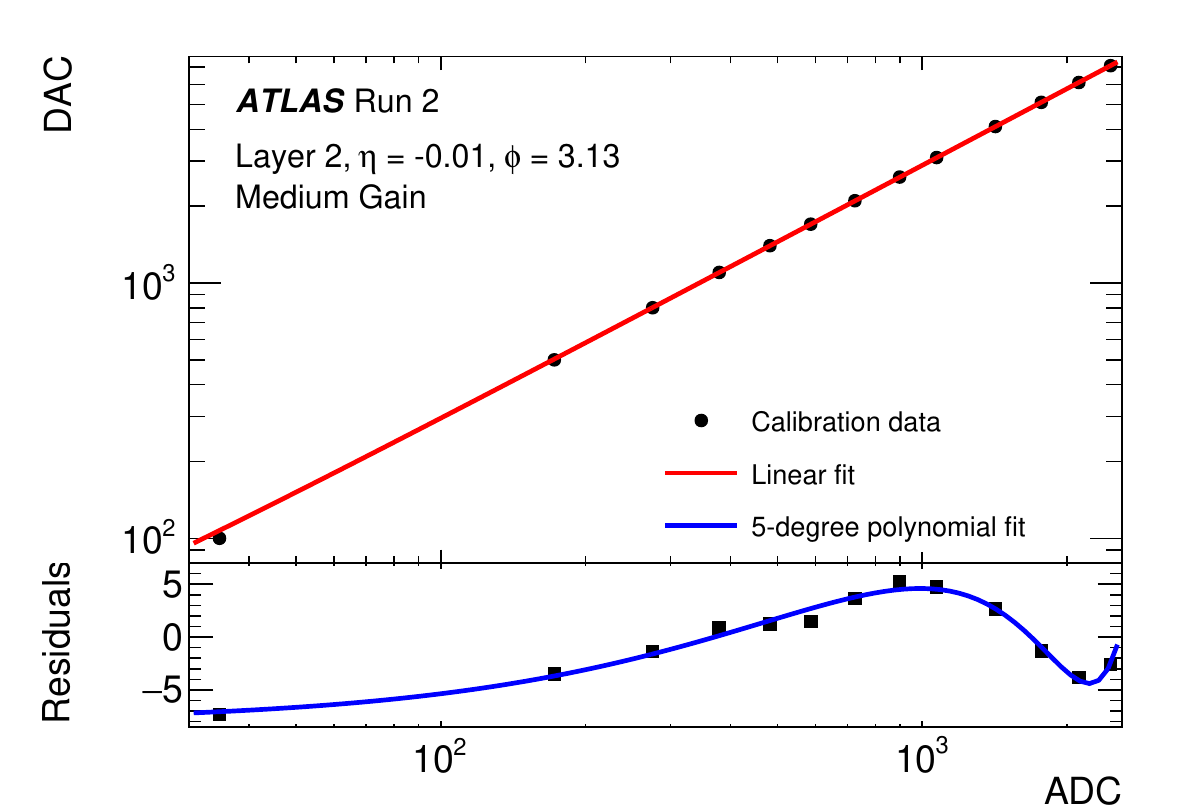}
\caption{Output of a calibration run performed in medium gain, for an example cell in the second layer of the EM calorimeter. The relation
between injected current, in DAC (digital-to-analogue converter) units, and ADC counts is assumed to be
linear (top panel). The evolution of the residuals (in DAC units) between the measurements and the
linear fit as a function of ADC counts (bottom panel) is parameterized with a fifth-order polynomial
and used to correct the cell energy.
}
\label{fig:adccorr}
\end{figure*}
 
\begin{figure*}[h]
\centering
\subfigure[]{\includegraphics[width=.49\textwidth]{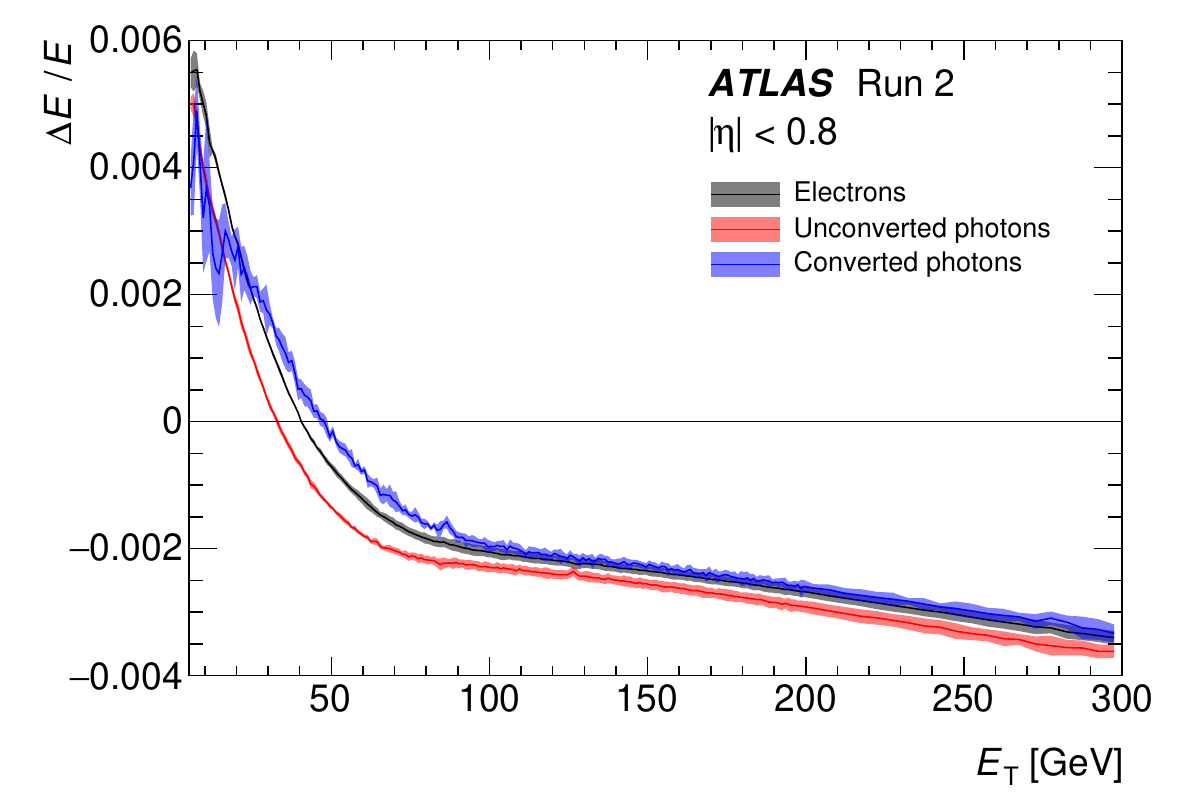}}
\subfigure[]{\includegraphics[width=.49\textwidth]{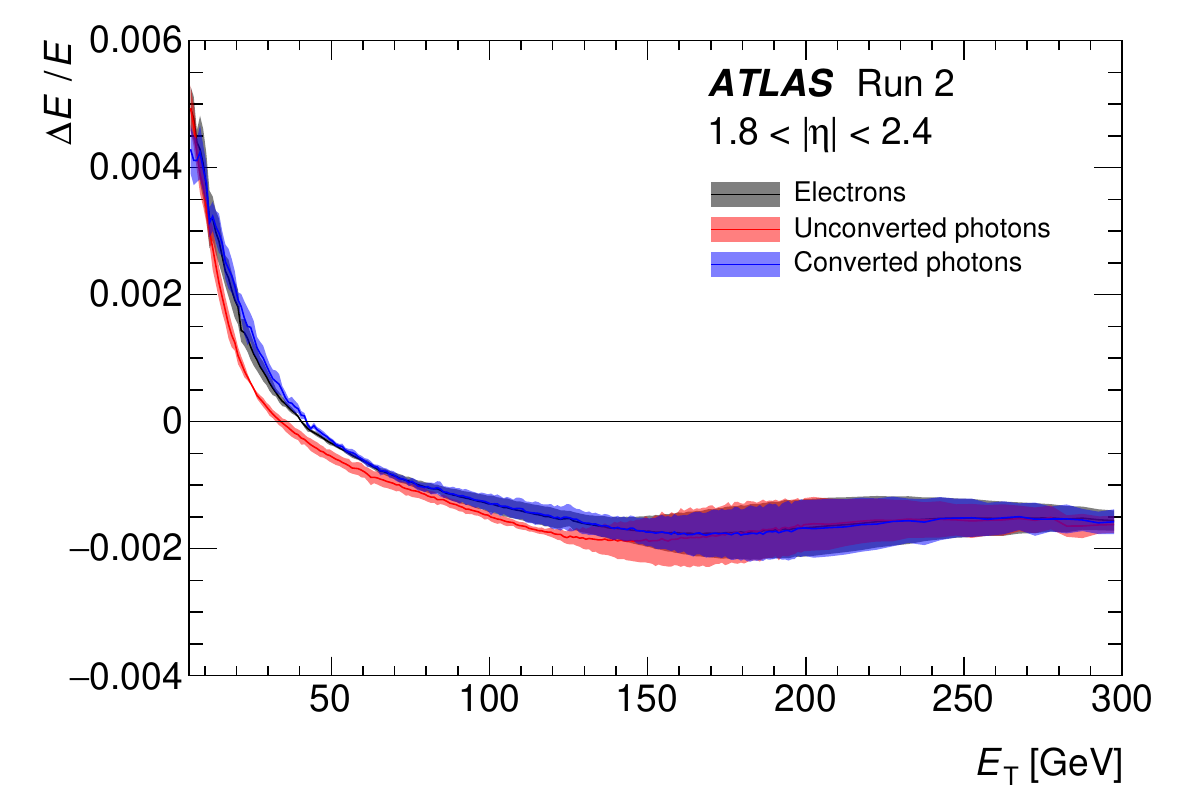}}
\caption{Relative cluster energy correction as a function of
\et, for (a) $|\eta| < 0.8$ and (b) $1.8 < |\eta| < 2.4$
for electrons (black), unconverted photons (red) and converted photons (blue).
For each particle type, the size of the envelope reflects the dependence of the correction on $\eta$, at a given \et.
}
\label{fig:adccorrImp}
\end{figure*}
 
In addition, non-linearities in the ADC-to-current conversion affect comparisons of the energy response in different readout gains. For a given energy, the measured ADC count and the residuals of the current conversion depend on the gain in which the cell is recorded, so that a direct comparison of the energy response in HG and MG is difficult to interpret. The ADC non-linearity correction described above removes this bias. The energy measurement in high and medium gains can then be compared directly, as described in the next section.

\subsection{Energy response in high and medium gain}
\label{sec:L2}
Non-linearity in the cell energy measurement introduces a dependence of the energy response on the energy of the reconstructed particle. As discussed above, the linearity of the readout electronics is better than a few per mille in each of the three gains used to digitize the calorimeter signals. The relative calibration of the high, medium and low readout gains is assumed to be perfect in the simulation, but is less well understood in data~\cite{PERF-2013-05,PERF-2017-03}. The HG and MG readouts are most relevant for the calibration studies presented in this paper, and discussed below.
 
In the standard configuration, the HG readout is used for the majority of cells in clusters from electrons in \Zee decays, especially in the barrel, where the transition to medium gain occurs for second-layer cell energies above 25~\gev, corresponding to electrons of \et = 50--60 \GeV. The relative calibration of the HG and MG readouts is studied using 0.3~\ifb of data recorded in 2017 and 2018 using special conditions. For these data, the threshold to switch from HG to MG readout for the cells in the second layer was lowered, typically by a factor of three, such that almost all electrons from $Z$-boson decays have at least the highest-energy cell in Layer~2 recorded in MG. The reconstructed dielectron invariant mass distribution in the special runs is compared with the distribution observed in 1.5~\ifb\ of data with the standard gain transition, and recorded around the same time. The procedure described in Section~\ref{sec:adclin} is used to calibrate the ADC-to-current conversion function in both gains. An example comparison is shown in Figure~\ref{fig:l2gaina} for electrons with $|\eta|<0.8$. The invariant mass distributions built in standard runs are parameterized with a sum of three Gaussian functions in $(i,j)$ categories, where $i$ and $j$ label the five $|\eta|$ regions\footnote{The definition of these regions is motivated by the calorimeter geometry.} where the electrons of the pair are reconstructed. The same shapes are used to describe the corresponding distributions for the data taken in the special runs, but with the means and widths of the Gaussian functions modified by a multiplicative factor $\sqrt{(1+\ag[,i])(1+\ag[,j])}$, where $\ag[,i]$ is the energy scale difference between the two datasets in the $|\eta|$ region $i$.
These \ag[,i] parameters are then extracted from a simultaneous fit of all $(i,j)$ regions.
 
The measured values of \ag are shown in Figure~\ref{fig:l2gainb} as a function of $|\eta|$. The uncertainties shown are statistical only; systematic uncertainties are negligible for this study. For perfectly intercalibrated HG and MG readouts, $\ag=0$. Instead, small but significant differences are observed, especially for $0.8<|\eta|<1.37$. Compared to the previous analysis of this effect in Ref.~\cite{PERF-2017-03}, \ag is found to be smaller by about a factor of two; this change is mostly driven by the improved ADC calibration using a residual correction discussed in Section~\ref{sec:adclin}.
 
The relative difference between the energy responses in HG and MG can be written as a function of the particle type, \et and $\eta$ as follows:
\begin{equation}
\frac{\Delta E}{E} = \ag(\eta) \cdot \frac{1}{\delta_Z(\eta)} \cdot \delta_{\mathrm{MG}}^{e,\gamma}(\eta,\et),
\label{eq:gainsyst}
\end{equation}
where
\begin{itemize}
\item $\delta_Z(\eta)$ quantifies the fractional change in energy for electrons from $Z$-boson decays between the data with lower and standard thresholds for a given change in the energy recorded in MG. This sensitivity factor is about 0.3 to 0.4 (0.2 to 0.25) in the barrel (endcap) calorimeter. It takes into account the fact that only a fraction of the electron energy is recorded in MG in the special-settings data taking, while in data recorded with normal settings  some cells in the second layer can be read out in MG. This is particularly true for the endcaps, where the electron energies are larger.
\item $\delta_{\mathrm{MG}}^{e,\gamma}(\eta,\et)$ quantifies, for a given particle, the fractional change in total energy for a given change in the energy recorded in MG, for standard gain thresholds. It is estimated using simulated single-particle samples. It is close to zero up to \et = 50 \gev\ for electrons and \et = 40 \gev\ for photons, and rises to reach a maximum value of about 0.7 for $\et\sim 400$~\gev in the barrel, and of about 0.5 for $\et\sim 200$~\gev in the endcap. Above these values, $\delta_{\mathrm{MG}}^{e,\gamma}(\eta,\et)$ decreases as the low-gain readout starts contributing.
\end{itemize}
 
The gain dependence of the energy response, $\Delta E / E$, is considered as a systematic uncertainty in the energy measurement; the size of the uncertainty is defined as the full size of the observed dependence. The effect increases with energy, and typically reaches 0.1\% in the barrel, and 0.4\% in the endcaps, with low dependence on the particle type.
 
For completeness, a corresponding, independent uncertainty is defined for the transition between the MG and low-gain (LG) readouts. The uncertainty is defined based on Eq.~(\ref{eq:gainsyst}), adapted using a dedicated parameterisation $\delta_{\mathrm{LG}}^{e,\gamma}(\eta,\et)$. It has a sub-dominant impact for the energy range considered in the present analysis.
 
\begin{figure*}[h]
\centering
\subfigure[]{\includegraphics[width=.49\textwidth]{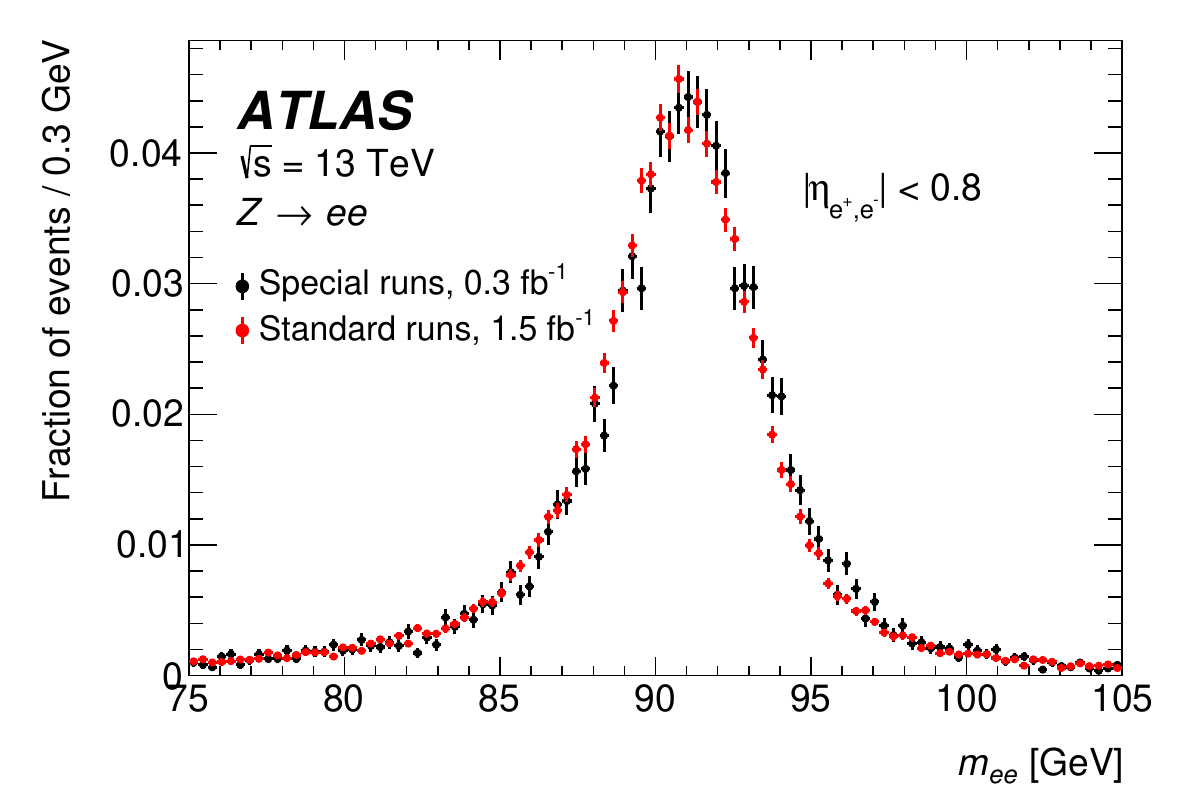}\label{fig:l2gaina}}
\subfigure[]{\includegraphics[width=.49\textwidth]{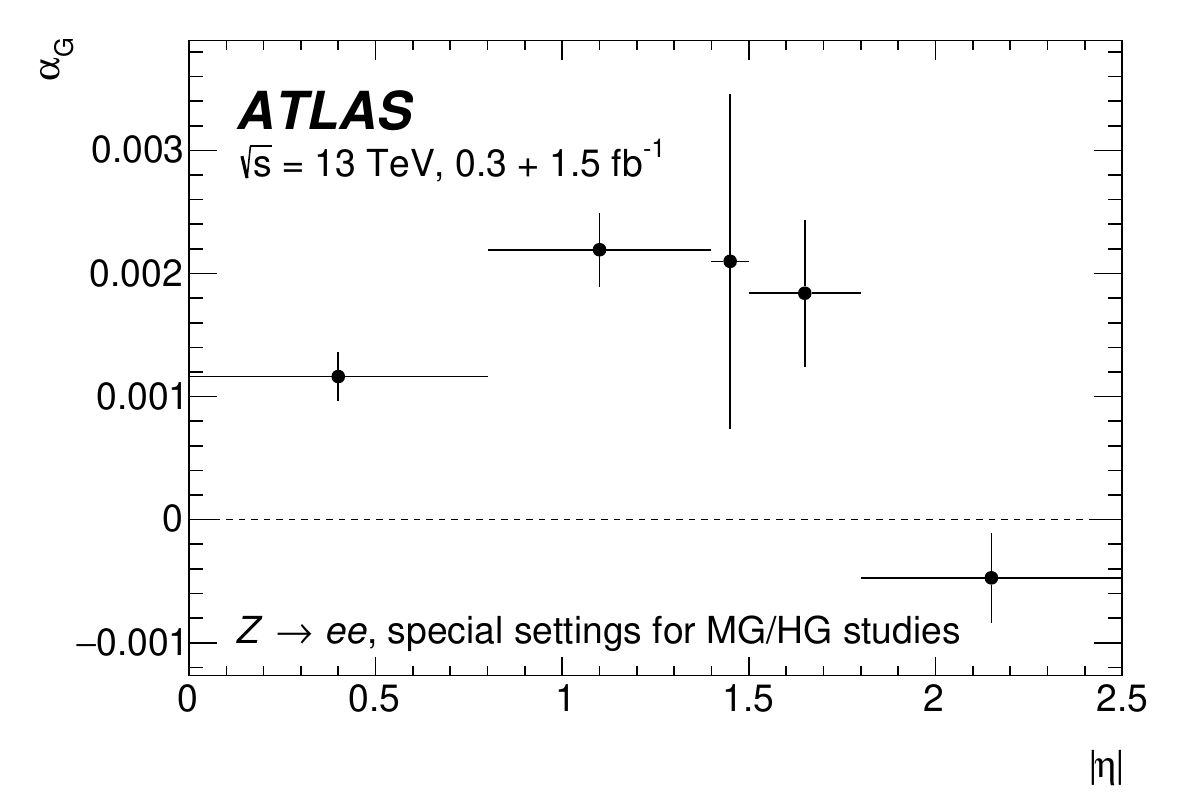}\label{fig:l2gainb}}
\caption{(a) Example dielectron invariant mass distributions, one from events collected in special runs and the other from standard runs,
and (b) corresponding energy scale factors and their statistical uncertainties as a function of $|\eta|$.}
\end{figure*}


\section{Intercalibration of the EM calorimeter layers}
\label{sec:layerInterCal}

\subsection{Presampler energy scale}
\label{sec:psscale}
The presampler energy scale \aps is defined as the ratio of the presampler energies in data and simulation.
The analysis presented in Refs.~\cite{PERF-2013-05,PERF-2017-03} used samples of electrons
and photons to measure \aps. However, the interplay between \aps and the amount of material in front of the
presampler produces a strong correlation between the corresponding uncertainties. In this paper, the determination of \aps is performed using
muon candidates selected in the low-\pileup data sample. The muon
energy deposits are insensitive to the material in front of the presampler, and therefore provide a direct measurement of the
presampler energies. A high-purity sample of $W \to \mu\nu$ and $Z \to \mu\mu$
events is selected by requiring one or two isolated muons, with additional criteria for the transverse energy and mass, and in the $Z$-boson case, the dimuon mass.
 
Although the signal-to-noise ratio for a typical energy deposit from a muon in a presampler cell
is rather low, the near absence of \pileup in low-\pileup data ensures it is still significant if the sample is large enough.
Figure~\ref{fig:pscalib1} shows examples of muon energy deposits
in data and MC samples. The noise distributions, as measured by the energy deposits in the neighbouring cells located
at $\Delta\eta = \pm 0.025$ from the crossed cells, are also shown for the simulation. The peak position in data and simulation matters for this study, while differences in the width, as observed in the endcaps, have no impact. The mean energy
deposits measured in a cell are $\sim 45, 100$ and 75~\mev\ at $\eta \sim 0, 1.2$ and 1.6, respectively.
\begin{figure*}[h]
\centering
\subfigure[]{\includegraphics[width=.49\textwidth]{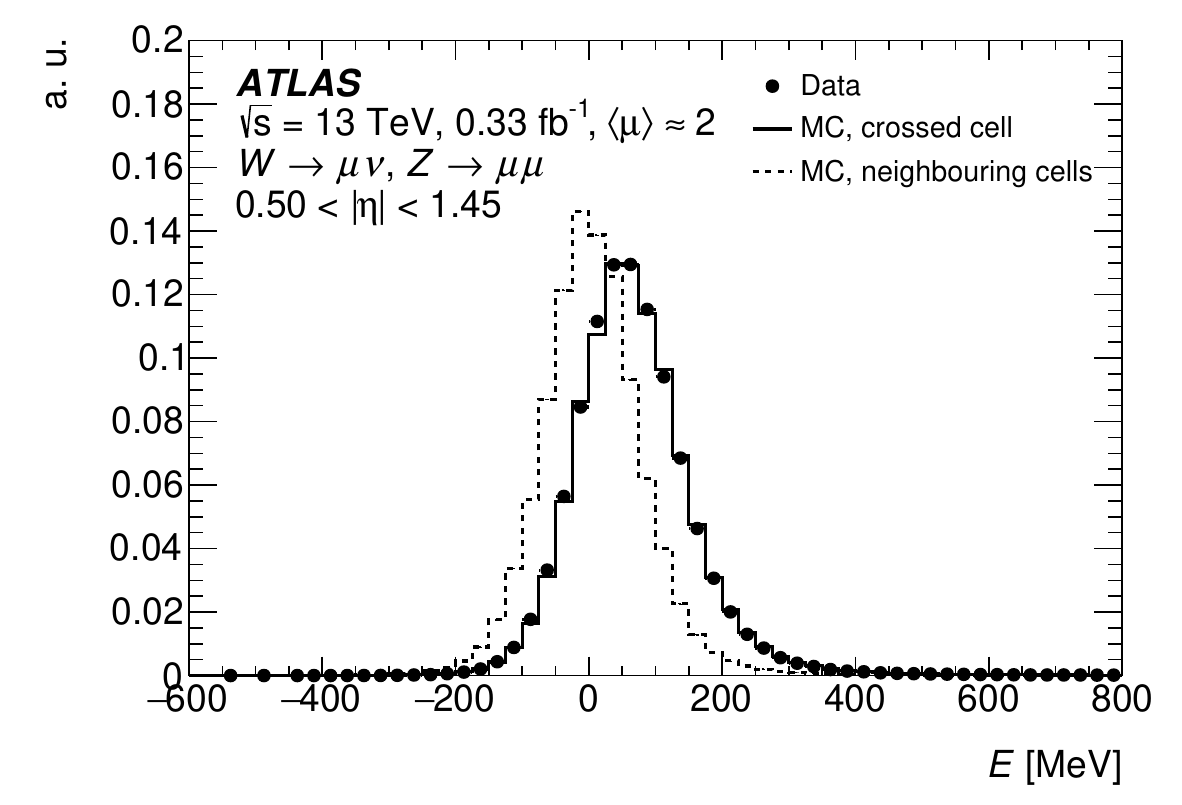}}
\subfigure[]{\includegraphics[width=.49\textwidth]{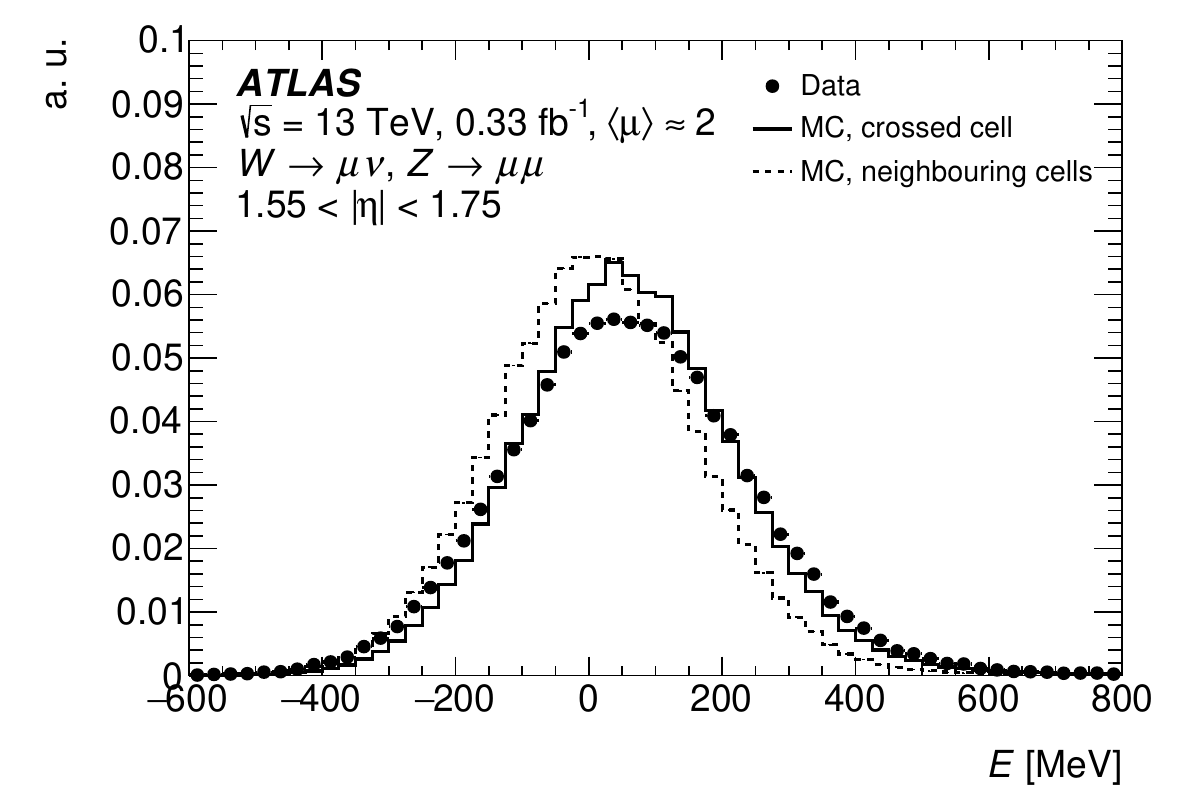}}
\caption{Distributions of muon energy deposits in  presampler
cells traversed by muon tracks, in two $|\eta|$ regions: (a) barrel and (b) endcap.
The corresponding distributions for the neighbouring cells in $\eta$
are shown for the simulation by the dashed histograms.}
\label{fig:pscalib1}
\end{figure*}
 
The mean values of these distributions are extracted within the interval $[-1.6, 1.6]$~\gev\
for data and simulation, and \aps is determined from their ratio. The measurement is performed in nine
pseudorapidity bins corresponding to the size of the presampler modules ($\Delta\eta~=~0.2$ up to
$|\eta| = 1.4$, $\Delta\eta~=~0.12$ in the last barrel module and $\Delta\eta~=~0.3$ in the endcaps).
 
Systematic uncertainties in the measurements of \aps are estimated as the envelope of the
difference between the nominal \aps values and alternative measurements obtained by
1) varying the muon candidate selection, i.e.\ selecting candidates crossing a cell close to its centre; 2) varying the
presampler  energy definition, i.e.\ using a cluster of three cells in $\eta$ instead of a single cell; 3)
subtracting a pedestal energy estimated from neighbouring cells in $\phi$; and 4) varying the interval
used to determine the mean of the $E_{\text{PS}}$ distributions. These changes contribute equally to the total systematic uncertainty. At a given $|\eta|$, the values of \aps are found to be compatible for positive and negative pseudorapidity, and are averaged. The results are shown in Figure~\ref{fig:pscalib2}.
The statistical uncertainties are at the percent level, while the systematic uncertainties vary between 2\% and 4\% depending on $|\eta|$. The results obtained here using muons are compatible with the earlier electron- and photon-based measurements.
 
\begin{figure*}[h]
\centering
\includegraphics[width=.49\textwidth]{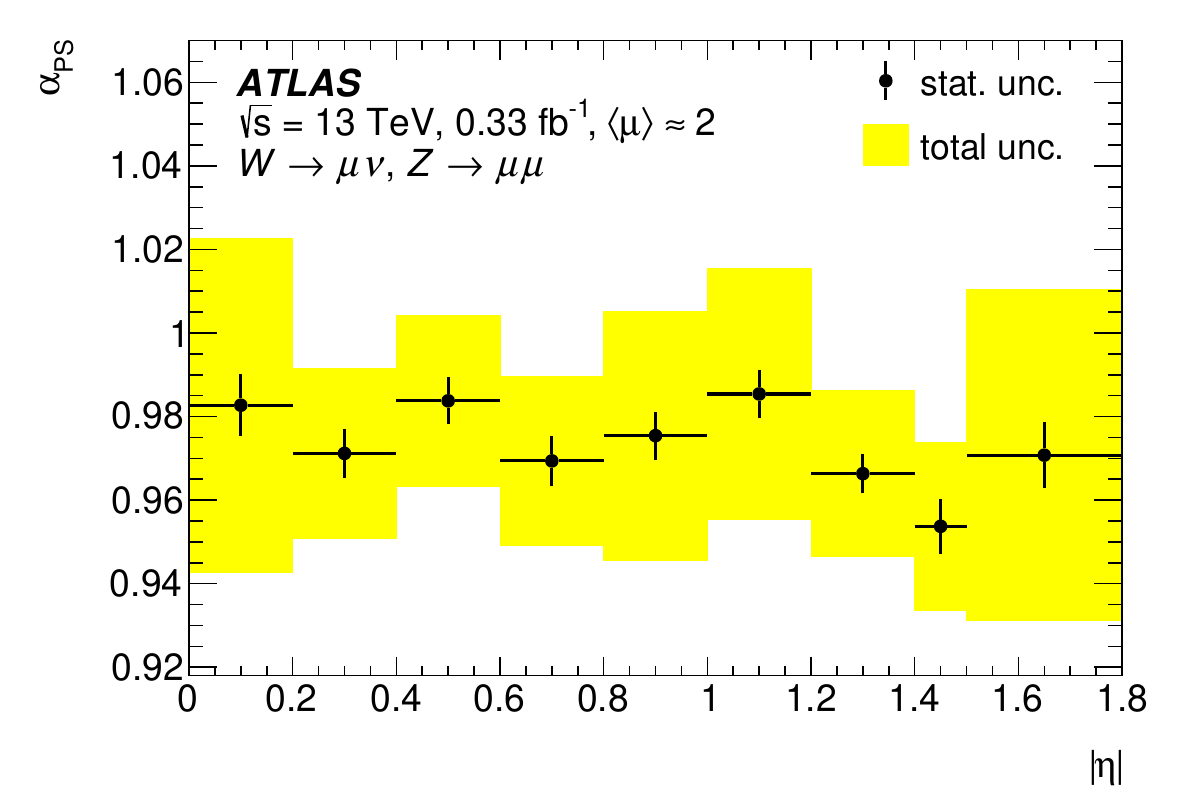}
\caption{Measured presampler energy scale \aps as a function of
$|\eta|$. The error bars represent the statistical uncertainty and
the yellow band shows the total uncertainty.}
\label{fig:pscalib2}
\end{figure*}


\subsection{Intercalibration of the first and second calorimeter layers}
\label{sec:e1e2}
The intercalibration of the first and second calorimeter layers is paramount in controlling
the linearity of the electron and photon responses. Differently from previous iterations, both the muon and electron candidates
from $Z$-boson decays are used here to measure the relative calibration of the two layers.
 
\subsubsection{Intercalibration using muons}
The measurements using muon candidates from $Z$-boson decays closely follow the analysis in
Ref.~\cite{PERF-2017-03}. The intercalibration factor is defined as
$\alot = (\avg{E_1}/\avg{E_2})^{\text{data}} / (\avg{E_1}/\avg{E_2})^{\text{MC}}$
where $\langle E_i\rangle$ is the mean value of the distribution of the energy
deposited in layer $i$. Two methods differing in the estimation of $\avg{E_i}$ are considered:
the \emph{most probable value} method extracts it from a fit of a Landau distribution
convolved with a noise template. The \emph{truncated mean} (TM) method uses the mean of the
distribution computed over a restricted window to minimize the sensitivity to the tails. In both methods,
these quantities are measured in intervals of \avgmu and the final \alot measurement is obtained by
extrapolating linearly to $\avgmu = 0$. The extrapolation parameters are determined from the measurements
in the interval $\avgmu \in [10,40]$. In the TM method, the extrapolation was validated by comparing
the nominal results with the \alot\ value estimated with the low-\pileup data sample, for which muon candidates from $W$-boson decays are considered in addition to candidates from $Z \to \mu\mu$ decays. This is illustrated for
$0.3 < |\eta| < 0.4$ in Figure~\ref{fig:TMa12a}; a comparison of the measurements
using the standard data sample and the low-\pileup one is shown in Figure~\ref{fig:TMa12b}.
For both methods, calibration uncertainties are derived by varying the muon selection cuts and the fitting or averaging ranges, as described in Ref.~\cite{PERF-2017-03}.
Agreement between the two measurements is excellent and well within the total uncertainty of
the nominal result.
The total uncertainty in the muon measurements varies from about 0.7\% in the barrel to about 2\% in the endcaps.
 
\begin{figure*}[h]
\centering
\subfigure[]{\includegraphics[width=.49\textwidth]{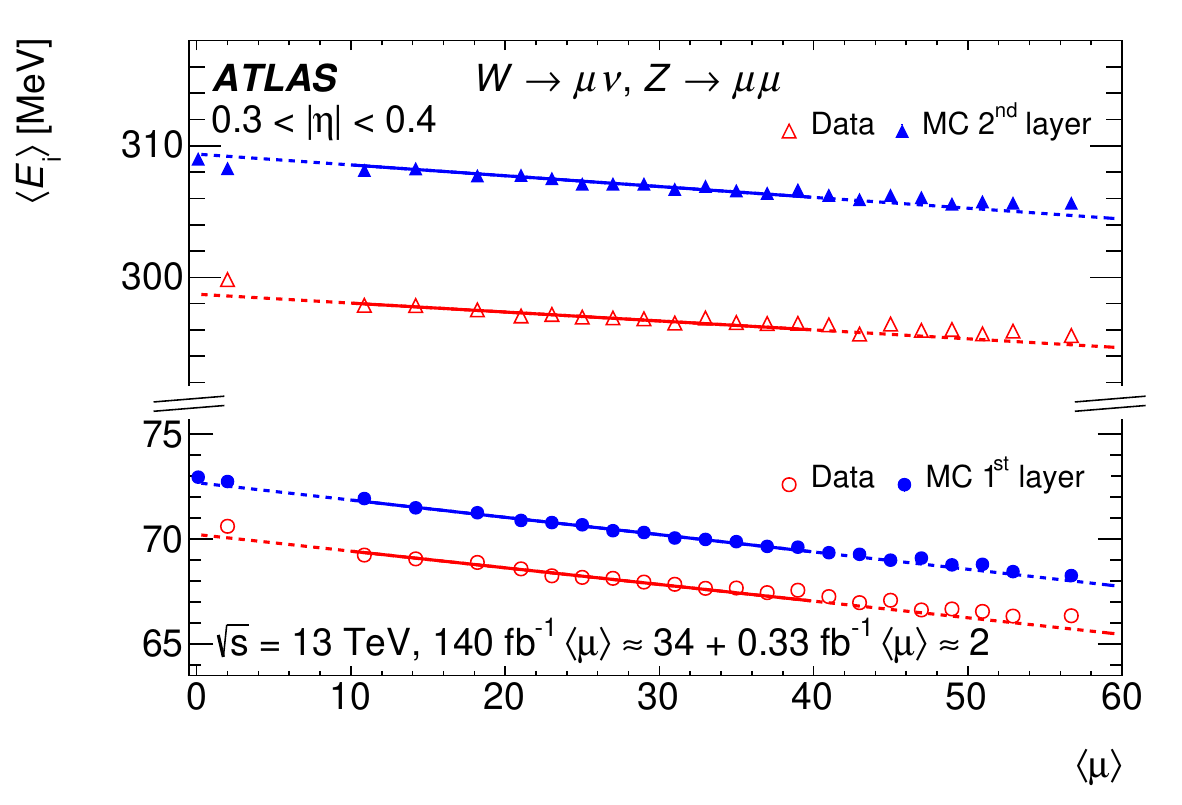}\label{fig:TMa12a}}
\subfigure[]{\includegraphics[width=.49\textwidth]{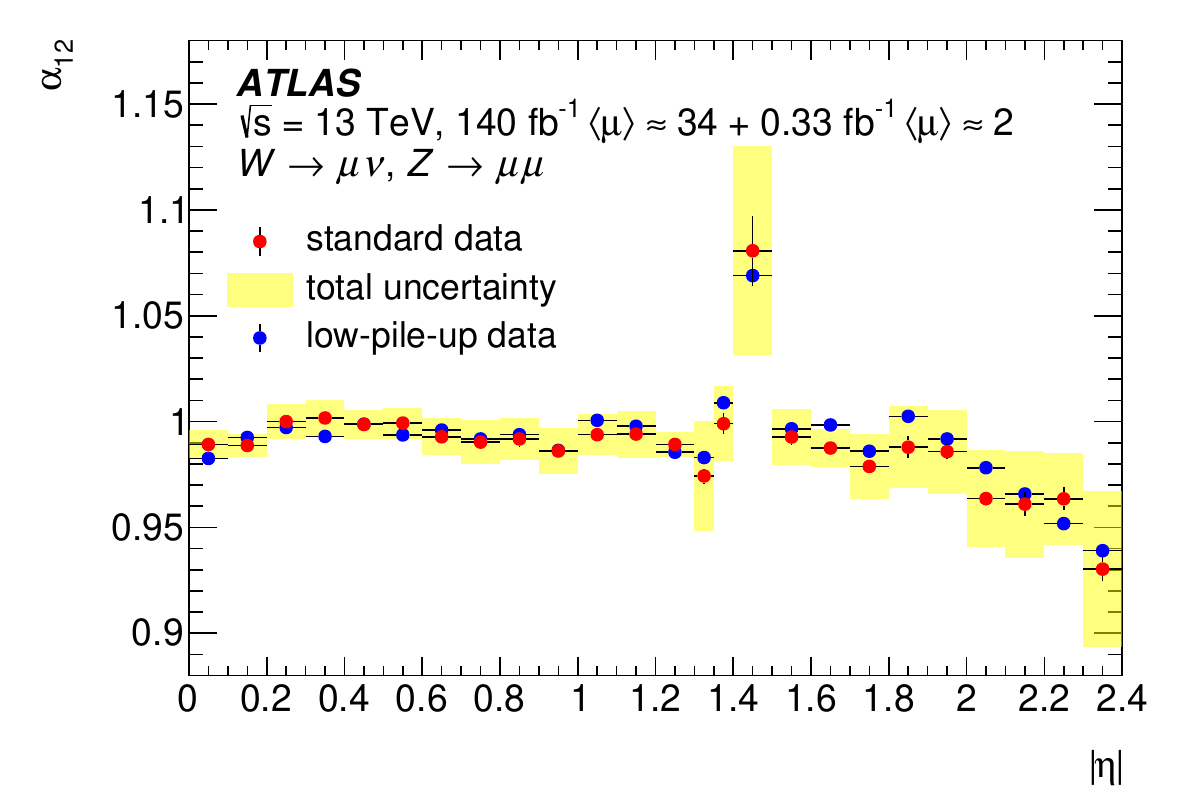}\label{fig:TMa12b}}
\caption{(a) Evolution of the truncated mean of the muon energy deposit distribution for $0.3 < |\eta| < 0.4$,
for the first (dots) and second (triangles) calorimeter layers, in data (open symbols) and in simulation (full symbols),
as a function of the average number of \pileup interactions per bunch crossing \avgmu. The values obtained from the low-\pileup
samples are also shown, together with the values extracted from a MC sample without \pileup. The lines show the result of linear
fits to the points for $\avgmu \in  [10, 40]$ and the dotted lines show the extrapolation to lower and higher \avgmu.
(b) Comparison of the evolution with $|\eta|$ of \alot\ obtained from extrapolation at $\mu=0$, for the low-\pileup and standard data samples,
for the TM method. The yellow band shows the total uncertainty of the nominal results, obtained from the standard dataset. The error bars on the data points are statistical only.}
\end{figure*}

\subsubsection{Intercalibration using electrons}
In the previous analyses~\cite{PERF-2013-05,PERF-2017-03}, electron probes were used only as a cross-check
of the layer intercalibration performed with muons. From the few discrepancies between electron and muon correction factors
observed in some pseudorapidity regions, systematic uncertainties were derived and applied to the muon result.
The method first described in Ref.~\cite{PERF-2013-05} was revisited with the full \RunTwo
data sample in order to better constrain the layer intercalibration using electrons, and to combine the electron-based measurement with the result obtained from muons.
 
The dependence of the energy response on the depth of the EM shower allows a direct extraction of \alot. This parameter is determined in 25 bins of $|\eta| \in [0,2.4]$, using the ratio of the measured energy $E$ and momentum $p$ of electron candidates, $E/p$, and the distribution of the invariant mass of electron pairs, $m_{ee}$.
The $E/p$ distribution is affected by bremsstrahlung from interactions between the electrons and the material in front of the calorimeter, and is modelled using a Crystal~Ball distribution~\cite{CB} in the range $0.9 < E/p < 1.3$. For this method, only the region $|\eta|< 1.37$ is considered since the ID momentum resolution deteriorates rapidly at large $|\eta|$. The invariant mass distribution is fitted in each ($|\eta|, E_1/E_2$) bin\footnote{Each $m_{ee}$ value is considered in the two relevant ($|\eta|, E_1/E_2$) bins, where $E_i$ is the cluster energy contained in  Layer $i$.}, in the range $[80, 100]$~\gev, using the convolution of a Crystal~Ball function and a Breit--Wigner distribution, while the small background in data is
modelled by a second-order Chebyshev polynomial. In both the $E/p$ and $m_{ee}$ methods, the most probable value of the fitted Crystal~Ball function is used as the estimator.
Examples of fits to the dielectron invariant mass and the $E/p$ distributions are illustrated in Figure~\ref{fig:meeoreopShapefit}.
\begin{figure*}[h]
\centering
\subfigure[]{\includegraphics[width=.49\textwidth]{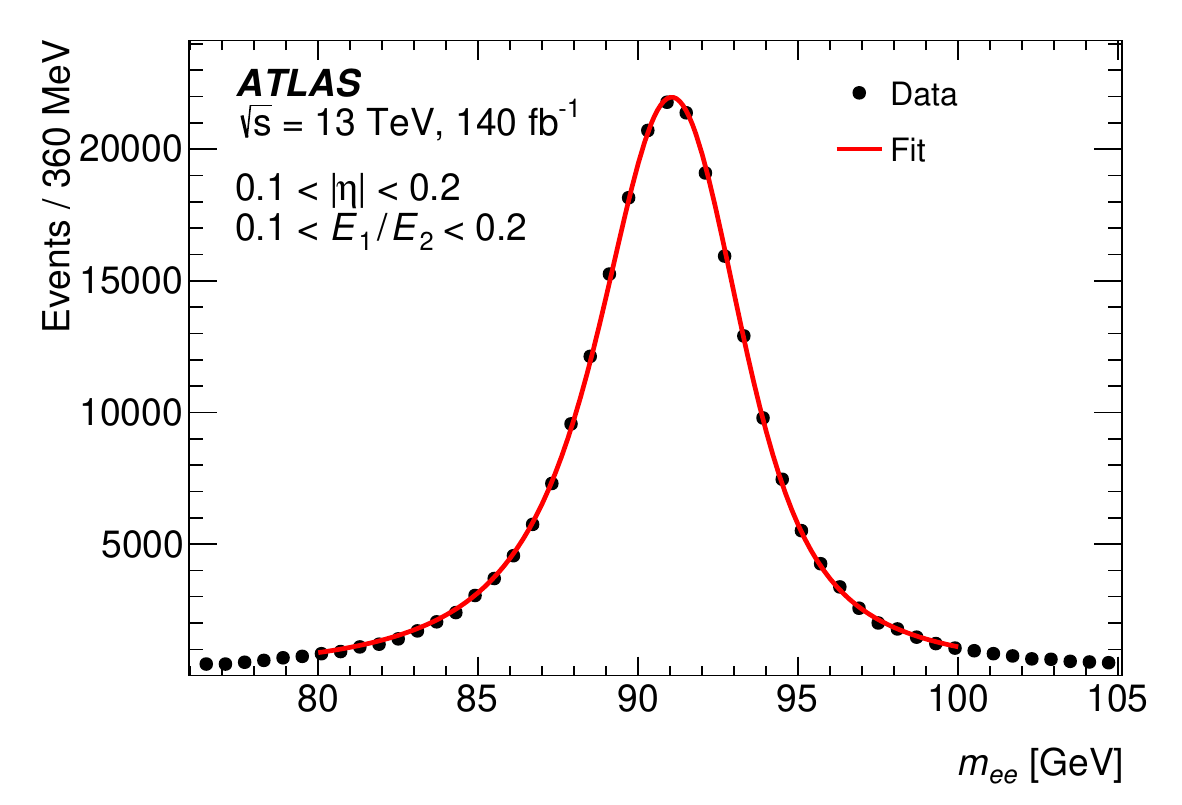}}
\subfigure[]{\includegraphics[width=.49\textwidth]{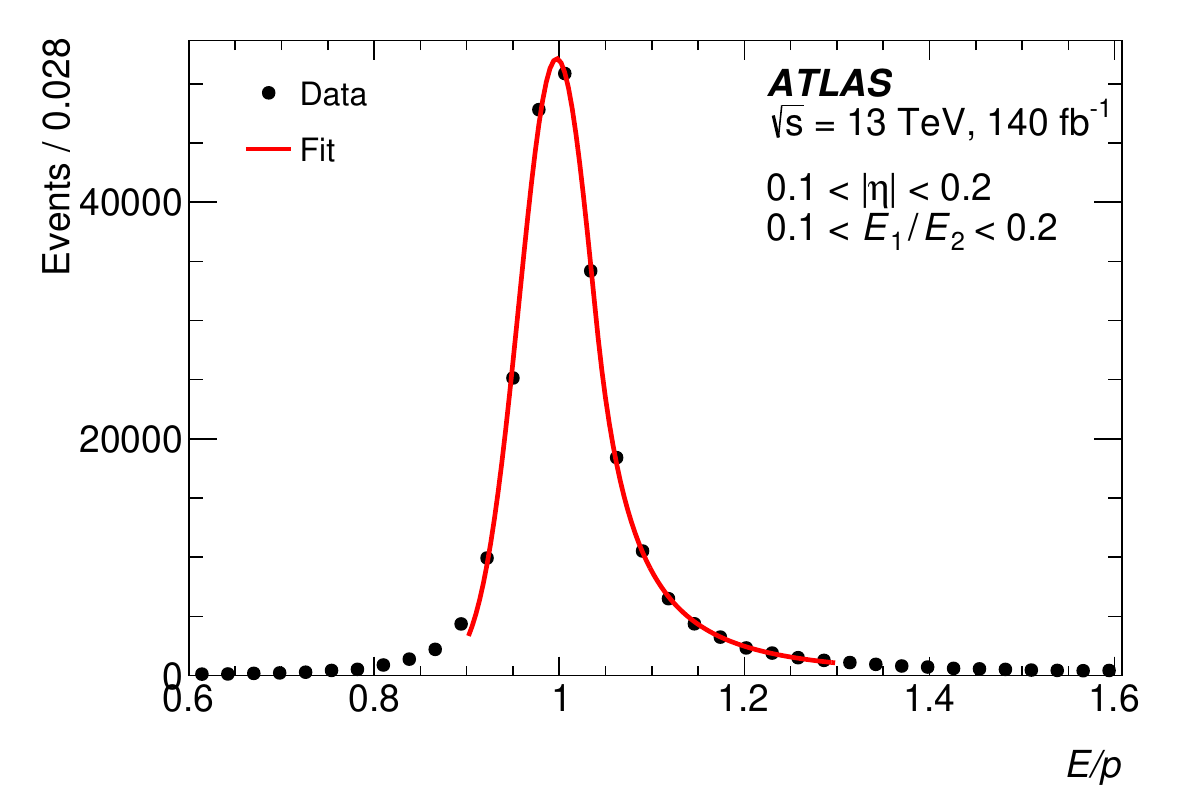}}
\caption{Distributions of (a) the dielectron invariant mass and (b) electron $E/p$ for
electrons with $0.1 < |\eta| < 0.2$ and $0.1 < E_1/E_2 < 0.2$. The fitted functions used
to estimate the most probable value are superimposed.}
\label{fig:meeoreopShapefit}
\end{figure*}

In a given $|\eta|$ bin, the dependence of the $E/p$ and $m_{ee}$ estimators on $E_1/E_2$ is determined in data and in simulation.
For a perfect layer intercalibration, the ratio of the estimator in data and the nominal simulation is expected to be constant, and any
mis-calibration induces a slope. A constant data-to-simulation ratio is recovered by rescaling $E_1$ in data and recomputing
$m_{ee}$ and $E/p$ accordingly, adjusting \alot\ to minimize the deviation of the ratio from a horizontal line.
Figures~\ref{fig:l1l2eleca} and~\ref{fig:l1l2elecb} show the dependence of $m_{ee}$ and $E/p$ on $E_1/E_2$ for data and simulation,
for $|\eta| \in [0.1, 0.2]$ ($m_{ee}$) and $|\eta| \in [1.0, 1.1]$ ($E/p$). Data-to-simulation ratios are shown in the bottom panels.
For $|\eta| \in [0.1, 0.2]$, the uncorrected ratio is compatible with a constant, indicating a good layer intercalibration, while for $|\eta| \in [1.0, 1.1]$,
a constant ratio is recovered after rescaling $E_1$ by 0.97. The measurements of \alot\ using $m_{ee}$ or $E/p$ are compatible in the whole $|\eta|$ range.
 
\begin{figure*}[h]
\centering
\subfigure[]{\includegraphics[width=.49\textwidth]{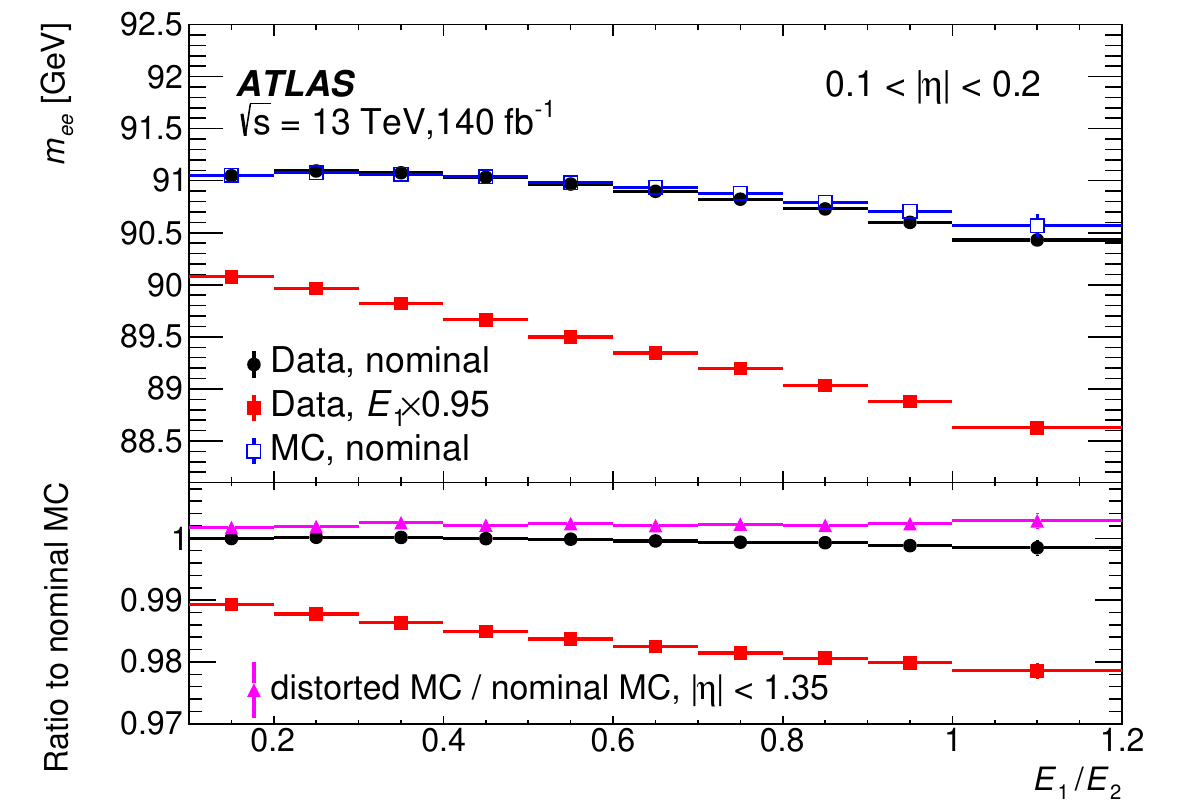}\label{fig:l1l2eleca}}
\subfigure[]{\includegraphics[width=.49\textwidth]{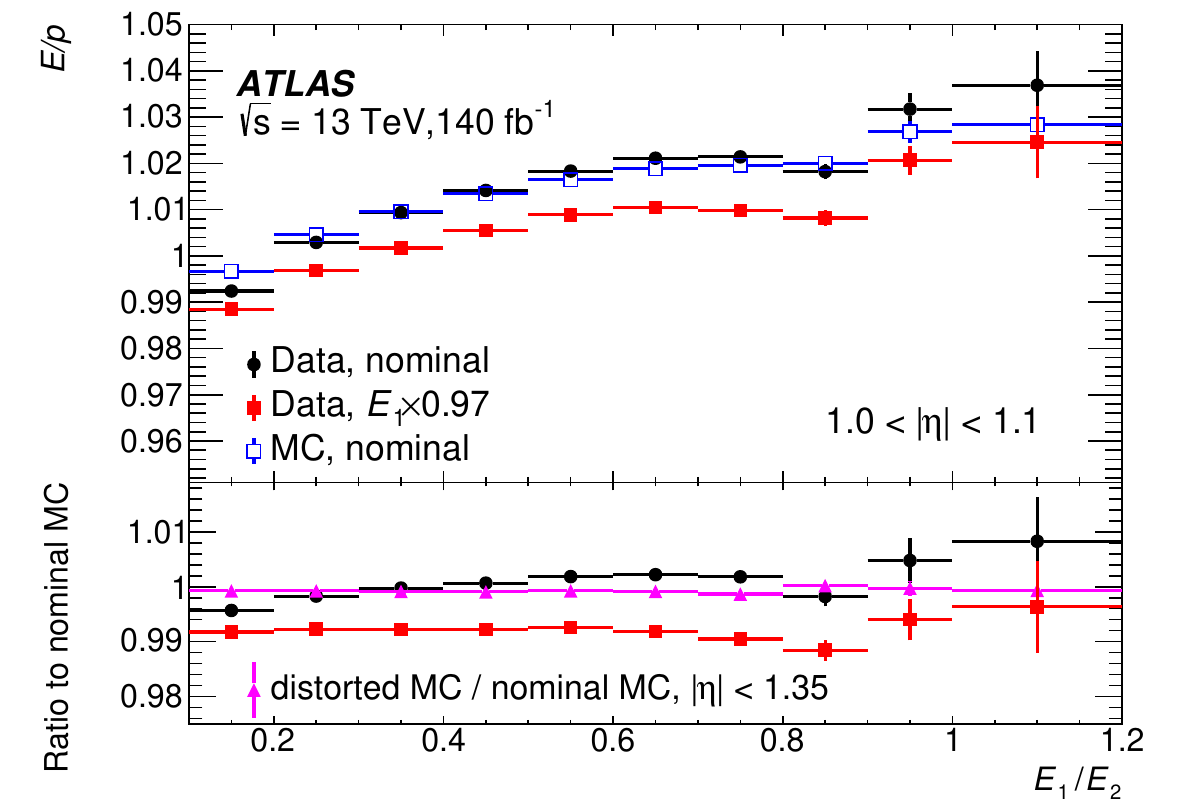}\label{fig:l1l2elecb}}
\caption{Evolution of (a) $m_{ee}$ and (b) $E/p$ estimators as a function of $E_1/E_2$ for the data (full dots), the data with a scaling of $E_1$ (red
squares), and the nominal simulation (open squares). The $m_{ee}$ and $E/p$ figures show values for electrons in the pseudorapidity bin $0.1 < |\eta| < 0.2$ and $1.0 < |\eta| < 1.1$
respectively. The bottom panels show the ratio to the nominal simulation. In addition,
the evolution for a simulation with additional material is also shown. For this sample, the range $|\eta| < 1.35$ is used.}
\label{fig:l1l2elec}
\end{figure*}
 
The following sources of systematic uncertainty are considered:
\begin{itemize}
\item The effect of uncertainties in the amount of material is estimated from simulations in which the passive material
in the barrel cryostat, the SCT and the endcaps of the TRT is increased according to the uncertainties described in Section~\ref{sec:passivemat}. The distorted MC samples are used as pseudo-data, and the deviation of \alot\ from 1 defines the associated uncertainty. As can be seen in ratio panels of Figure~\ref{fig:l1l2elec}, the method is rather insensitive to passive material variations (the constant ratio of the distorted MC to nominal MC values in the range $|\eta|<1.35$ is also observed in finer $|\eta|$ bins). The uncertainty ranges from ${\sim}0.5\%$ at low $|\eta|$ to ${\sim}5\%$ for $|\eta| \in [1.4, 1.8]$.
\item The intervals used to fit the energy response are
modified to $[70, 100]$~\gev\ or $[84, 98]$~\gev\ for the $m_{ee}$
method, and to $[0.95, 1.5]$ for the $E/p$ method. The difference between the
\alot values measured in the nominal and alternative intervals is
used as an uncertainty.
\item The uncertainty in the presampler energy scale also impacts
the measurement. The corresponding uncertainty in \alot
can be as large as 1\% at the end of the barrel for the $E/p$ method.
\item Variations of the electron identification working point,
the residual bias of the method, and the limited size of the \Zee sample
contribute with much smaller uncertainties.
\end{itemize}
The precision of the \alot measurement with electrons ranges from 0.7\% to 2\% in the barrel, and from 1.5\% to 6.2\% in the endcap calorimeters. In the barrel--endcap transition regions and in the first half
of the endcap calorimeters, the distributions in data and simulation differ significantly, regardless of \alot, leading to poor convergence
of the minimization procedure and increased uncertainties.

\subsubsection{Combination}
The two muon measurements and the two electron measurements are combined using the BLUE prescription~\cite{Valassi:2003mu}.
In each channel, the statistical correlation between the two methods is ignored, as systematic uncertainties dominate; the latter are found to be largely uncorrelated as a function of $|\eta|$.
First, the muon and electron measurements are combined separately.
The results are shown as the blue and red open dots in Figure~\ref{fig:l1l2combia}.
The uncertainty in the combined muon measurement includes the systematic uncertainty induced by the extrapolation
of this calibration to electron and photon showers~\cite{PERF-2013-05}. This uncertainty accounts for the uncertainties
in the simulation of the ionization current induced by muons, and how these uncertainties affect EM showers.
These two combined results are then combined to provide the final measurement, illustrated in Figure~\ref{fig:l1l2combia}.
When the $\chi^2$ of the combination is larger than one, the combined uncertainties are
rescaled by a factor $\sqrt{\chi^2}$. The total uncertainty is presented in Figure~\ref{fig:l1l2combib}, and varies between 0.6\%
in the central part of the barrel to 3\% at $|\eta| \sim 2.4$. The inclusion of the electron measurement allows
the uncertainty to be reduced by a factor of ${\sim}1.8$ in the first half of the barrel. In the endcaps, the combined uncertainties are dominated by the differences between the electron and muon results.
 
\begin{figure*}[h]
\centering
\subfigure[]{\includegraphics[width=.49\textwidth]{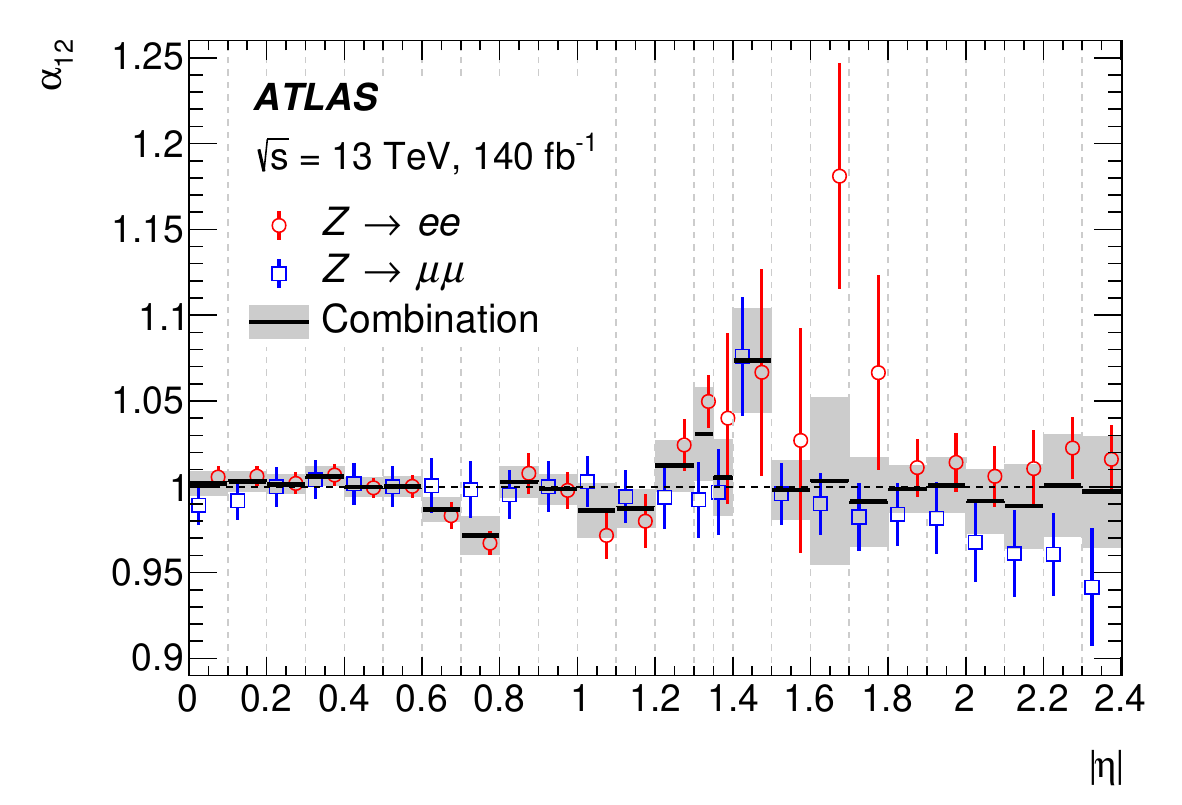}\label{fig:l1l2combia}}
\subfigure[]{\includegraphics[width=.49\textwidth]{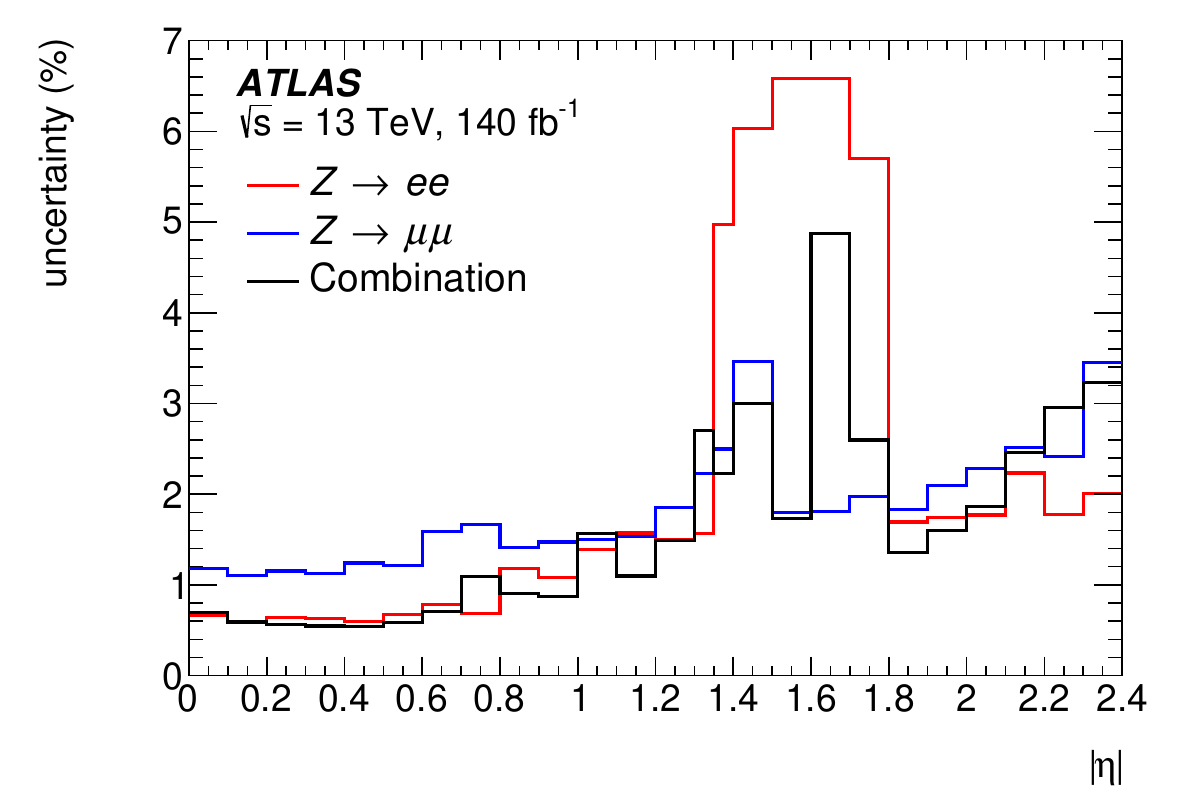}\label{fig:l1l2combib}}
\caption{(a) Relative calibration scale factor \alot of the first and second EM calorimeter layers as a function of $|\eta|$. Open blue squares correspond to the results obtained from the study of muon energy deposits in $Z \rightarrow \mu\mu$ events, combining the \emph{truncated mean} and \emph{most probable value} methods. Open red circles show the extracted combined values obtained from the study of the dependence of the dielectron invariant mass $m_{ee}$ and the $E/p$ ratio as a function of $E_1/E_2$ in \Zee events. The final scale factors, combining electron and muon results, are shown as the black solid lines. (b) Corresponding total uncertainties, taking into account rescaling in the case of large $\chi^2$.}
\end{figure*}


\section{Determination of the energy scale and resolution with \Zee events}

\label{sec:insitu}
Electron energy scale and resolution corrections are determined using electron pairs from $Z$-boson decays. For electrons reconstructed in $\eta$ regions labelled $i$ and $j$, the difference between the positions of the resonance in data and simulation is used to determine invariant mass scale corrections $\alpha_{ij}$, defined by $m_{ij}^{\mathrm{corr}} = m_{ij}/(1 + \alpha_{ij})$. Similarly, a correction to the mass resolution is parameterized as $({\sigma_m}/m)_{ij}^{\mathrm{corr}} = ({\sigma_m}/m)_{ij} \oplus c_{ij}$. The values of $\alpha_{ij}$ and $c_{ij}$ are those which give the best agreement between the invariant mass distributions in data and simulation, separately for each $(i,j)$ category.
 
The analysis in Section~\ref{sec:psscale} measures an absolute energy scale for the presampler, but the calibration procedure for the first and second accordion layers
presented in Section~\ref{sec:e1e2} only determines their relative response. Since the contribution of the third accordion layer to the energy measurement is negligible for electrons from $Z$-boson decays, the only remaining degree of freedom is the overall energy scale of the accordion calorimeter. Accordion energy\footnote{The accordion energy is the sum of the energies in the three layers of the accordion calorimeter: $E^\text{acc} = \sum_\text{layer=1}^{3} E_\text{layer}$.} scale corrections for single electrons, defined by $E^\text{acc,corr}_{i}=E^\text{acc}_{i}/(1 + \alpha^\text{acc}_i)$, are at first order related to the invariant mass scale factors $\alpha_{ij}$ following
\begin{equation*}
\alpha_{ij}= \frac{f^\text{acc}_i \alpha^\text{acc}_i+f^\text{acc}_j \alpha^\text{acc}_j}{2},
\end{equation*}
where $f^\text{acc}_i$, shown in Figure~\ref{fig:facc}, is determined from simulation. It represents the sensitivity of the total calibrated electron energy to the energy measured in the accordion calorimeter, for electrons in $\eta$ bin~$i$. It is expected to be smaller than one since part of the electron energy is deposited
in the presampler (for $|\eta| < 1.8$) and in the scintillators in the transition regions between the barrel and endcap calorimeters ($1.4 < |\eta| < 1.6$).
The accordion energy scale corrections are applied only to the data.
 
\begin{figure*}[h]
\centering
\includegraphics[width=.49\textwidth]{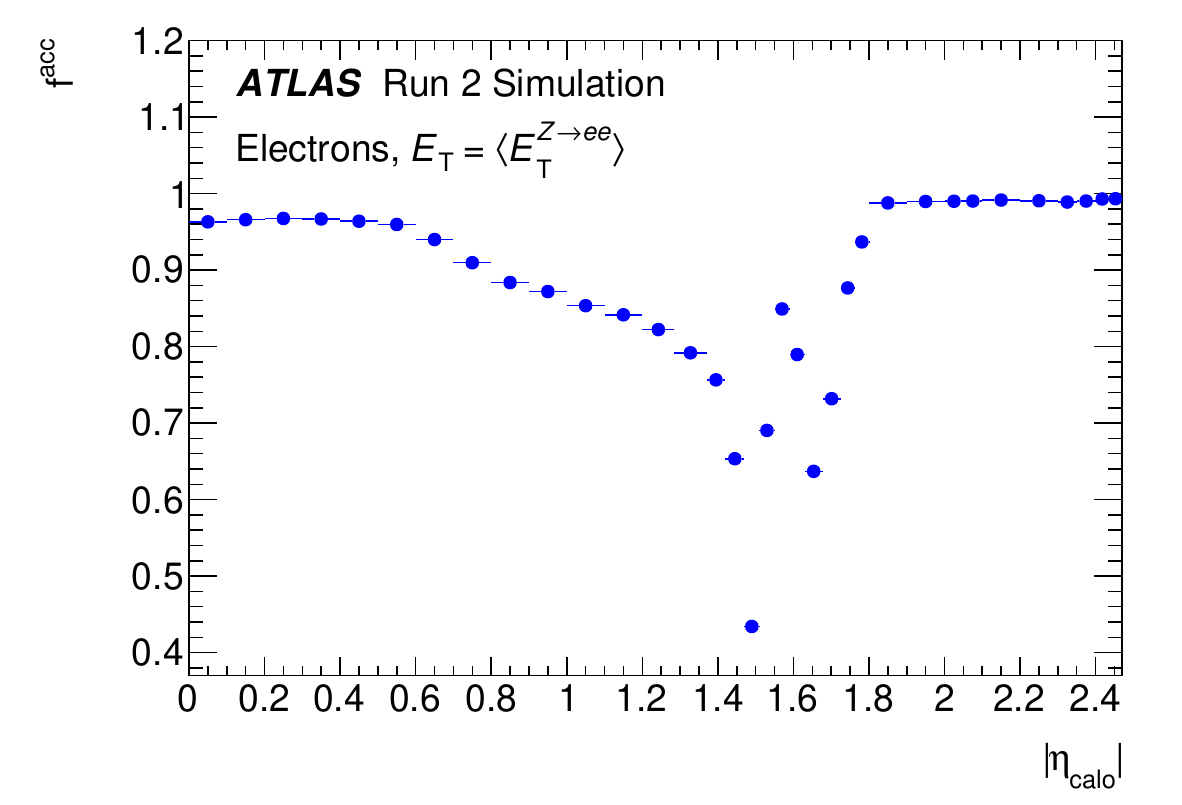}
\caption{Sensitivity of the calibrated electron energy to the energy measured in the accordion calorimeter, $f^\text{acc}$, as a function of $|\eta|$ for electrons from $Z$-boson decays.}
\label{fig:facc}
\end{figure*}
 
The invariant mass resolution correction $c_{ij}$ can be expressed in terms of single-electron energy resolution corrections, $c_i$, as $c_{ij} = (c_i \oplus c_j)/2$. The resolution correction is applied to the reconstructed energy in simulation.
 
The invariant mass window considered for the determination of $\alpha_{ij}$ and $c_{ij}$ is $80<m_{ee}<100~\gev$, and the $\alpha^\text{acc}_i$ and $c_i$ parameters are extracted from simultaneous fits of all categories. Possible biases from imperfect modelling of $Z$-boson production are removed by reweighting the simulated two-dimensional $Z$-boson transverse momentum and rapidity distribution to the data. Figure~\ref{fig:ZCalib_alphaAndc} shows the results for (a) $\alpha^\text{acc}_i$ and (b) $c_i$ derived in 68 and 24 $\eta$ intervals, respectively, separately for the 2015, 2016, 2017 and 2018 data samples. Changes in the $\alpha^\text{acc}_i$ values between successive years are mainly due to variations of the LAr temperature and the instantaneous luminosity. The temperature variations induce changes in the charge/energy collection, affecting the
energy response by about $-2\%$/K~\cite{deLaTaille:686091}. Increases in luminosity during \RunTwo imply that more energy is deposited in the liquid-argon gap, which creates a higher current in the high-voltage lines, effectively reducing
the high voltage applied to the gap and changing the response by up to 0.1\% in the endcap regions. Given the small size of the observed dependence, dedicated energy scale corrections for each data-taking year provide adequate stability for the energy measurement.
 
\begin{figure*}[htbp]
\centering
\subfigure[]{\includegraphics[width=.6\textwidth]{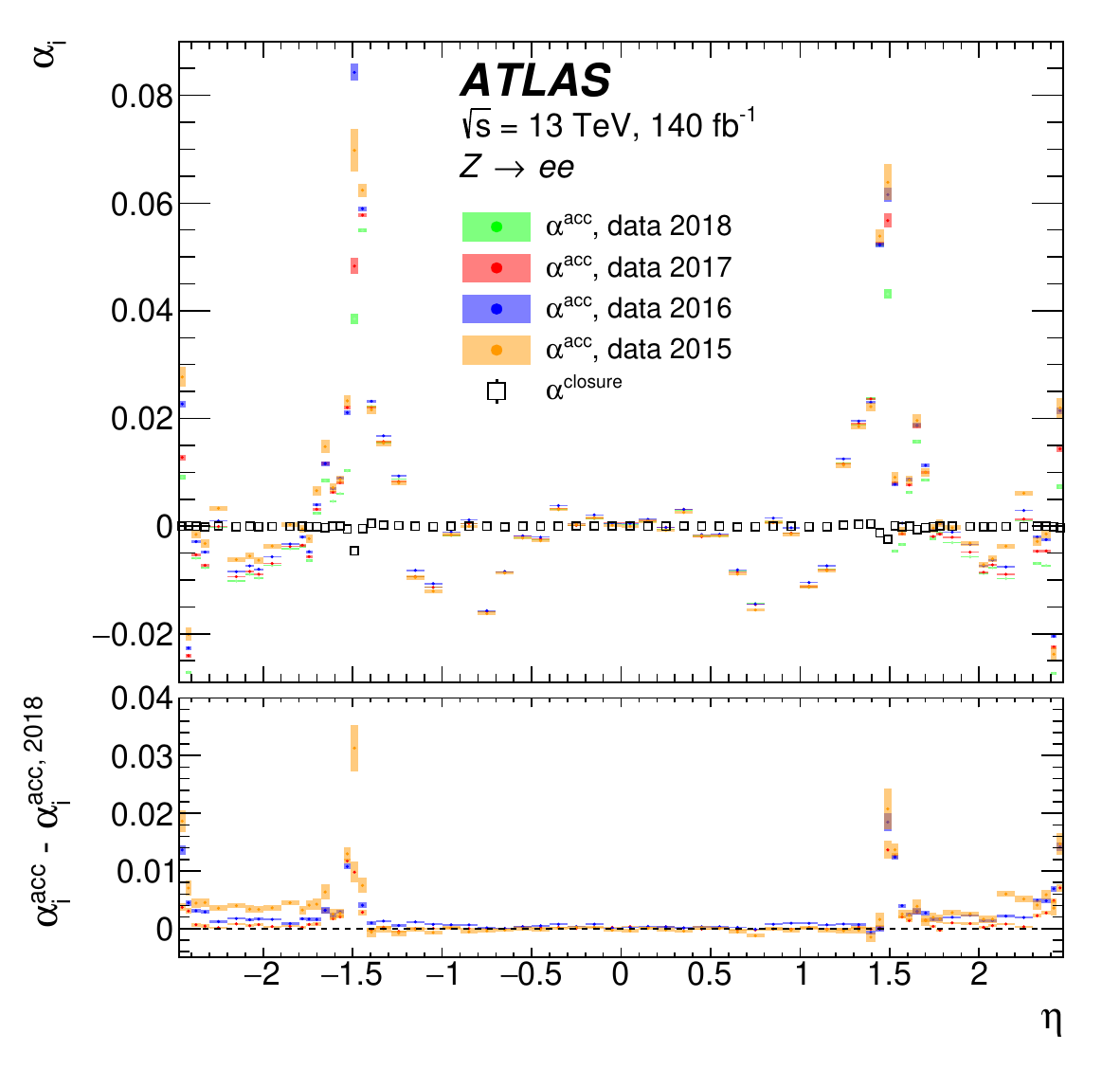}\label{fig:ZCalib_alphaAndca}} \\
\subfigure[]{\includegraphics[width=.6\textwidth]{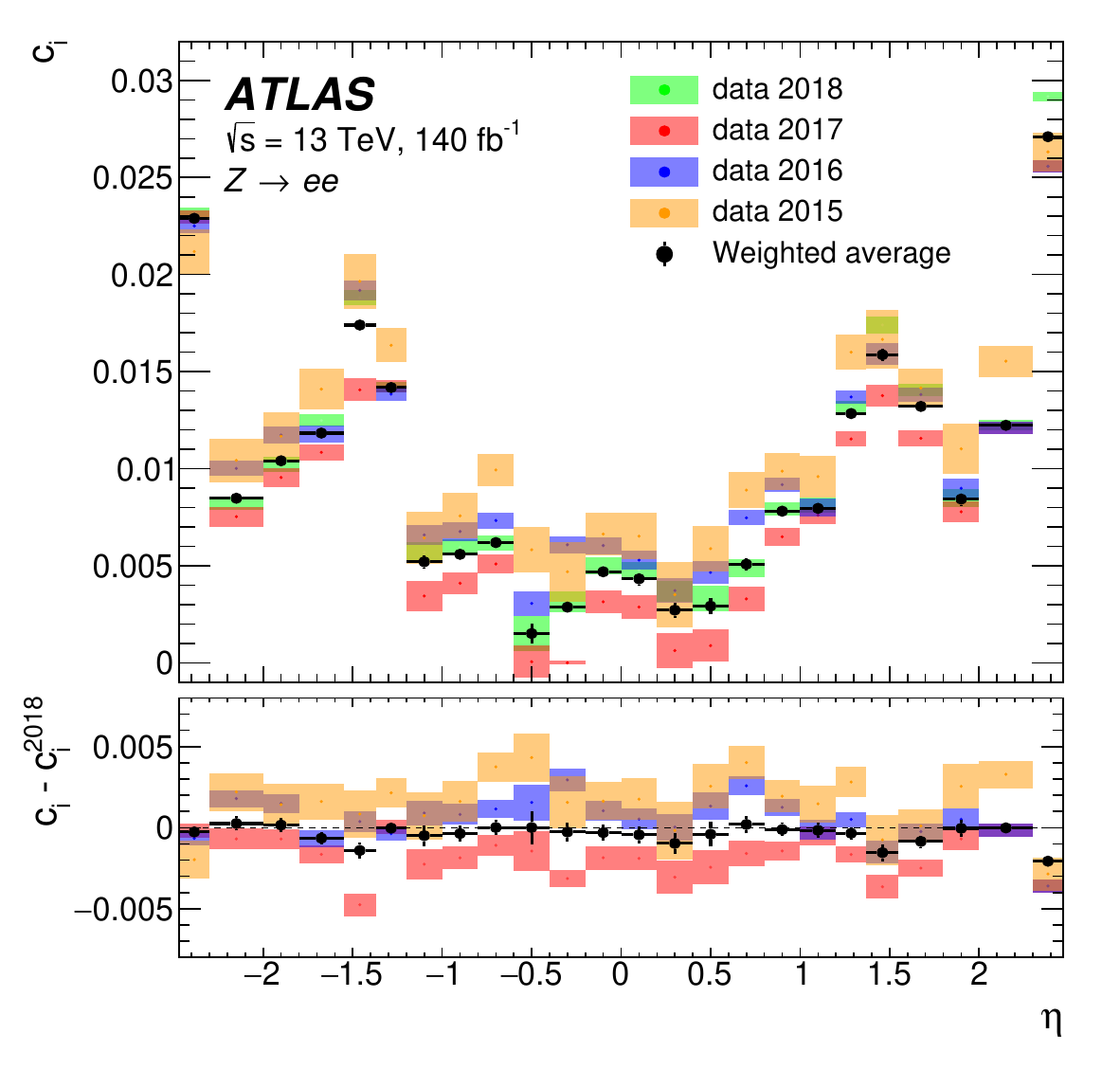}}
\caption{(a) Energy scale calibration factors $\alpha^\text{acc}_i$ and $\alpha^\text{closure}_i$, and (b) the additional
constant term $c_i$, as a function of $\eta$. The shaded areas
correspond to the statistical uncertainties. The bottom panels
show the differences between (a) $\alpha^\text{acc}_i$ and (b) $c_i$
measured in a given data-taking period and the measurements
using 2018 data.}
\label{fig:ZCalib_alphaAndc}
\end{figure*}
 
The energy scale correction is  applied as an overall correction to the energy measured in the three accordion layers, and the $Z$-based calibration fit is repeated to verify that the procedure converges. The residual energy scale factors $\alpha^\text{closure}_i$ obtained from this iteration are also shown in Figure~\ref{fig:ZCalib_alphaAndca}. Their values are below $10^{-4}$ everywhere except in the transition regions, $1.37<|\eta|<1.52$, where scintillators installed between the barrel and endcap cryostats contribute to the energy measurement but are not the subject of a specific calibration. The residual non-closure is applied as a final correction to the reconstructed energy.
 
For the constant term resolution corrections $c_i$, a dependence on the \pileup level is observed through the different values obtained from the 2015--2018 data.
The dependence of the $c_i$ values on the amount of \pileup is explained by the larger \pileup noise predicted by the simulation,
compared with that observed in the data, for a given value of $\mu$~\cite{EGAM-2018-01}.
A weighted average of the $c_i$ values for the different years is applied in the analyses of the complete dataset. The additional constant term in the energy resolution is typically less than 1\% in most of the barrel and between 1\% and 2\% in the endcaps.
 
Systematic uncertainties in the determination of $\alpha^\text{closure}_i$ and $c_i$ are assessed using variations of the event selections and fitting range. The event selection variations are a change of the electron identification criterion from Medium to Tight, removal of the isolation criterion, and addition of a cut on the momentum fraction lost by bremsstrahlung inside the ID. The mass window used for the fit is varied from $80<m_{ee}<100~\gev$ to $87<m_{ee}<94.5~\gev$. The combined effect of these variations is a systematic uncertainty of about 0.05\% in $\alpha^\text{closure}_i$ over most of the calorimeter acceptance, but reaching 0.5\% in the transition regions between the barrel and endcap calorimeters and at the edge of the calorimeter ($|\eta|>2.3$). The systematic uncertainty in the $c_i$ corrections ranges between 0.1\% and 0.2\% across the detector acceptance. The dominant contribution to the systematic uncertainty comes from the mass window variation.
 
Figure~\ref{fig:ZCalib_lineshape} shows the invariant mass distribution of \Zee\ candidates in data and in simulation
after applying the energy scale correction to the data and the resolution correction to the simulation.
Background contamination is not taken into account in this comparison, but it is expected to be no more than 1\% in the shown mass range. The uncertainty band
corresponds to the propagation of the uncertainties in the $\alpha^\text{closure}_i$ and $c_i$ factors. Within these uncertainties, the data and simulation are in fair agreement.
 
\begin{figure*}[h]
\centering
\includegraphics[width=.6\textwidth]{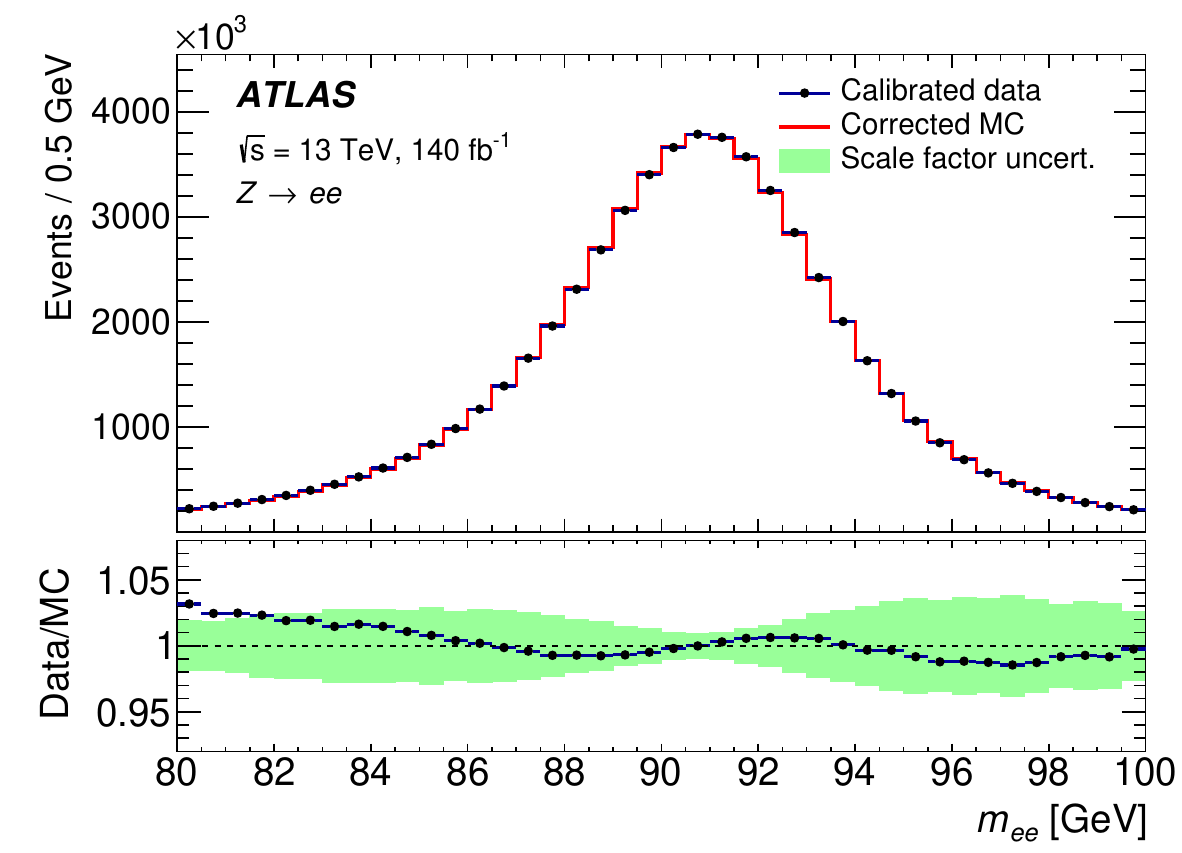}
\caption{Comparison of the invariant mass distributions of the electron pair in the selected \Zee\
candidates in data and simulation, after the calibration and resolution corrections are
applied. The total number of events in the simulation is
normalized to that in data. The ratio between data and simulation is shown in the bottom panel, where the uncertainty band represents the impact of the uncertainties in the calibration and resolution correction factors.}
\label{fig:ZCalib_lineshape}
\end{figure*}


\section{Photon-specific calibration}

\label{sec:phSpec}
 
\subsection{Modelling of the photon reconstruction classification}
\label{sec:convmis}
 
The simulation-based energy reconstruction procedure, discussed in Section~\ref{sec:overview}, is optimized and applied separately for unconverted photons and converted photons. Therefore, a difference between data and simulation in the
rates of classifying photons as converted or unconverted generates a bias in the photon energy scale. Misclassifications arise from inefficiencies in the conversion-finding algorithm and from the incorrect classification of genuine unconverted photons as converted photons by matching the cluster to \pileup-induced tracks. The rates of correct and incorrect classification are measured using a sample of photons selected from radiative $Z$ events. As described in Ref.~\cite{PERF-2017-02}, these rates are evaluated, in both data and simulation, using the ratio of the energies deposited in the first and the second layers of the calorimeter to discriminate between genuine converted and unconverted photons. The uncertainty in the energy scale is evaluated, as a function of $|\eta|$ and \et, by reweighting the conversion fractions in a sample of simulated single photons according to the values obtained from the radiative $Z$ sample in simulation and data, respectively, and it is taken to be the relative difference of the energy responses.
 
For photons with $\et = 60$~\gev, the uncertainty in the energy scale for unconverted photon candidates is about 0.02\% in the barrel and 0.02\%--0.13\% in the endcaps. For converted photon candidates, it is about 0.12\% in the barrel and smaller than 0.01\% in the endcaps. For photons with lower energy the uncertainty increases significantly: for $\et = 15$~\gev, it amounts to 0.18\% in the barrel and 0.08\%--0.67\% in the endcaps for unconverted photons. For converted photons, it becomes 0.69\%--1.31\% in the barrel and 0.01\%--0.1\% in the endcaps. This systematic uncertainty is considered as a single source, correlated between converted and unconverted photons.
 
\subsection{Out-of-cluster energy leakage mis-modelling}
\label{sec:leak}
Electrons and photons deposit about 1\% to 6\% of their energy outside of the cluster used in the reconstruction,
depending on \et, $\eta$ and the particle type. This effect is corrected for by the MC-based energy response calibration. However,
a bias in the reconstructed energy could appear in data if this lateral leakage is mis-modelled by the simulation.
For electrons, the global energy scale correction (Section~\ref{sec:insitu}) absorbs any potential discrepancy at
$\avg{\et} \approx 40~\gev$.\footnote{The energy dependence of the difference between data and simulation for electron
lateral leakage was studied and found to be small, below 0.1\%,  and is accounted for as a (\et,$|\eta|$)-dependent
systematic uncertainty.} To take into account possible differences between electron and photon showers related to
the different probabilities for interaction with the material in front of the calorimeter, the lateral energy leakage in the
calorimeter outside the area of the cluster is studied directly in data and simulation.
 
The $Z \to \ell\ell\gamma$ and \Zee\ event samples are used to estimate this difference. The lateral energy leakage, $l$, is defined
by comparing the energy collected in the second-layer cells belonging to the supercluster, $E^{\text{\tiny{L2}}}_{\text{nom}}$, with the energy deposited in second-layer cells in a larger rectangular window of size $7\times 11$ in $\eta\times \phi$
around it, $E^{\text{\tiny{L2}}}_{7\times11}$:
\begin{equation*}
l =\frac{E^{\text{\tiny{L2}}}_{7\times11}}{E^{\text{\tiny{L2}}}_{\text{nom}}}-1.
\end{equation*}
The difference between data and simulation is presented in Figure~\ref{fig:leak} as a function of \et and in two pseudorapidity
regions for electrons, unconverted photons and converted photons.
\begin{figure*}[h]
\centering
\subfigure[]{\includegraphics[width=.49\textwidth]{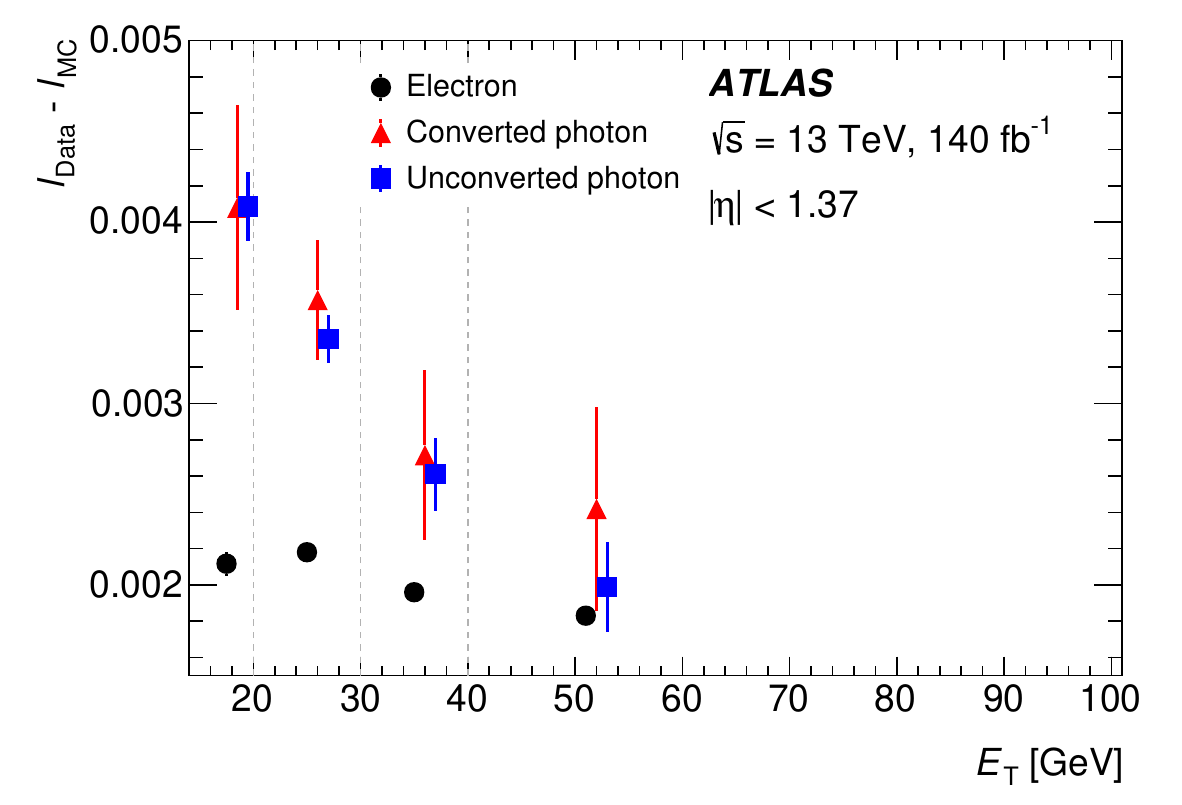}}
\subfigure[]{\includegraphics[width=.49\textwidth]{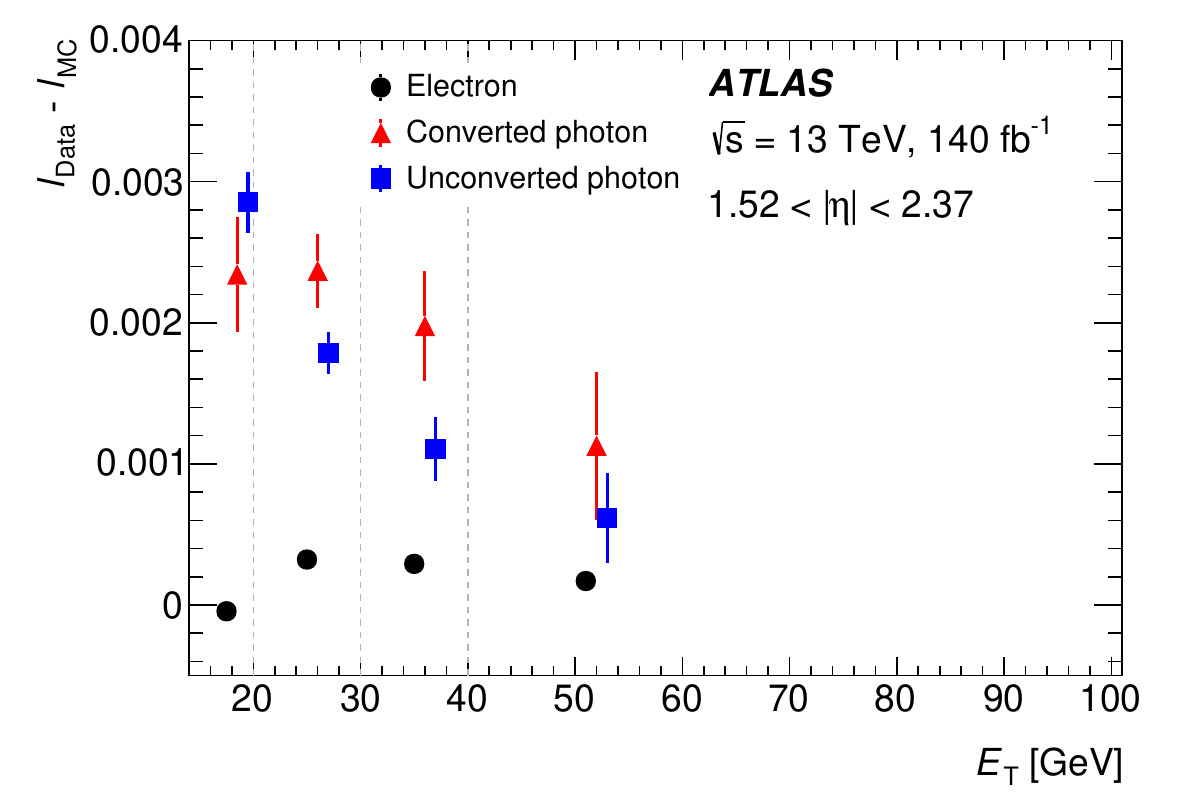}}
\caption{The difference between the leakage fractions in data and simulation, in (a) the barrel
and (b) the endcaps as a function of \et and the particle type. A variable bin size was chosen to make optimal use of the available samples.
The last bin covers the range $\et > 40$~\gev, and the corresponding markers are set near the \et-average in that bin.}
\label{fig:leak}
\end{figure*}
 
The double difference between electrons and photons in data and simulation, $\alpha_l$, defined as :
\begin{equation*}
\alpha_l = \left(l_e-l_{\gamma}\right)^{\text{data}} -\left(l_e-l_{\gamma} \right)^{\text{MC}},
\end{equation*}
is estimated separately for converted and unconverted photons and is used to correct the photon energy scale. The reconstruction of photon conversions and the classification into converted photon and unconverted photon categories have an impact on $\alpha_l$ that is
estimated by reweighting the simulation to match the data (as detailed in Section~\ref{sec:convmis}), and the difference
between the nominal  $\alpha_l$ values and the values obtained without reweighting is taken as a systematic uncertainty.
The $\alpha_l$ values vary from $-$(0.3--0.2)\% at low \et to $-$(0.1--0.05)\% at high
\et, with an absolute uncertainty ranging from 0.01\% to 0.07\% depending on the kinematic bins and the particle type.
 
Discrepancies between leakage modelling and data were observed in previous publications~\cite{PERF-2013-05,PERF-2017-03},
and the full size of the discrepancies was considered as a systematic uncertainty in the photon energy calibration.
The statistical power of the \RunTwo data allows an improved mapping of the $|\eta|$ and \et dependence of this effect, and a correction is derived.
The systematic variations, defined as above, reduce the residual uncertainty and allow a decrease in the corresponding calibration uncertainty
by a factor of about two.


\section{Electron and photon energy scale uncertainties}

\label{sec:uncer}
 
The complete systematic uncertainty model contains 64 and 67 independent uncertainty variations
for the electron and photon energy scales, respectively.
 
Uncertainties in the material upstream of the calorimeter are derived in Refs.~\cite{PERF-2013-05,PERF-2017-03,EGAM-2018-01} from a combination of detector construction information and \textit{in situ} measurements, and are evaluated for up to nine $|\eta|$ regions.
 
The uncertainty model for the cell readout non-linearity has been revised with respect to Ref.~\cite{PERF-2017-03}. The present analysis considers separate sources of uncertainty for the transitions between High and Medium gain, and between Medium and Low gain in Layer~2. In the previous analyses, a single uncertainty was used to cover both gain transitions. In addition, because of the improved strategy for the \alot measurement (now also using an electron sample), the size of the uncertainty associated with the transition between High and Medium gain in Layer~1 was re-estimated for the current analysis. The corresponding sources of uncertainty are considered fully correlated in pseudorapidity.
 
The presampler calibration and the intercalibration of the first and second accordion layers are estimated in nine and seven $|\eta|$ regions, respectively, with corresponding sources of uncertainty. Uncertainties corresponding to the $Z$ scale calibration, photon reconstruction and classification, \pileup modelling, and lateral shower shape development are all fully correlated in pseudorapidity. For the last category, an exception is the energy scale dependence on the shower width in $\eta$, where the region $1.52<|\eta|<1.82$ is considered to be independent of the rest of the pseudorapidity interval.
 
The full list of systematic uncertainty sources is summarized in Table~\ref{table:systematics_list}. The impact of the most significant ones is illustrated as a function of the electron
or photon \et in Figure~\ref{fig:calibESUncvseta}, for two pseudorapidity values.
 
\begin{table}
\begin{center}
\caption{List of the independent systematic uncertainties affecting the energy calibration
and their division into a number of $|\eta|$ regions, between which the uncertainties are not correlated.}
\resizebox{\textwidth}{!}{
\begin{tabular}{lllc}
\toprule
Source of uncertainty & Methodology & Description & Number of \\
& & & $|\eta|$ regions \\
\midrule
ID material                                      & \RunOne detector construction                           & \cite{PERF-2013-05,EGAM-2018-01}         & 4 \\
& Pixel services description                            & \cite{PERF-2017-03,EGAM-2018-01}         & 2 \\
Material presampler to calorimeter               & \RunOne measurement with unconv.\ photon                 & \cite{PERF-2013-05,EGAM-2018-01}         & 10\phantom{0} \\
($|\eta|<1.8$)                                   & Simulation of long.\ shower shape unconv.\ photon       & \cite{PERF-2013-05}                      & 2 \\
Material ID to presampler                        & \RunOne measurement with electrons                      & \cite{PERF-2013-05,EGAM-2018-01}         & 9  \\
($|\eta|<1.8$)                                   & Simulation of long.\ shower shape electrons            & \cite{PERF-2013-05}                      & 2 \\
Material ID to calorimeter                       & \RunOne measurement with electrons                     & \cite{PERF-2013-05,EGAM-2018-01}         & 3 \\
($|\eta|>1.8$)                                   & Simulation of long.\ shower shape electrons            & \cite{PERF-2013-05}                      & 1 \\
All material ID to calorimeter                   & Variations of \GEANT physics list                     & \cite{PERF-2013-05}                      & 1 \\
\midrule
Cell readout non-linearity                       & ADC non-linearity                                     & Section~\ref{sec:adclin}                 & 1 \\
& Medium gain/High gain Layer~2                         & Section~\ref{sec:L2}                     & 1 \\
& Low gain/Medium gain Layer~2                          & Section~\ref{sec:L2}                     & 1 \\
& Medium gain/High gain Layer~1                         & \cite{PERF-2017-03}                      & 1 \\
& \Pileup shift                                         & \cite{PERF-2017-03}                      & 1 \\
\midrule
Presampler calibration                           & \aps measurement                                      & Section~\ref{sec:psscale}                &  9 \\
Layer~1/Layer~2 calibration                      & \alot measurement                                     & Section~\ref{sec:e1e2}                   & 7 \\
Barrel--endcap gap scintillator                  & Scintillator calibration                              & \cite{PERF-2017-03}                      & 3 \\
($1.4<|\eta|<1.6$) & &  \\
\midrule
$Z\rightarrow ee$ calibration                    & Statistical uncertainty                               & Section~\ref{sec:insitu}                 & 1 \\
& Systematic uncertainty                                & Section~\ref{sec:insitu}                 & 1 \\
\midrule
Conversion reconstruction                        & Classification (efficiency and fake rate)             & Section~\ref{sec:convmis}                & 1 \\
& Radius dependence of conversion reconstruction        & \cite{PERF-2013-05}                      & 1 \\
\midrule
Lateral shower shape modelling                   & Dependence on shower $\eta$ width                     & \cite{PERF-2017-03}                      & 2 \\
& Lateral leakage for electrons                         & Section~\ref{sec:leak}                   & 1 \\
& Lateral leakage for unconv.\ photons                   & Section~\ref{sec:leak}                   & 1 \\
& Lateral leakage for conv.\ photons                     & Section~\ref{sec:leak}                   & 1 \\
\midrule
\Pileup modelling                                & Mis-modelling of \pileup noise vs $\avg{\mu}$        & \cite{EGAM-2018-01}                      & 1 \\
\bottomrule
\end{tabular}
}
\label{table:systematics_list}
\end{center}
\end{table}
\begin{figure*}[htbp]
\centering
\subfigure[]{\includegraphics[width=.49\textwidth]{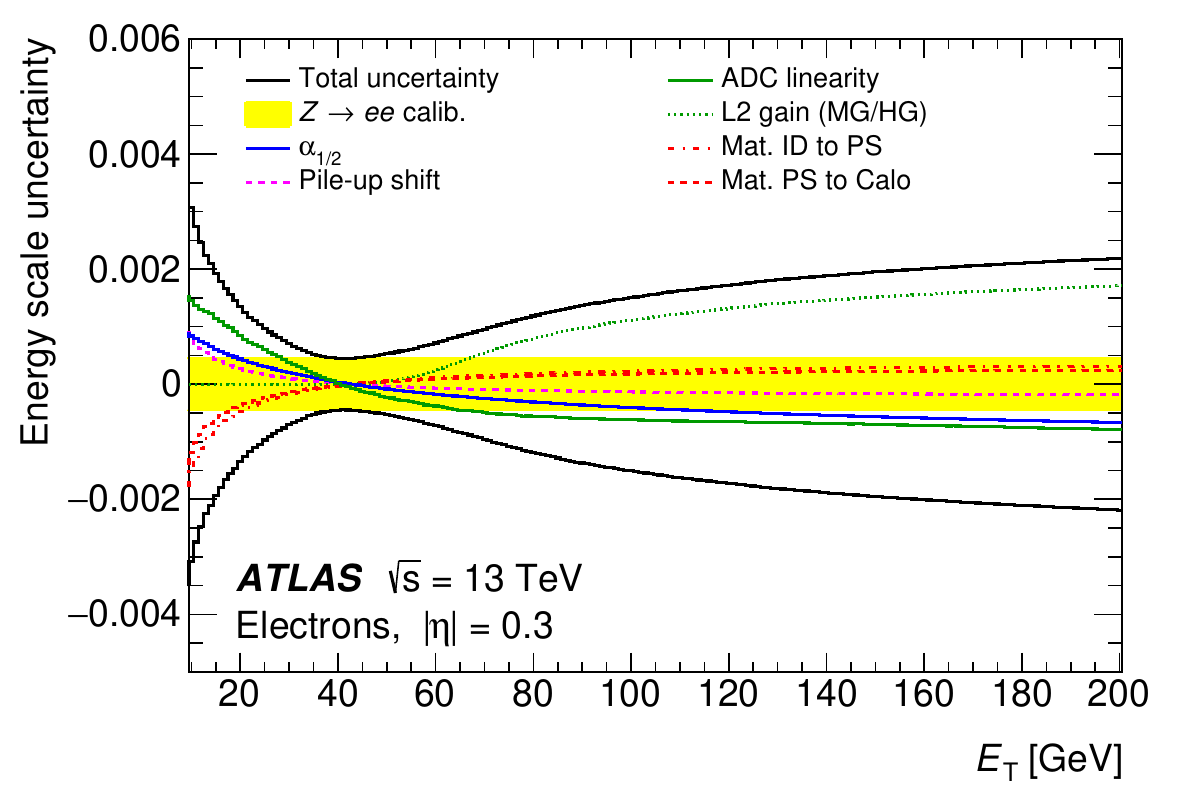}}
\subfigure[]{\includegraphics[width=.49\textwidth]{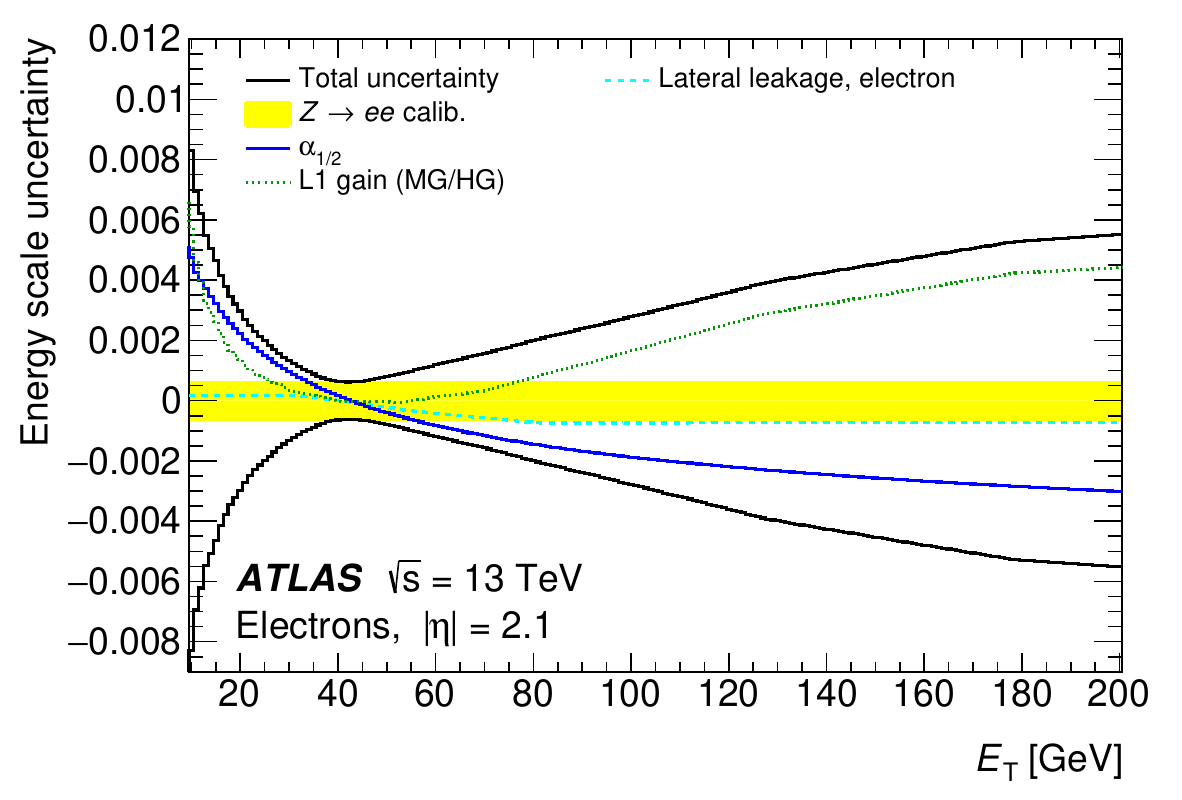}}
\subfigure[]{\includegraphics[width=.49\textwidth]{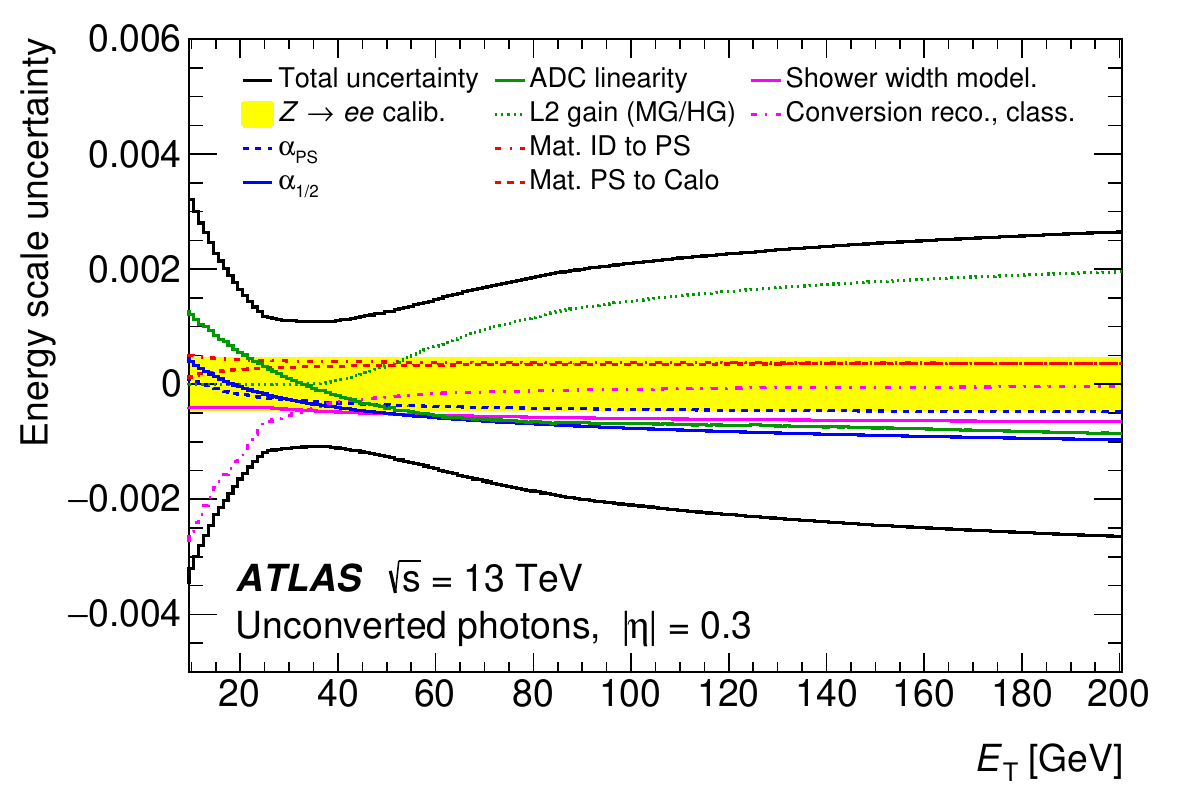}}
\subfigure[]{\includegraphics[width=.49\textwidth]{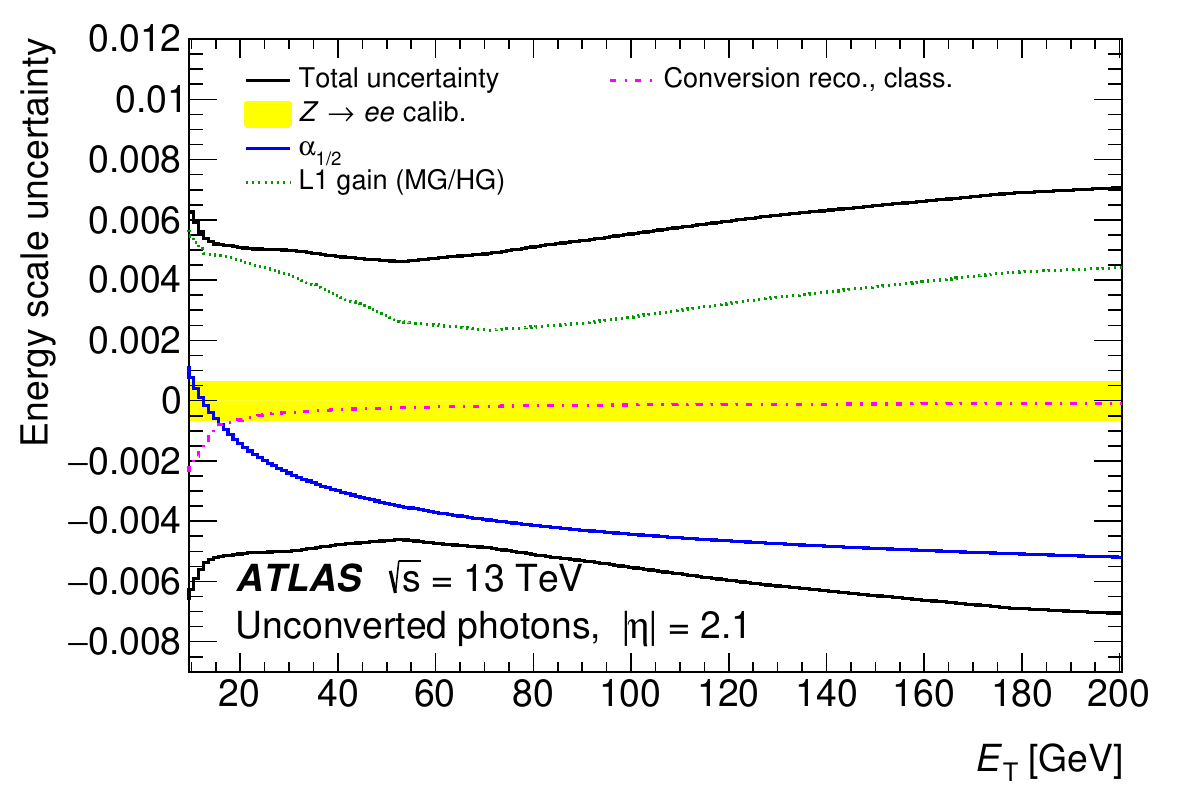}}
\subfigure[]{\includegraphics[width=.49\textwidth]{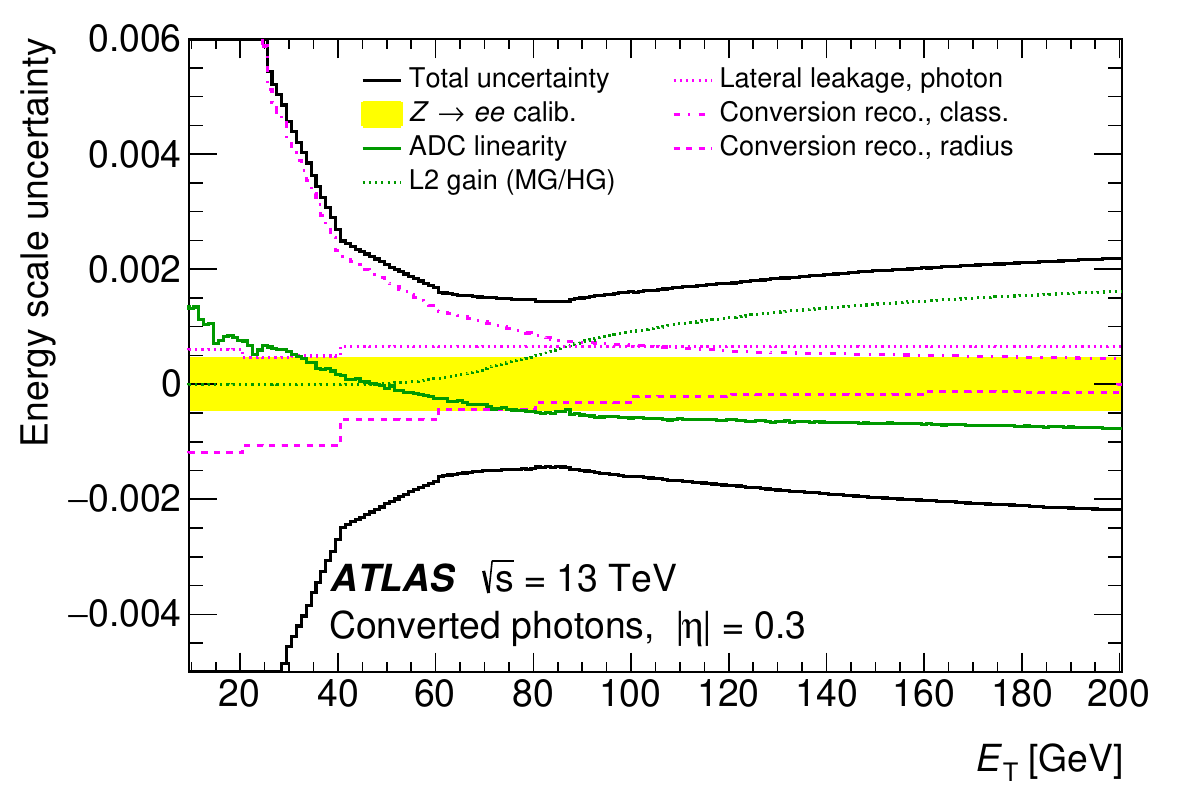}}
\subfigure[]{\includegraphics[width=.49\textwidth]{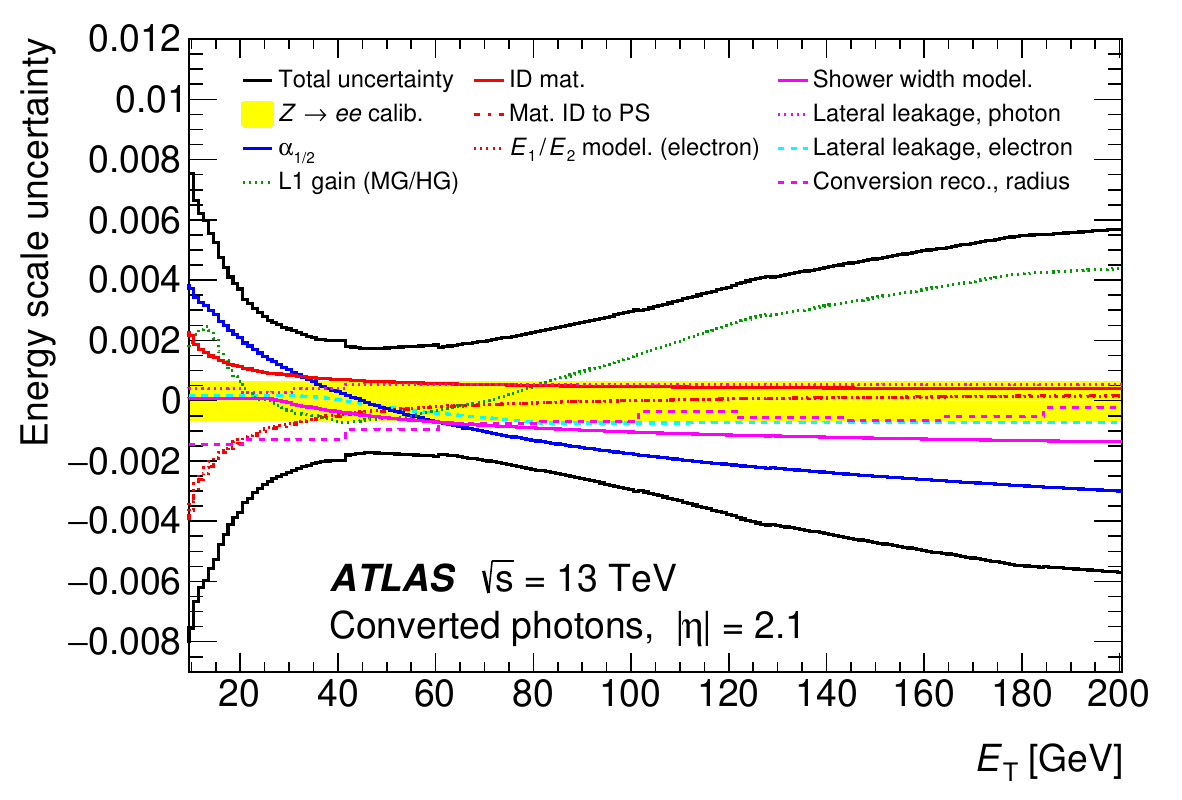}}
\caption{Relative energy scale calibration uncertainty for (a, b)
electrons, (c, d) unconverted photons and (e, f) converted photons, as a
function of \et for (a, c, e) $|\eta| = 0.3$ and (b, d, f) $|\eta| = 2.1$.
For converted photons at $|\eta| = 0.3$, the relative uncertainty reaches 2.1\% at \et=10~GeV.
The total uncertainty is shown along with the main contributions,
which are represented by the signed impact of a one-sided
variation of the corresponding uncertainty.
Only a one-sided variation for each uncertainty source is shown for clarity, except for the uncertainty
related to the \textit{in situ} global energy scale determination with \Zee\ candidate events.}
\label{fig:calibESUncvseta}
\end{figure*}
 
The uncertainty in the energy resolution is shown in Figure~\ref{fig:calibResoUncvsET} as a function
of \et, for electrons and unconverted photons, at $|\eta| = 0.3$. Only the uncertainty related
to the resolution correction (Section~\ref{sec:insitu}) was updated in this analysis
with respect to the descriptions provided in Refs.~\cite{PERF-2013-05,PERF-2017-03,EGAM-2018-01}.
At high \et, the resolution uncertainties are larger than those presented in Ref.~\cite{PERF-2017-03}, where fixed-size clusters were used for electron and photon reconstruction. The present clustering algorithm is more sensitive to \pileup fluctuations. This effect is larger for electrons than for photons.
\begin{figure*}[h]
\centering
\subfigure[]{\includegraphics[width=.49\textwidth]{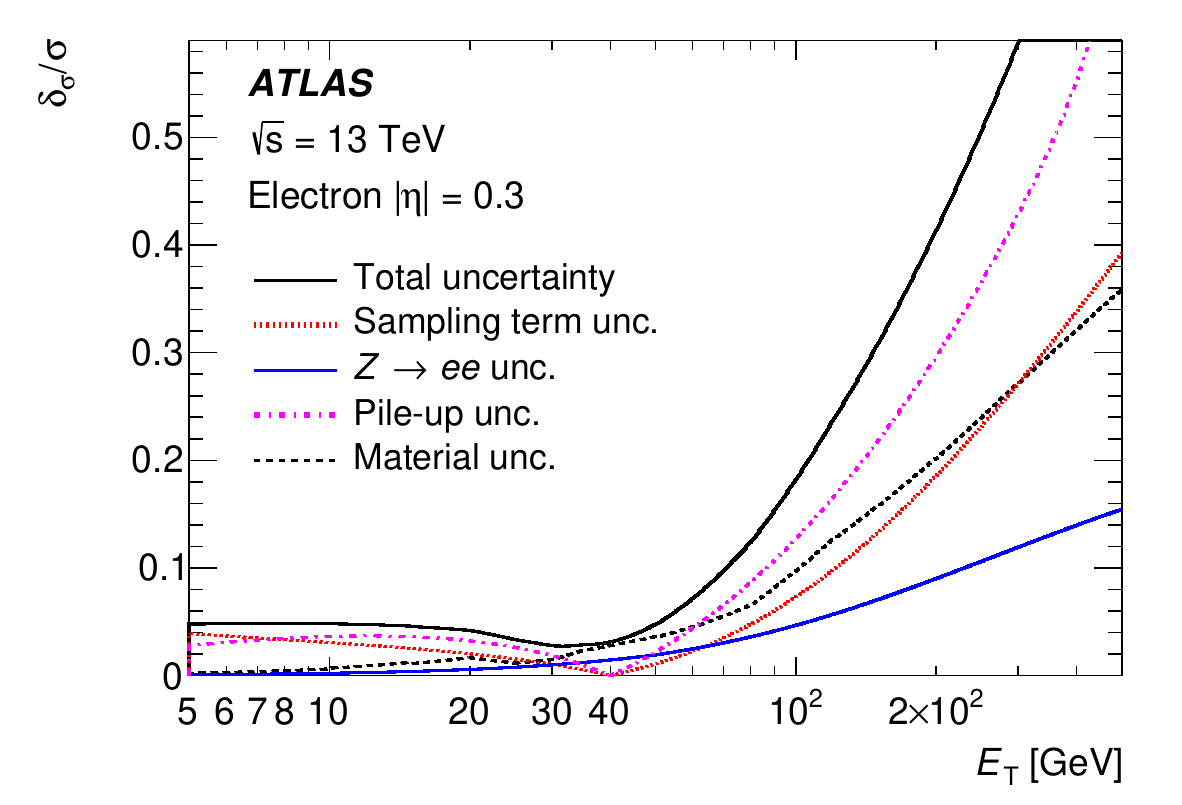}}
\subfigure[]{\includegraphics[width=.49\textwidth]{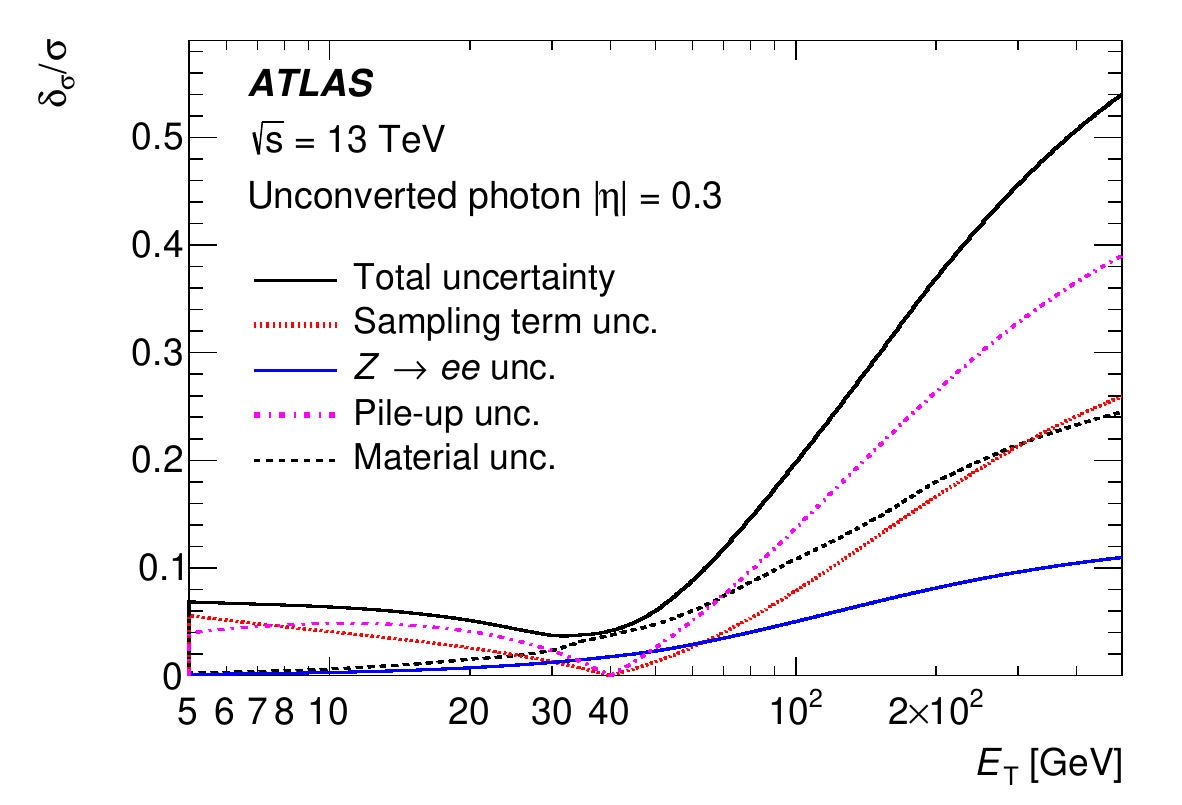}}
\caption{Relative uncertainty in the energy resolution, $\delta_{\sigma}/\sigma$ as a function
of \et for (a) electrons and (b)  unconverted photons at $|\eta| = 0.3$. The total uncertainty is
shown along with the breakdown into the different contributions.}
\label{fig:calibResoUncvsET}
\end{figure*}


\section{Energy linearity and constraints on the calibration uncertainties}

\label{sec:linearity}
 
\subsection{Energy linearity measurement}
In order to test the \et dependence of the energy scale corrections $\alpha_i$, the procedure in Section~\ref{sec:insitu} is repeated in bins of $|\eta|$ and \et. Specifically, an extended definition of the calibrated energy is introduced as
\begin{equation*}
E^\text{data,corr}=E^\text{data}/[(1 + \alpha_i)(1+\alpha^\prime_j)],
\end{equation*}
where the energy scale factors $\alpha_i$ are left at their values determined above, and $\alpha^\prime_j$ quantifies the energy dependence of the energy scale (in the absence of energy dependence, $\alpha^\prime=0$). The index $j$ labels a two-dimensional mapping of the electron phase space, with the following bin boundaries:
\begin{itemize}
\item 0, 0.6, 1.0, 1.37, 1.55, 1.82, 2.47 in $|\eta|$;
\item 27, 33, 38, 44, 50, 62, 100, $\sqrt{s}/2$ in \et[\ensuremath{\text{G\electronvolt}}].
\end{itemize}
Events are assigned to a bin $j$ if either one of the two decay electrons falls in this bin, and $\alpha^\prime_j$ is varied to obtain the best agreement between the invariant mass distributions in data and simulation, separately for each bin $j$. Since the second electron is distributed randomly, iterations are required and the procedure is repeated until convergence is reached.
 
The $\alpha^\prime$ coefficients are shown in Figure~\ref{fig:linearity} as a function of $|\eta|$ and \et. The observed energy dependence is significant, and is measured with a typical precision of 0.03\%, including the effect of systematic variations as described in Section~\ref{sec:insitu}. In the regions $1.37<|\eta|<1.55$ and $1.55<|\eta|<1.82$, the precision is about 0.3\%. The measurement precision is better than the calibration uncertainty derived in the previous sections, as can be seen from the outer uncertainty band in Figure~\ref{fig:linearity}. The energy linearity measurement can thus be used to further constrain the uncertainty model, as described in the following.
\begin{figure*}[h]
\centering
\includegraphics[width=.49\textwidth]{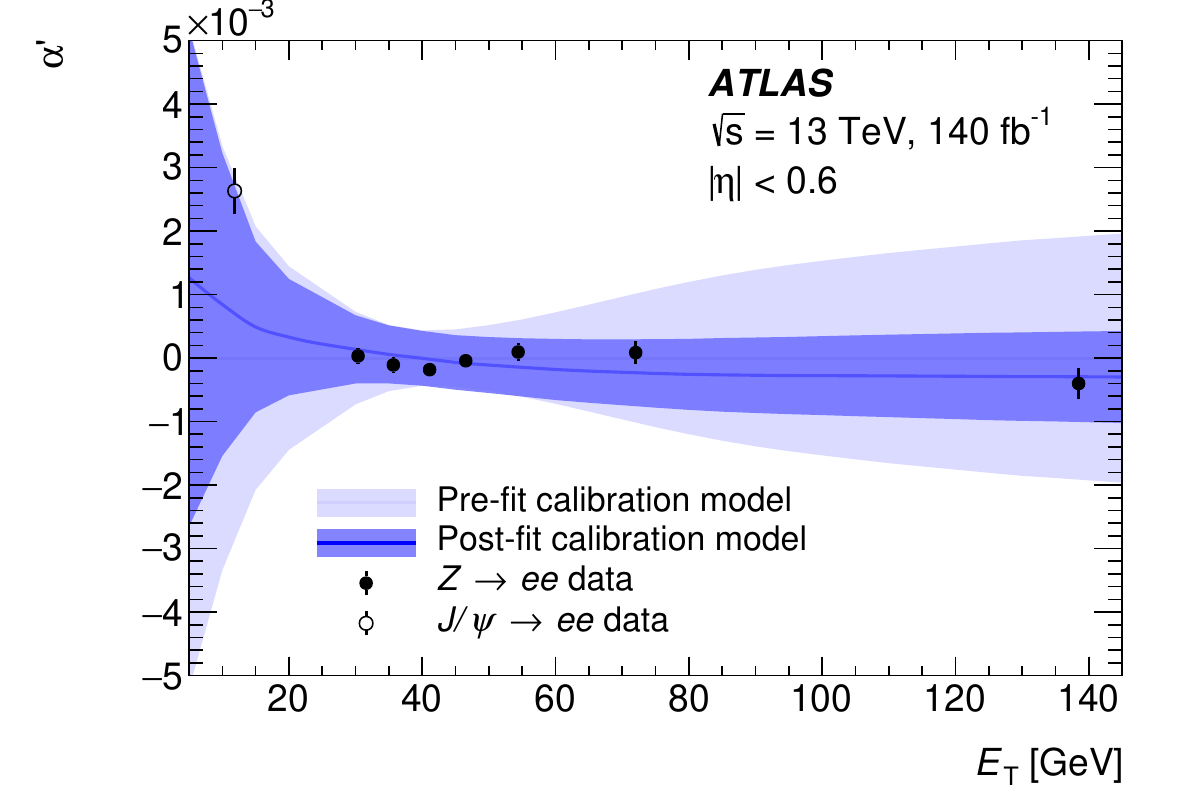}
\includegraphics[width=.49\textwidth]{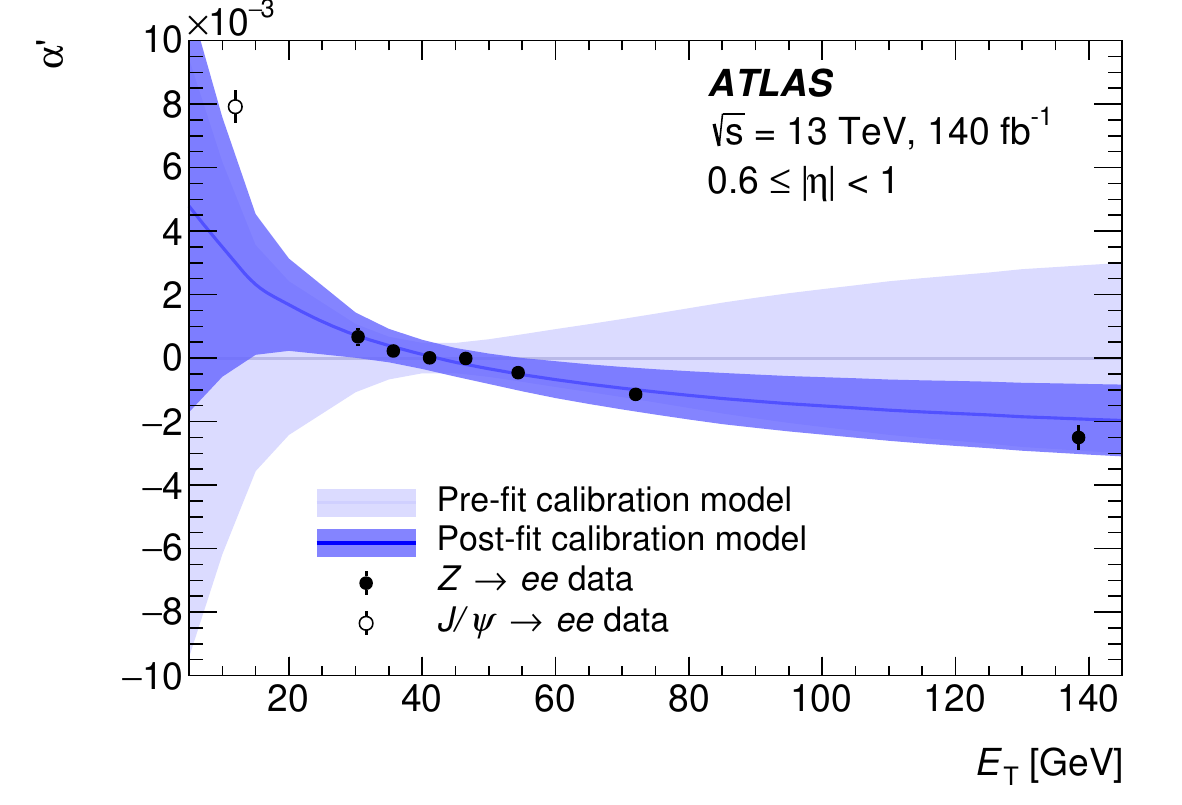}
\includegraphics[width=.49\textwidth]{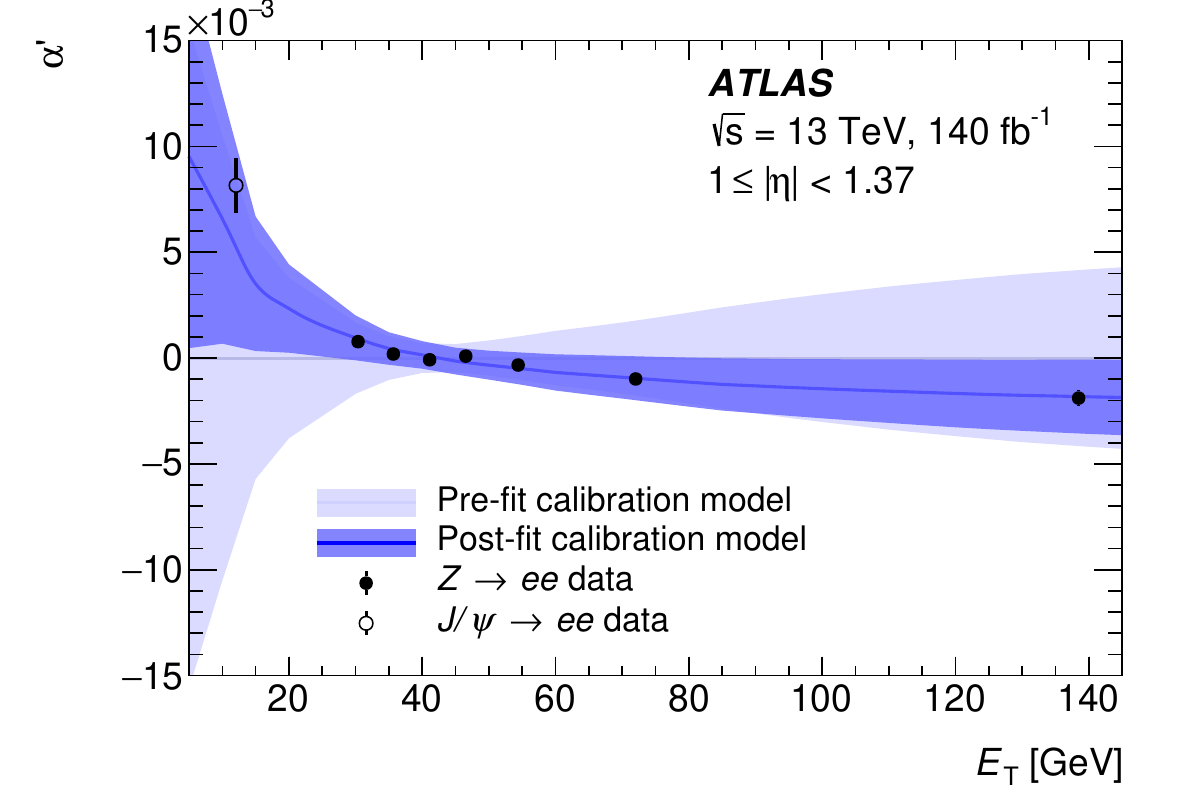}
\includegraphics[width=.49\textwidth]{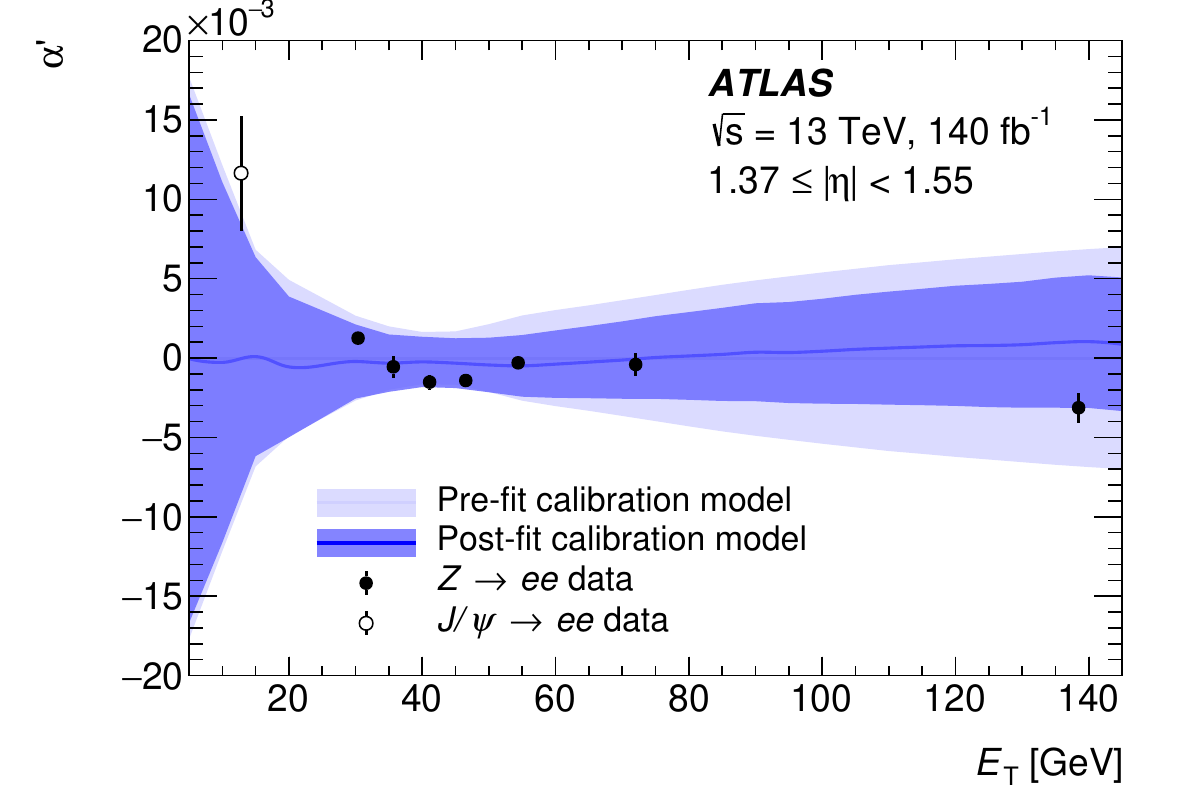}
\includegraphics[width=.49\textwidth]{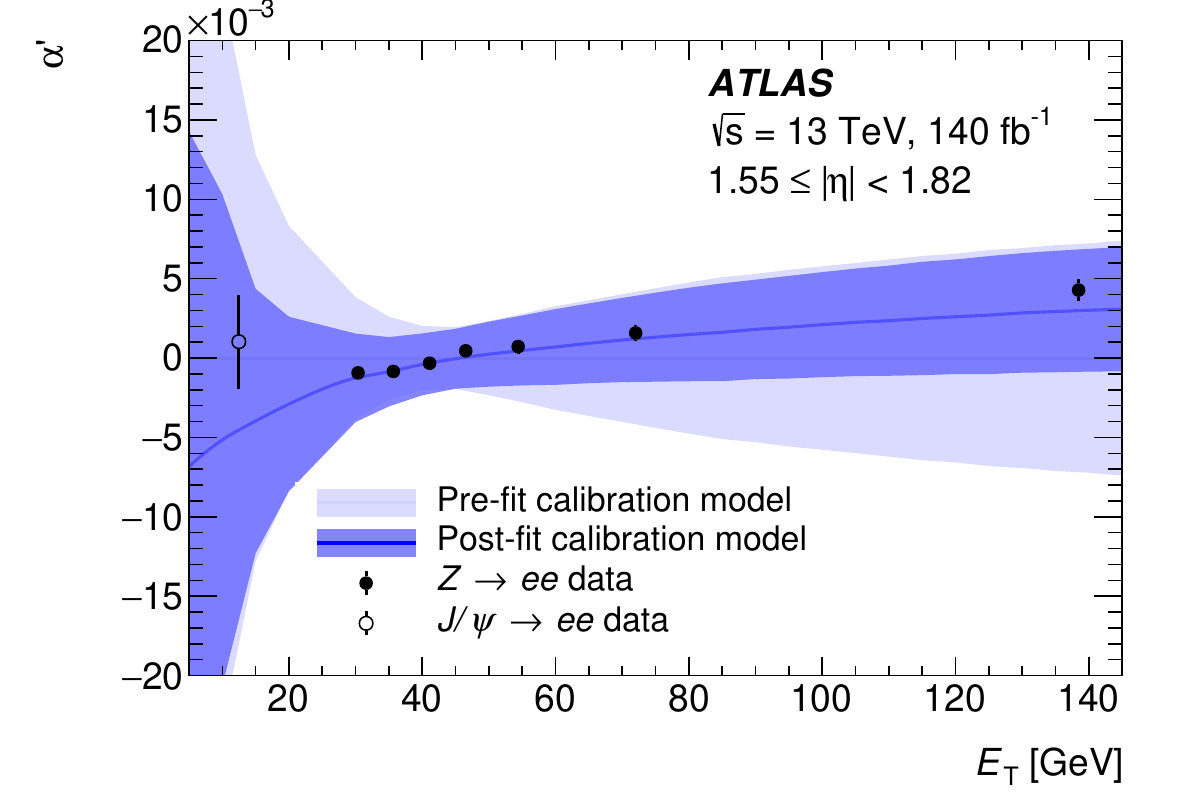}
\includegraphics[width=.49\textwidth]{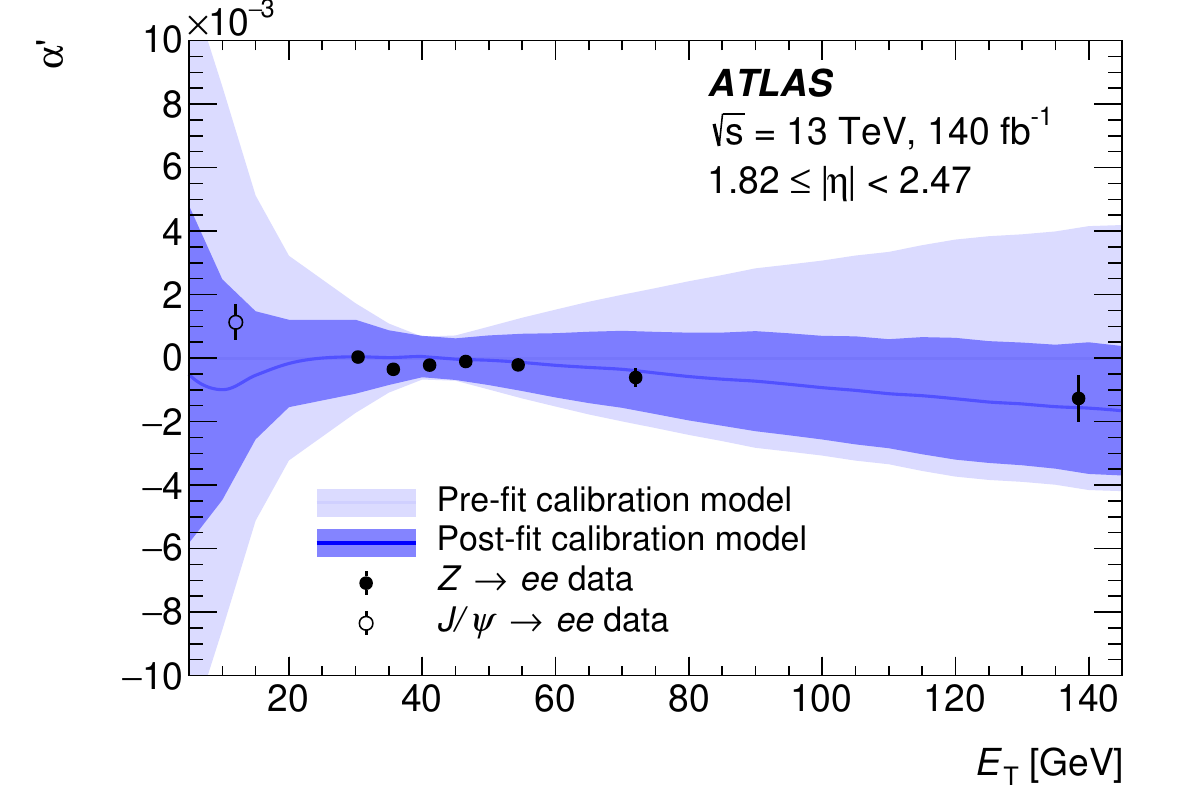}
\caption{Comparison of the measured values of $\alpha^\prime$ with the pre-fit and post-fit linearity models.
The full lines represent the nominal pre- and post-fit models, and the bands represent the corresponding uncertainties.
The measurements in the $|\eta|$ range
corresponding to the barrel--endcap transition regions, $1.37 \leq |\eta| < 1.55$, and from the $J/\psi \to ee$ analysis
are not included in the energy linearity fit. The analysis described in Section~\ref{sec:jpsi} was
repeated with the present $\eta$ binning to obtain the open dots in these figures.}
\label{fig:linearity}
\end{figure*}
\FloatBarrier
\subsection{Constraints on the calibration systematic uncertainties}
Accounting for the energy scale calibration using \Zee events, varying a given source of uncertainty by one standard deviation affects the reconstructed energy of an electron as follows:
\begin{equation*}
\delta_{\text{rel}} E_k(\et, \eta) = \Delta_{\text{rel}} E_k(\et,\eta) - \Delta_{\text{rel}} E_k(\avg{\et},\eta),
\end{equation*}
where $k$ labels the source of uncertainty (see Table~\ref{table:systematics_list}), $\Delta_{\text{rel}} E_k(\et,\eta)$ represents its fractional impact on the reconstructed energy, and $\avg{\et} \approx 40~\gev$ is the average transverse energy for electrons produced in $Z$-boson decays. The quantities $\Delta_{\text{rel}} E_k(\et,\eta)$ are estimated by varying the corresponding sources of uncertainty in the reconstruction or detector simulation. The $Z$-based calibration absorbs the systematic uncertainty effect for electrons with $\et=\avg{\et}$ and leaves the residual effect $\delta_{\text{rel}} E_k(\et, \eta)$. In this model, the total effect of all systematic variations on the linearity can be parameterized as
\begin{equation*}
\alpha^\prime_{\text{mod},j} = \sum_k \delta_{\text{rel}} E_{jk} \theta_k,
\end{equation*}
where $\delta_{\text{rel}} E_{jk}$ is the average of $\delta_{\text{rel}} E_k(\et, \eta)$ over \et and $\eta$ in bin $j$, and $\theta_k$ is the normally distributed nuisance parameter (NP) associated with the source $k$. The calibration model is fitted to the data by adjusting the nuisance parameters $\mathbf\theta$, minimizing the following $\chi^2$:
\begin{equation}\label{eq: likelihood}
\chi^2 = \sum_{j_1,j_2} \left[\alpha^\prime_{j_1} - \alpha^\prime_{\text{mod},j_1}(\mathbf{\theta})\right] C^{-1}_{j_1,j_2} \left[\alpha^\prime_{j_2} - \alpha^\prime_{\text{mod},j_2}(\mathbf{\theta})\right] + \sum_k \theta^2_k,
\end{equation}
where $C$ is the covariance of the $\alpha^\prime$ measurements, calculated from the statistical and systematic uncertainties in the measurement procedure. The model contains 46 nuisance parameters\footnote{The full model contains 64 nuisance parameters relevant for the electron energy scale (Section~\ref{sec:uncer}). Two of them related to the final energy scale determination (Section~\ref{sec:insitu}) are not part of the $\theta_k$ and 16 are removed from the fit by a pruning procedure.} with the corresponding constraint terms, and 7$\times$5 measured values of $\alpha^\prime$. The transition regions between the barrel and endcap calorimeters are not included in the fit.
The nuisance parameters represent uncertainties in the passive material model, electromagnetic shower shape development, readout electronics calibration and layer intercalibration;
each of these uncertainty classes may be subdivided into up to 12 pseudorapidity regions.
 
The $\chi^2$ minimization is performed analytically. Denoting the fitted values of the nuisance parameters by $\hat{\theta}_k$ and their post-fit covariance by $V$, the post-fit linearity and its uncertainty are defined as:
\begin{equation*}
\hat\alpha^\prime_{\text{mod},j} = \sum_k \delta_{\text{rel}} E_{jk} \, \hat\theta_k \,\,\,\, , \,\,\,\,
\delta\hat\alpha^\prime_{\text{mod},j} = \left[ \sum_{k_1,k_2} \delta_{\text{rel}} E_{j k_1}  \, \delta_{\text{rel}} E_{j k_2} \, V_{k_1 k_2} \right]^{1/2}.
\end{equation*}
For comparison, the pre-fit expressions are $\alpha^\prime_{\text{mod},j}=0$ and $\delta\alpha^\prime_{\text{mod},j} = \left[ \sum_k \delta_{\text{rel}} E^2_{jk} \right]^{1/2}$. The minimization result is sensitive to the details of the covariance matrix $C$ assigned to the linearity measurements. A global goodness-of-fit of $\chi^2/N = 90/35$ is obtained, assuming full correlation of the $\alpha^\prime_j$ systematic uncertainties across \et bins within each $|\eta|$ bin, but ignoring correlations across $|\eta|$ bins.
The MC statistical uncertainty is accounted for in the evaluation of the systematic uncertainties.
For $|\eta|$ bins with a partial $\chi^2$ per degree of freedom $\chi_\mathrm{bin}^2/N_\mathrm{bin}$ greater than one, the $\alpha^\prime$ measurement uncertainties are rescaled by a factor $\sqrt{\chi_\mathrm{bin}^2/N_\mathrm{bin}}$. This scaling typically increases the fit uncertainties by 5\%. The final goodness-of-fit is $\chi^2/N = 41/35$, corresponding to a $p$-value of 0.22.
 
The results of the analysis are illustrated in Figures~\ref{fig:linearity} and~\ref{fig:pulls}. With few exceptions, the measured values of $\alpha^\prime_j$ are well within the initial calibration uncertainties. The fit $\chi^2$ in Eq.~(\ref{eq: likelihood}) captures the measured non-linearities, and reduces the uncertainties by up to a factor of two
for $\et\ <50~\gev$, and up to three
for $\et \sim 150~\gev$.
As can be seen in Figure~\ref{fig:pulls}, the reduction in uncertainty is mostly driven by the nuisance parameters associated with the cell-level non-linearity (ADC corrections, and HG/MG transition) and the shower development (lateral leakage and shower width). Most other nuisance parameters are typically constrained by 5\%--10\%. All nuisance parameter shifts are within the initially assigned uncertainties, with the exception of a 1.5$\sigma$ effect observed for the presampler calibration in one specific $|\eta|$ region.
 
\begin{figure*}[h]
\centering
\includegraphics[width=\textwidth]{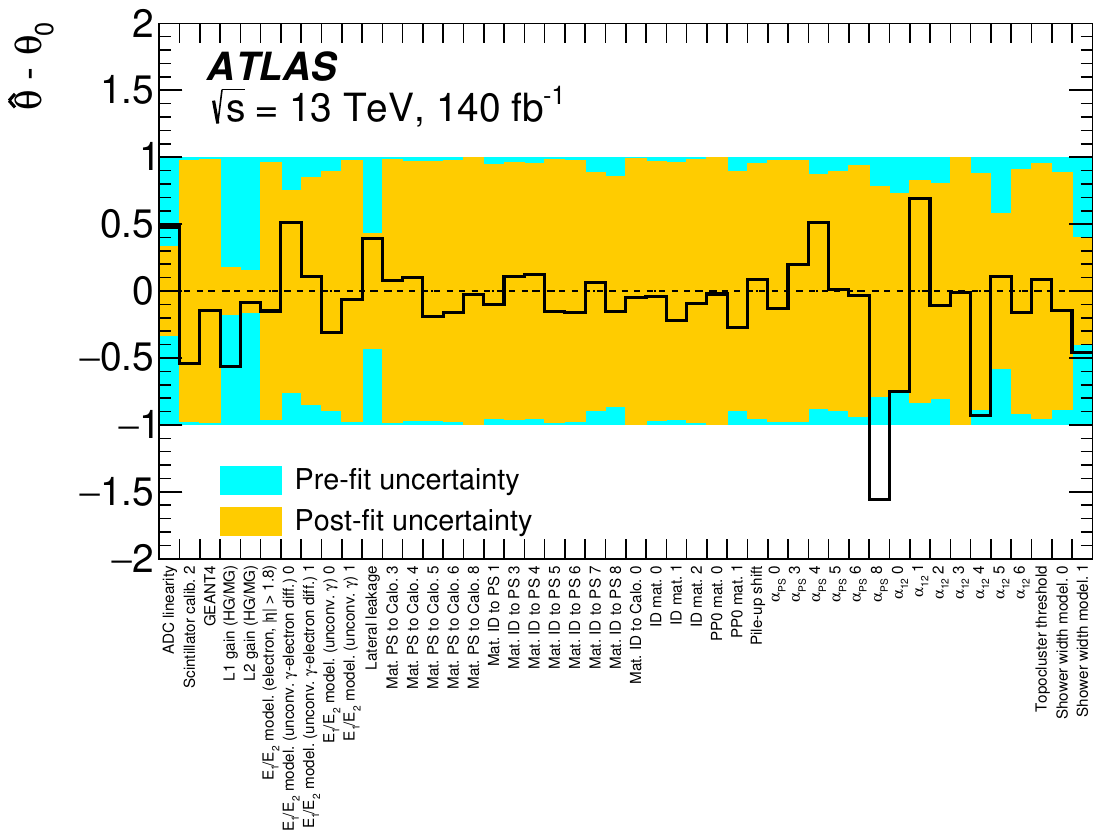}
\caption{Shifts and constraints on the nuisance parameters of the systematic uncertainty model from the energy linearity fit. A digit after the NP name represents a given $\eta$ range (\textit{a priori} different for different NP sources).}
\label{fig:pulls}
\end{figure*}
 
The impact of the linearity fit on the electron energy scale uncertainty is illustrated in Figure~\ref{fig:calibESUncvsETa} as a function of \et and $|\eta|$.
The precision for electrons with $\et\sim 40~\gev$ is mostly unchanged, since these particles are typical of on-shell $Z$-boson decays and their calibration is essentially determined by the energy scale analysis in Section~\ref{sec:insitu}. Energy scale uncertainties for electrons with $\et=10~\gev$ or $\et=1~\tev$ are typically reduced by 30\%--50\%, and vary from 0.2\%--0.3\% for $|\eta|<1$ and $|\eta|>1.8$ to between 0.5\% and 1\% for  $1<|\eta|<1.8$.
 
The impact of the present analysis on photon calibration uncertainties is shown in Figure~\ref{fig:calibESUncvsETb} for converted and unconverted photons, and for $\et=60~\gev$, which is typical for photons from Higgs boson decays. Uncertainties for converted photons, which are experimentally close to electrons, are only moderately reduced for this energy. For unconverted photons, the energy calibration uncertainty is typically reduced by 30\% in the barrel, and by up to a factor of two in the endcaps.
 
\begin{figure*}[h]
\centering
\subfigure[]{\includegraphics[width=.49\textwidth]{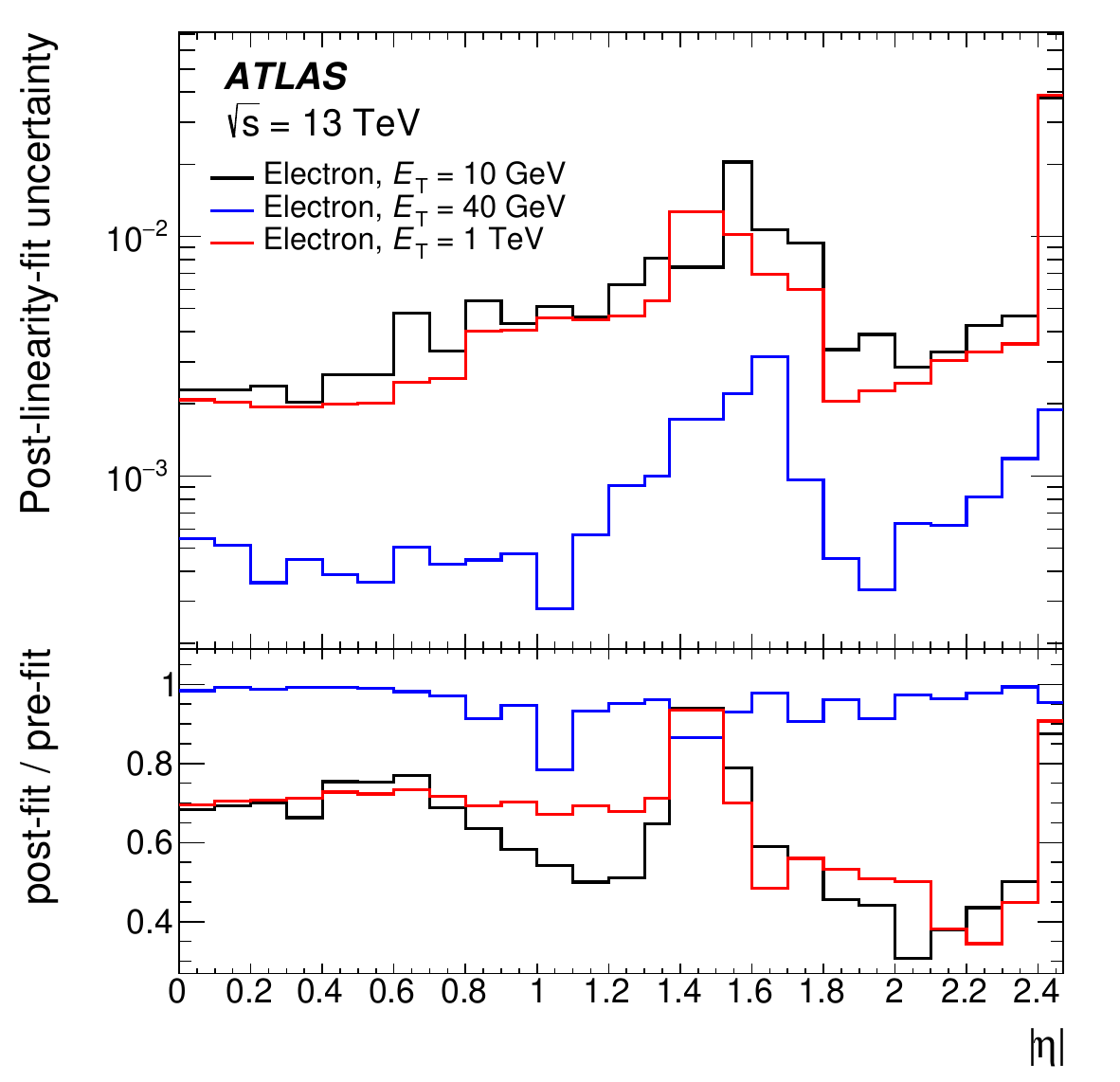}\label{fig:calibESUncvsETa}}
\subfigure[]{\includegraphics[width=.49\textwidth]{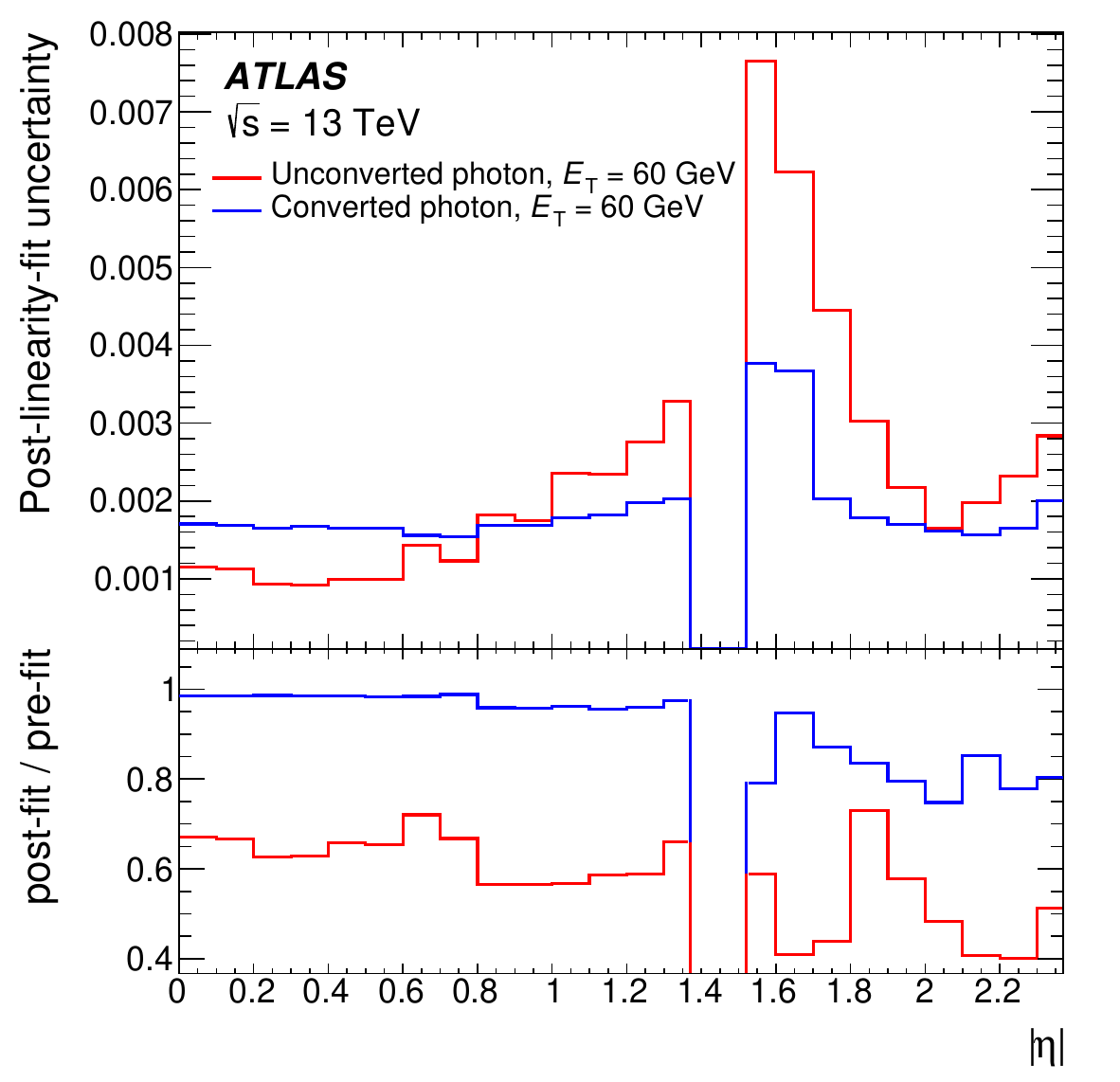}\label{fig:calibESUncvsETb}}
\caption{Total relative systematic uncertainty in the energy scale as a function of $|\eta|$ for (a) electrons with
$\et =10~\gev, 40~\gev$ or 1~\tev\ and (b) photons with $\et = 60~\gev$, after the constraints from the linearity fit.
The bottom panels show the ratio of the post-fit to pre-fit uncertainties.}
\end{figure*}


\section{Calibration cross-checks}

\label{sec:xcheck}
\subsection{Checks using $J/\psi \rightarrow ee$ events}
\label{sec:jpsi}
The known mass of the $J/\psi$ resonance provides a completely independent check of
the energy calibration for electrons with transverse energy in the range from 5 to 30~\gev. For this, the full calibration procedure discussed in the previous
sections, including the energy scale derived from $Z\rightarrow ee$ events, is applied. The difference between
data and simulation for $J/\psi \rightarrow ee$ events is then quantified using residual energy scale factors 
extracted from the peak positions of the reconstructed invariant mass.
If the energy calibration is correct, the residual energy scale factors should be
consistent with zero within the combined uncertainties of the $J/\psi\rightarrow ee$ measurement and the systematic
uncertainty of the energy calibration.
 
The event selection requires two tightly identified, loosely isolated, opposite-sign electron candidates with $\et > 5~\gev$ and $|\eta|<2.4$.
The primary vertex must be located within $|z| = 150$~mm of the IP, and the dielectron invariant mass must be in the range [1, 5]~\gev.
 
The residual energy scale factors are denoted by $\alpha_i$, where $i$ labels the kinematic bin, and determined as follows:
\begin{itemize}
\item $J/\psi$ particles can be produced promptly or in $b$-hadron decays. The hadronic activity surrounding the decay electrons differs in both cases, biasing the energy-scale determination if the relative event fractions are not modelled accurately.
The prompt fraction is extracted from a fit to the proper decay-time distribution of the data and is found to be between 76\% and 82\% depending on the leading electron's \et, with uncertainties of up to 4\%. The simulated prompt and non-prompt samples are then combined  using the measured fractions.
\item The data are divided into categories depending on the $\eta$ values of the two selected electrons.
The $\eta$ bin boundaries are $-2.4$, $-1.52$, $-1.37$, $-1.0$, $-0.8$, $-0.4$, 0, 0.4, 0.8, 1.0, 1.37, 1.52, and 2.4.
The region $1.37 < |\eta| < 1.52$ is not considered for the nominal results.
The measurement can also be performed as a function of \et and integrated in $\eta$.
In this case, the \et bin boundaries are 5, 10, 15, 20, and 30~\gev.
\item Comparison  of the dielectron invariant mass distributions in data and simulation
requires the background contributions to be subtracted from the data sample. The subtraction is performed by fitting a signal+background distribution to the data, separately in each category. The total distribution is expressed as 
the sum of two double-sided Crystal~Ball functions to describe the $J/\psi$ and $\psi(2S)$ resonances, and a second-order Chebyshev polynomial to represent the continuum background.
The parameters describing the resonance distributions are fixed to values determined from simulation, with the exception of the peak positions, which are parameterized using energy scales $\alpha_i$ and $\alpha_j$.
An example of an invariant mass fit is shown in Figure~\ref{fig:xcheck_jpsi_mass}.
\begin{figure*}[h]
\centering
\includegraphics[width=.49\textwidth]{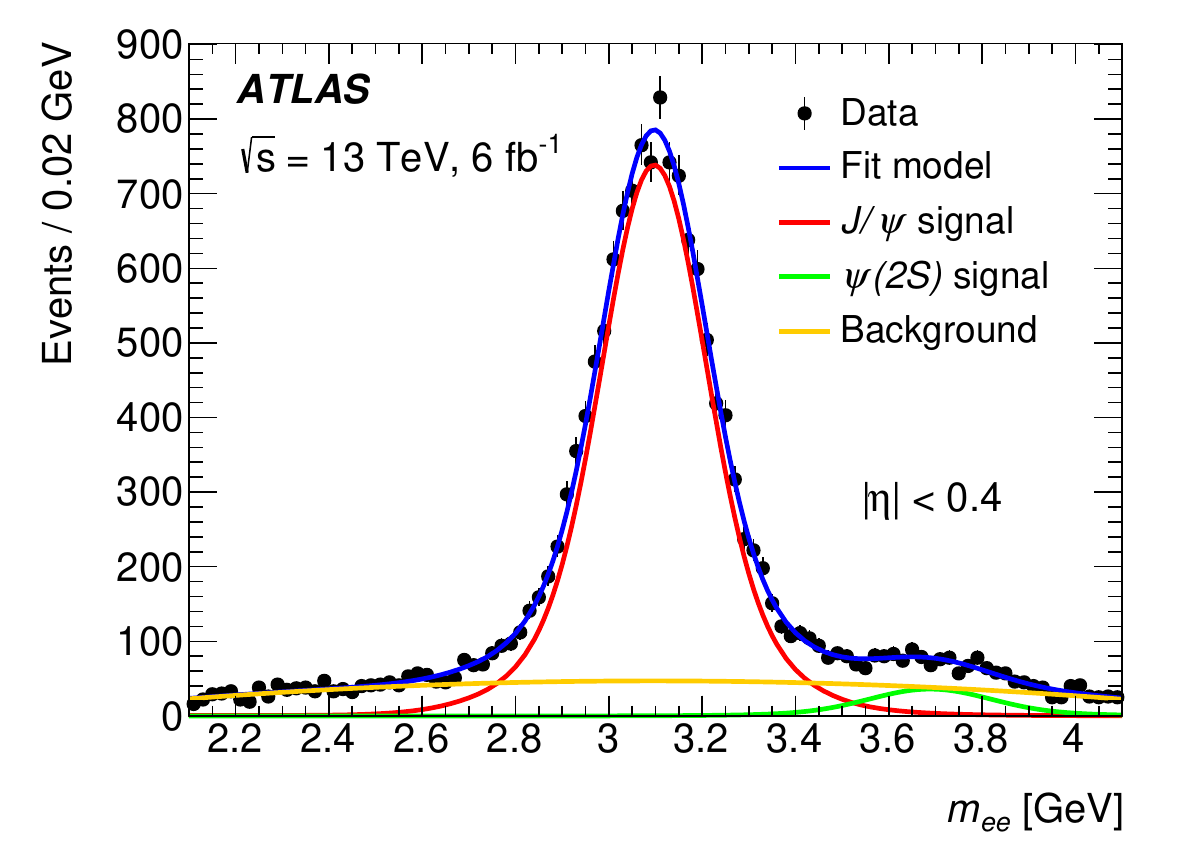}
\caption{An example dielectron invariant mass distribution with the fitted signal, background and $\psi(2S)$ contributions.
Both electrons are required to have $|\eta| < 0.4$.}
\label{fig:xcheck_jpsi_mass}
\end{figure*}
\item A simultaneous fit to all categories is performed to extract the residual energy scale factors. The considered systematic uncertainties are related to the modelling of signal and background, the fitted mass range, the uncertainties in the prompt fraction, and the modelling of the pseudorapidity distribution of electrons in simulation. The fit is repeated, varying each of the above sources of uncertainty in turn, and the deviations from the nominal $\alpha$ value are added in quadrature to obtain the final uncertainty in the measurement.
\end{itemize}
 
The results are given in Figure~\ref{fig:xcheck_jpsi}, where the evolution of $\alpha$ is shown as a function of (a) $\eta$ and (b) \et, before and after including the constraints from the linearity fit. The residual post-fit scale factors are below 0.5\% and are compatible with zero within the total calibration uncertainty.
\begin{figure*}[h]
\centering
\subfigure[]{\includegraphics[width=.49\textwidth]{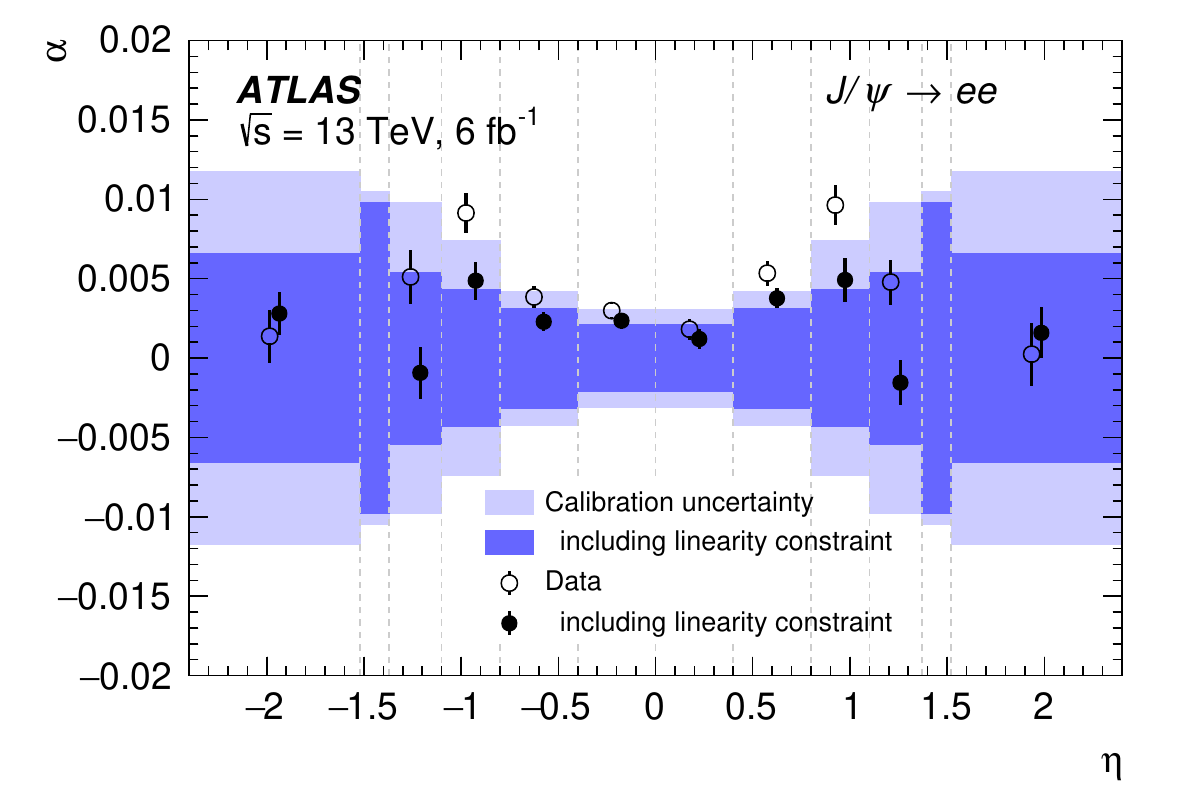}}
\subfigure[]{\includegraphics[width=.49\textwidth]{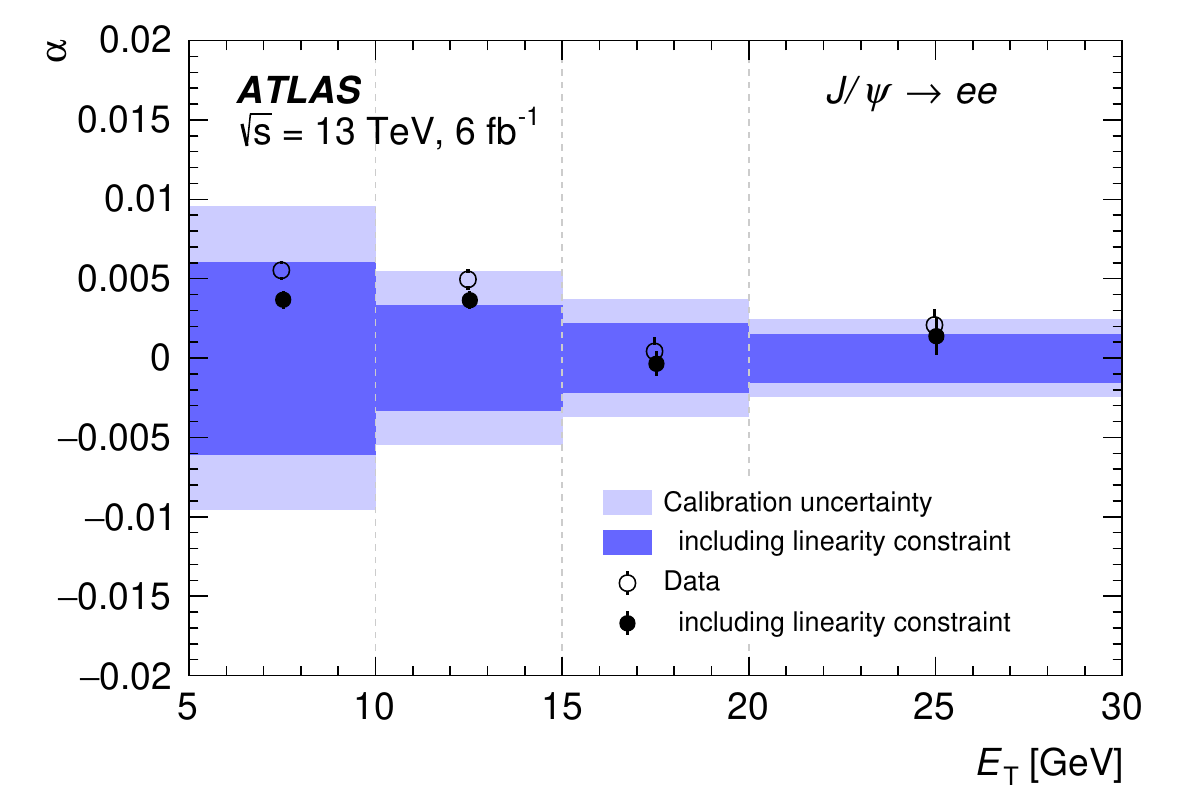}}
\caption{Variation of the residual energy scale as a function of (a) $\eta$ and (b) \et, as measured with $J/\psi \rightarrow ee$ events. The data points and uncertainty bands are shown for both the pre- and post-linearity-fit energy scale models. The uncertainty bands correspond to the energy calibration uncertainty for the energy range of the $J/\psi \to ee$ decays.}
\label{fig:xcheck_jpsi}
\end{figure*}
 
\subsection{Checks using $Z \rightarrow \ell\ell\gamma$ events}
Radiative $Z$-boson decay events provide a way to investigate the validity of the final photon energy scale.
The selection requires a $Z$-boson candidate decaying into two opposite-sign electrons or muons and a photon from final-state radiation. Electrons (muons) must meet medium identification criteria with $\et > 18~(15)~\gev$. Tightly identified photons with $\et >10~\gev$ are selected. Electrons and photons in the barrel--endcap transition regions are not considered.
Loose isolation requirements are applied to all objects.
The invariant mass of the dilepton+photon system is required to be in the range $80 < m_{\ell\ell\gamma} < 100~\gev$ and the dilepton invariant mass must be in the range $40 < m_{\ell\ell} < 80$~\gev. Figure~\ref{fig:xcheck_Zllg_mass} shows the $m_{\ell\ell\gamma}$ distributions for the (a) dielectron and (b) dimuon channels. The inclusive residual photon energy scale factors are measured to be $(3.3\pm2.0)\times 10^{-3}$ and $(1.4\pm1.1)\times 10^{-3}$ in the $Z\to ee\gamma$ and $Z\to\mu\mu\gamma$ samples, respectively. This is consistent with the larger offset observed in Figure~\ref{fig:xcheck_Zllg_massa} than in Figure~\ref{fig:xcheck_Zllg_massb}. The residual scales measured in the two channels agree within one standard deviation.
The uncertainties combine the statistical uncertainty and the systematic uncertainty originating from the uncertainty in the
lepton-energy calibration (see Ref.~\cite{MUON-2022-01} for a description of the muon momentum scale determination).
 
To extract the residual photon energy scale, MC templates are compared with the data distribution through a $\chi^2$ test, for the electron and muon channels independently.
These templates are built by modifying the photon \et by a factor $(1+\alpha)$, where $\alpha$ is varied from $-0.0200$ to 0.0200 in steps of 0.0004, and the overall energy scale residual is given by the template providing the lowest $\chi^2$. The two channels are compatible and are statistically combined.
\begin{figure*}[h]
\centering
\subfigure[]{\includegraphics[width=.49\textwidth]{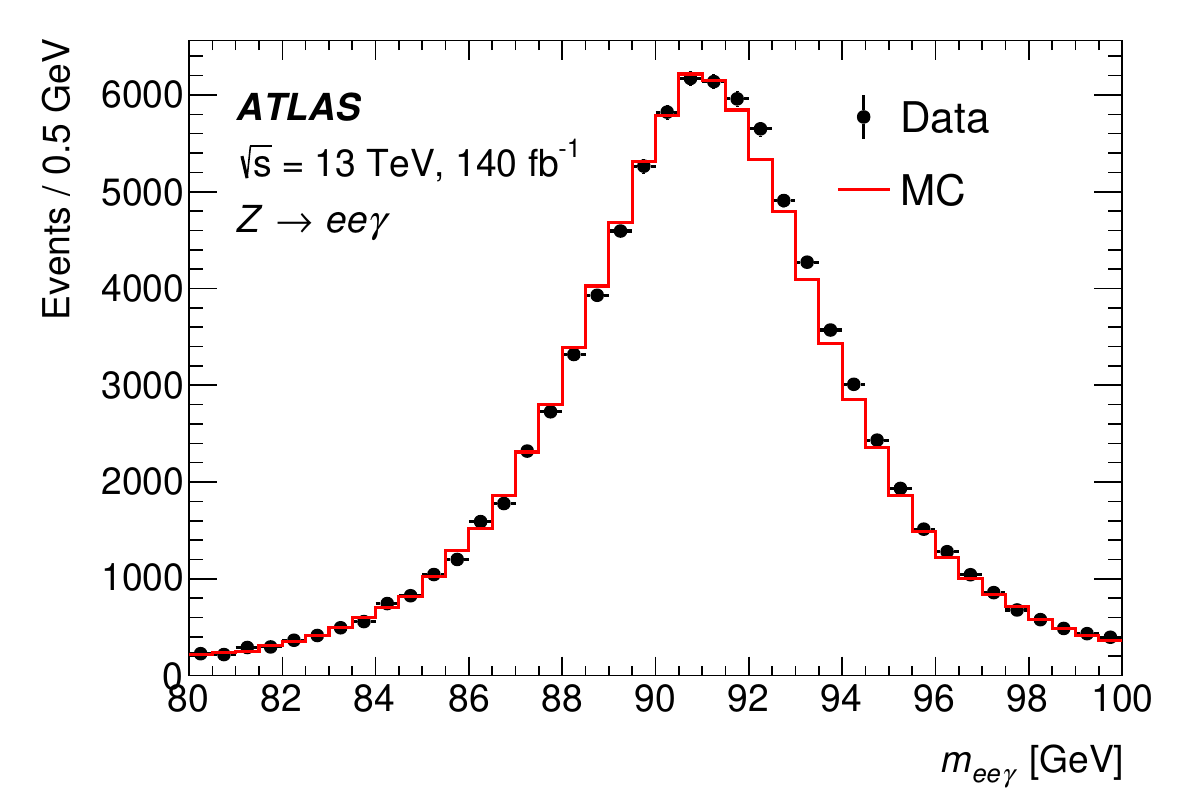}\label{fig:xcheck_Zllg_massa}}
\subfigure[]{\includegraphics[width=.49\textwidth]{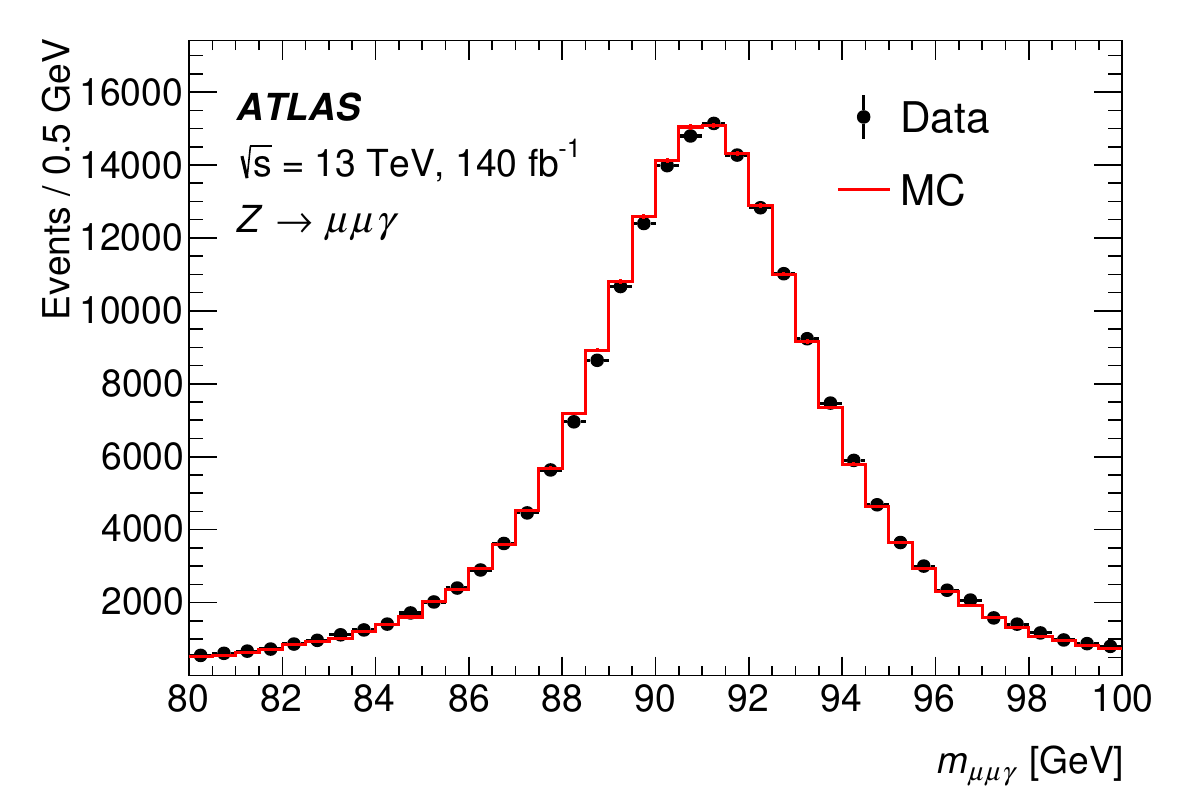}\label{fig:xcheck_Zllg_massb}}
\caption{Comparison of the (a) $ee\gamma$ and (b) $\mu\mu\gamma$ invariant mass distributions in data and simulation.}
\label{fig:xcheck_Zllg_mass}
\end{figure*}
 
The dependence of $\alpha$ on the energy and pseudorapidity of the photon is illustrated in Figure~\ref{fig:xcheck_zllg}, separately for converted photons and unconverted photons. The residual photon energy scales are compared with the total energy calibration uncertainty for photons from $Z \to \ell\ell\gamma$ decays, before and after including the constraints from the linearity fit. The error bars assigned to the data are typically 0.15\%, and are dominated by the uncertainty in the lepton-energy calibration. As in the $J/\psi$ analysis, the linearity-constrained calibration tends to reduce the values of the residual calibration factors, which are found to be within the corresponding calibration uncertainties. Mild tension is observed for $|\eta|>1.8$, for unconverted photons; this effect is driven by low-energy photons, and disappears at high \et as indicated by Figure~\ref{fig:xcheck_zllgb}.
\begin{figure*}[h]
\centering
\subfigure[]{\includegraphics[width=.49\textwidth]{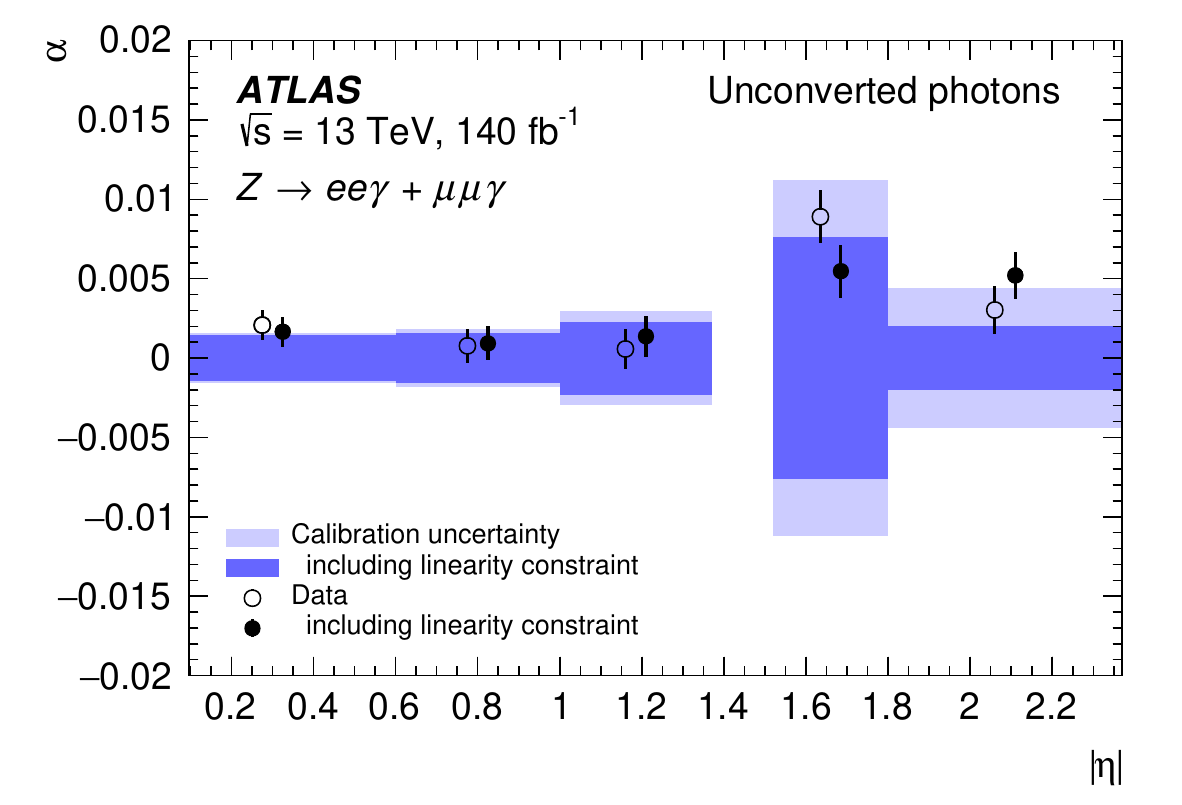}}
\subfigure[]{\includegraphics[width=.49\textwidth]{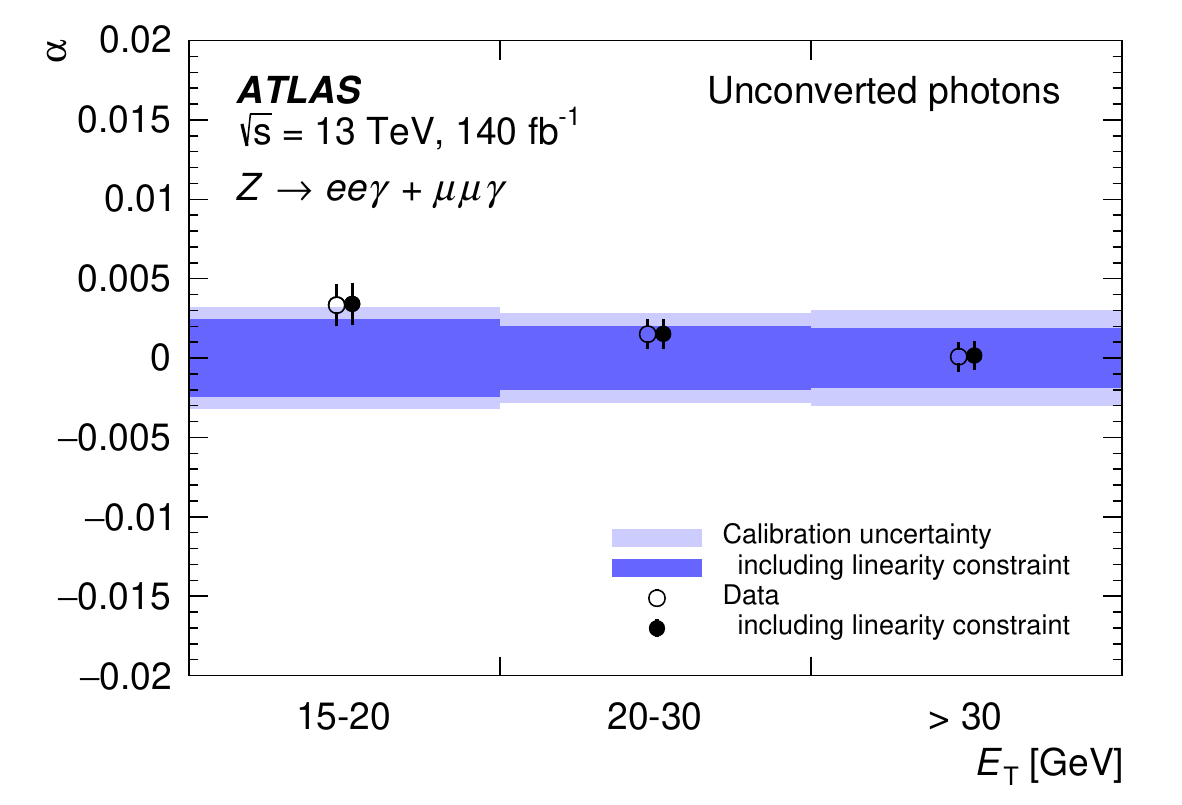}\label{fig:xcheck_zllgb}}
\subfigure[]{\includegraphics[width=.49\textwidth]{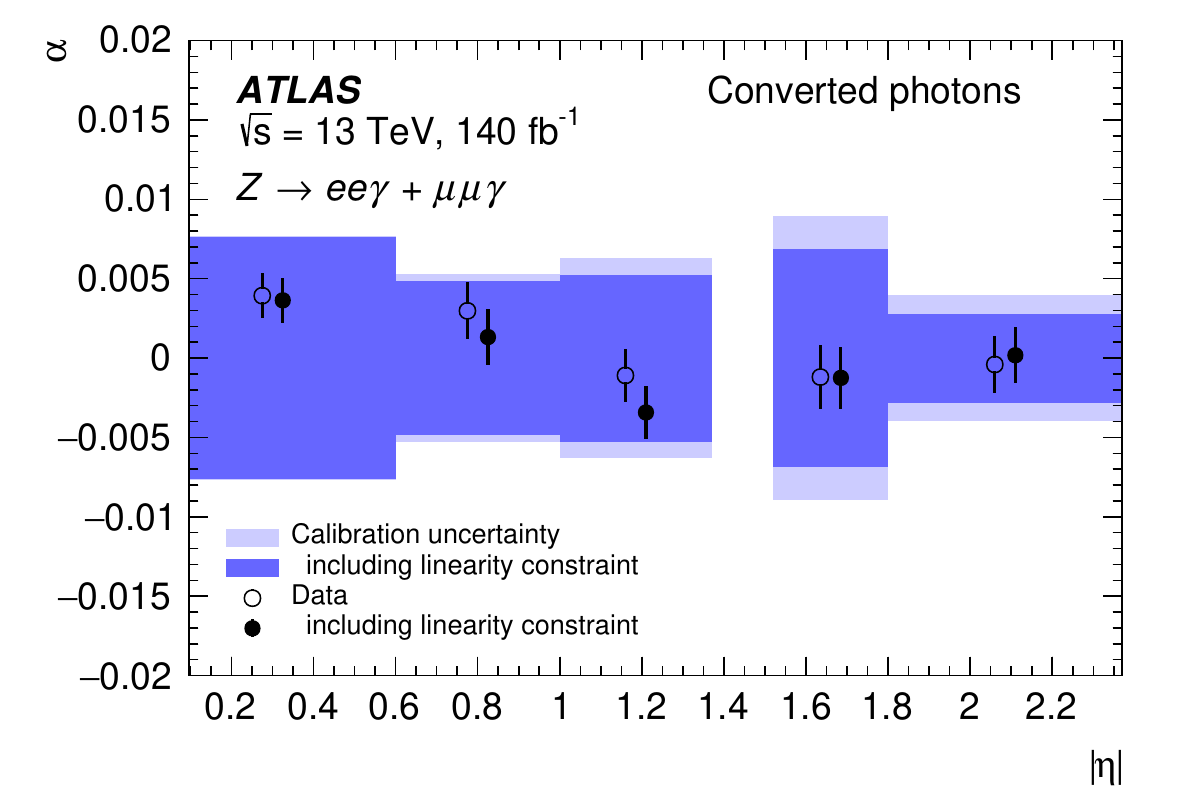}}
\subfigure[]{\includegraphics[width=.49\textwidth]{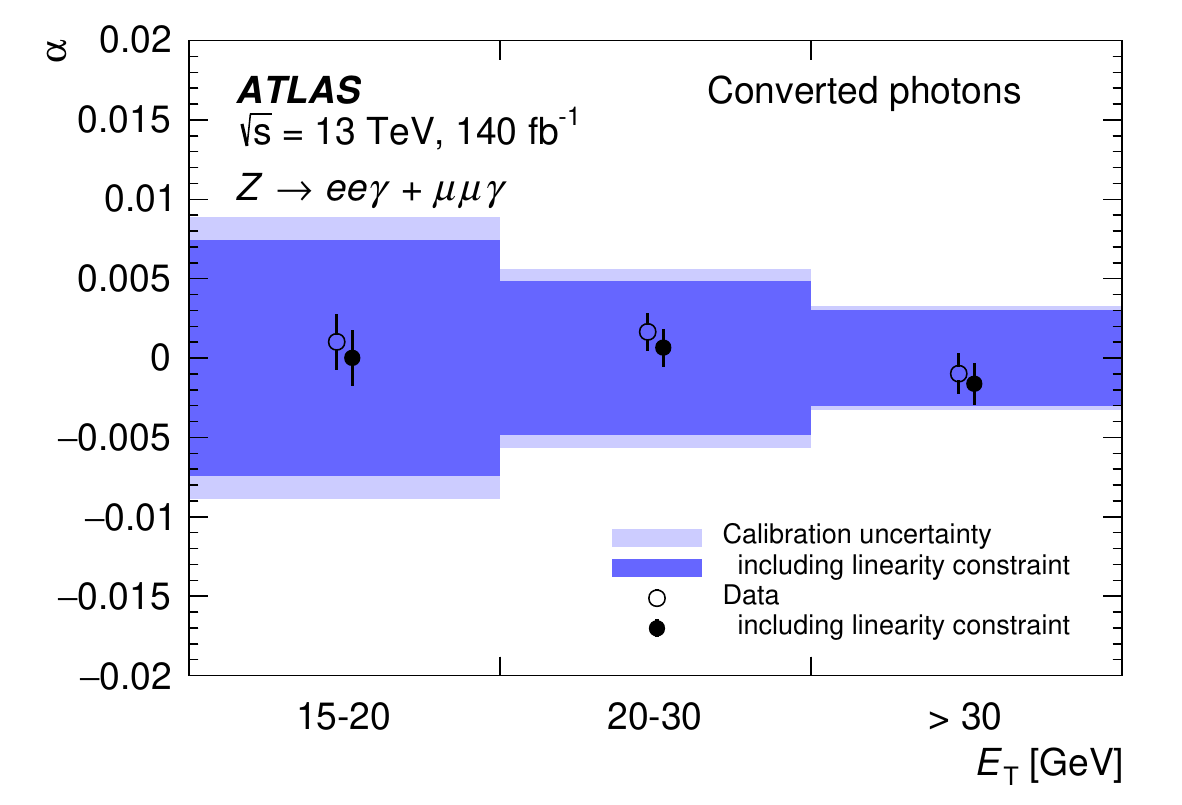}}
\caption{Variation of the residual energy scale for (a, b) unconverted and (c, d) converted photons as a function of (a, c) $|\eta|$ and (b, d) \et, as measured with $Z\rightarrow \ell\ell\gamma$ events. The data points and uncertainty bands are shown for both the pre- and post-linearity-fit energy scale models. The uncertainty bands correspond to the energy calibration uncertainty for photons
from $Z \to \ell\ell\gamma$ decays.}
\label{fig:xcheck_zllg}
\end{figure*}
\FloatBarrier


\section{Conclusion}
 
This paper presents the energy calibration for electrons and photons reconstructed in 140~fb$^{-1}$ of 13~\TeV\ proton--proton collision data recorded by ATLAS during \RunTwo of the LHC.
All of the major sources of uncertainty have been re-evaluated since the previous publications, and new methods are introduced to further reduce their impact.
Improved methods to calibrate energies in the calorimeter cells and layers, and an improved measurement of
lateral energy leakage from reconstructed electron and photon energy clusters,
reduce the \textit{a priori} calibration uncertainty by about 30\%. In addition, a precise measurement of the energy linearity, using electrons from $Z$-boson decays, provides a further reduction by about a factor of two. The overall calibration uncertainty is reduced by a factor of 2--3, depending on the particle type, pseudorapidity and energy.  The achieved calibration
uncertainties are typically 0.05\% for electrons from $Z$-boson decays, 0.4\% at $\et\sim 10$~\GeV, and 0.3\% at $\et\sim 1$~\TeV; for photons at $\et\sim 60$~\GeV, they are 0.2\% on average. These improvements are validated using independent samples of $J/\psi\to ee$ decays and radiative $Z$-boson decays.
The achieved precision is adequate for high-precision measurements of fundamental parameters such as the masses and properties of the Higgs, $W$ and $Z$ bosons, and is expected to improve the sensitivity of searches and measurements at the weak scale.

\section*{Acknowledgements}


We thank CERN for the very successful operation of the LHC and its injectors, as well as the support staff at
CERN and at our institutions worldwide without whom ATLAS could not be operated efficiently.
 
The crucial computing support from all WLCG partners is acknowledged gratefully, in particular from CERN, the ATLAS Tier-1 facilities at TRIUMF/SFU (Canada), NDGF (Denmark, Norway, Sweden), CC-IN2P3 (France), KIT/GridKA (Germany), INFN-CNAF (Italy), NL-T1 (Netherlands), PIC (Spain), RAL (UK) and BNL (USA), the Tier-2 facilities worldwide and large non-WLCG resource providers. Major contributors of computing resources are listed in Ref.~\cite{ATL-SOFT-PUB-2023-001}.
 
We acknowledge the support of ANPCyT, Argentina; YerPhI, Armenia; ARC, Australia; BMWFW and FWF, Austria; ANAS, Azerbaijan; CNPq and FAPESP, Brazil; NSERC, NRC and CFI, Canada; CERN; ANID, Chile; CAS, MOST and NSFC, China; Minciencias, Colombia; MEYS CR, Czech Republic; DNRF and DNSRC, Denmark; IN2P3-CNRS and CEA-DRF/IRFU, France; SRNSFG, Georgia; BMBF, HGF and MPG, Germany; GSRI, Greece; RGC and Hong Kong SAR, China; ISF and Benoziyo Center, Israel; INFN, Italy; MEXT and JSPS, Japan; CNRST, Morocco; NWO, Netherlands; RCN, Norway; MEiN, Poland; FCT, Portugal; MNE/IFA, Romania; MESTD, Serbia; MSSR, Slovakia; ARRS and MIZ\v{S}, Slovenia; DSI/NRF, South Africa; MICINN, Spain; SRC and Wallenberg Foundation, Sweden; SERI, SNSF and Cantons of Bern and Geneva, Switzerland; MOST, Taipei; TENMAK, T\"urkiye; STFC, United Kingdom; DOE and NSF, United States of America.
 
In addition, individual groups and members have received support from BCKDF, CANARIE, CRC and DRAC, Canada; PRIMUS 21/SCI/017 and UNCE SCI/013, Czech Republic; COST, ERC, ERDF, Horizon 2020, ICSC-NextGenerationEU and Marie Sk{\l}odowska-Curie Actions, European Union; Investissements d'Avenir Labex, Investissements d'Avenir Idex and ANR, France; DFG and AvH Foundation, Germany; Herakleitos, Thales and Aristeia programmes co-financed by EU-ESF and the Greek NSRF, Greece; BSF-NSF and MINERVA, Israel; Norwegian Financial Mechanism 2014-2021, Norway; NCN and NAWA, Poland; La Caixa Banking Foundation, CERCA Programme Generalitat de Catalunya and PROMETEO and GenT Programmes Generalitat Valenciana, Spain; G\"{o}ran Gustafssons Stiftelse, Sweden; The Royal Society and Leverhulme Trust, United Kingdom.
 
Individual members wish to acknowledge support from Chile: Agencia Nacional de Investigación y Desarrollo (FONDECYT 1190886, FONDECYT 1210400, FONDECYT 1230987); China: National Natural Science Foundation of China (NSFC - 12175119, NSFC 12275265); European Union: European Research Council (ERC - 948254), Horizon 2020 Framework Programme (MUCCA - CHIST-ERA-19-XAI-00), Italian Center for High Performance Computing, Big Data and Quantum Computing (ICSC, NextGenerationEU), Marie Sklodowska-Curie Actions (EU H2020 MSC IF GRANT NO 101033496); France: Agence Nationale de la Recherche (ANR-20-CE31-0013, ANR-21-CE31-0022), Investissements d'Avenir Idex (ANR-11-LABX-0012), Investissements d'Avenir Labex (ANR-11-LABX-0012); Germany: Baden-Württemberg Stiftung (BW Stiftung-Postdoc Eliteprogramme), Deutsche Forschungsgemeinschaft (DFG - CR 312/5-1); Italy: Istituto Nazionale di Fisica Nucleare (FELLINI G.A. n. 754496, ICSC, NextGenerationEU); Japan: Japan Society for the Promotion of Science (JSPS KAKENHI 22H01227, JSPS KAKENHI JP21H05085, JSPS KAKENHI JP22H04944); Netherlands: Netherlands Organisation for Scientific Research (NWO Veni 2020 - VI.Veni.202.179); Norway: Research Council of Norway (RCN-314472); Poland: Polish National Agency for Academic Exchange (PPN/PPO/2020/1/00002/U/00001), Polish National Science Centre (NCN 2021/42/E/ST2/00350, NCN OPUS nr 2022/47/B/ST2/03059, NCN UMO-2019/34/E/ST2/00393, UMO-2020/37/B/ST2/01043); Slovenia: Slovenian Research Agency (ARIS grant J1-3010); Spain: BBVA Foundation (LEO22-1-603), Generalitat Valenciana (Artemisa, FEDER, IDIFEDER/2018/048), La Caixa Banking Foundation (LCF/BQ/PI20/11760025), Ministry of Science and Innovation (RYC2019-028510-I, RYC2020-030254-I), PROMETEO and GenT Programmes Generalitat Valenciana (CIDEGENT/2019/023, CIDEGENT/2019/027); Sweden: Swedish Research Council (VR 2022-03845); Switzerland: Swiss National Science Foundation (SNSF - PCEFP2\_194658); United Kingdom: Leverhulme Trust (Leverhulme Trust RPG-2020-004); United States of America: Neubauer Family Foundation.


\printbibliography
 
\clearpage
 
\begin{flushleft}
\hypersetup{urlcolor=black}
{\Large The ATLAS Collaboration}

\bigskip

\AtlasOrcid[0000-0002-6665-4934]{G.~Aad}$^\textrm{\scriptsize 102}$,
\AtlasOrcid[0000-0002-5888-2734]{B.~Abbott}$^\textrm{\scriptsize 120}$,
\AtlasOrcid[0000-0002-1002-1652]{K.~Abeling}$^\textrm{\scriptsize 55}$,
\AtlasOrcid[0000-0001-5763-2760]{N.J.~Abicht}$^\textrm{\scriptsize 49}$,
\AtlasOrcid[0000-0002-8496-9294]{S.H.~Abidi}$^\textrm{\scriptsize 29}$,
\AtlasOrcid[0000-0002-9987-2292]{A.~Aboulhorma}$^\textrm{\scriptsize 35e}$,
\AtlasOrcid[0000-0001-5329-6640]{H.~Abramowicz}$^\textrm{\scriptsize 151}$,
\AtlasOrcid[0000-0002-1599-2896]{H.~Abreu}$^\textrm{\scriptsize 150}$,
\AtlasOrcid[0000-0003-0403-3697]{Y.~Abulaiti}$^\textrm{\scriptsize 117}$,
\AtlasOrcid[0000-0002-8588-9157]{B.S.~Acharya}$^\textrm{\scriptsize 69a,69b,m}$
% ...
\bigskip
\\

% ...
$^{29}$Physics Department, Brookhaven National Laboratory, Upton NY; United States of America.\\
% ...
$^{35}$$^{(a)}$Facult\'e des Sciences Ain Chock, R\'eseau Universitaire de Physique des Hautes Energies - Universit\'e Hassan II, Casablanca;$^{(b)}$Facult\'{e} des Sciences, Universit\'{e} Ibn-Tofail, K\'{e}nitra;$^{(c)}$Facult\'e des Sciences Semlalia, Universit\'e Cadi Ayyad, LPHEA-Marrakech;$^{(d)}$LPMR, Facult\'e des Sciences, Universit\'e Mohamed Premier, Oujda;$^{(e)}$Facult\'e des sciences, Universit\'e Mohammed V, Rabat;$^{(f)}$Institute of Applied Physics, Mohammed VI Polytechnic University, Ben Guerir; Morocco.\\
% ...
$^{49}$Fakult\"{a}t Physik , Technische Universit{\"a}t Dortmund, Dortmund; Germany.\\
% ...
$^{55}$II. Physikalisches Institut, Georg-August-Universit\"{a}t G\"ottingen, G\"ottingen; Germany.\\
% ...
$^{69}$$^{(a)}$INFN Gruppo Collegato di Udine, Sezione di Trieste, Udine;$^{(b)}$ICTP, Trieste;$^{(c)}$Dipartimento Politecnico di Ingegneria e Architettura, Universit\`a di Udine, Udine; Italy.\\
% ...
$^{102}$CPPM, Aix-Marseille Universit\'e, CNRS/IN2P3, Marseille; France.\\
% ...
$^{117}$Department of Physics, New York University, New York NY; United States of America.\\
% ...
$^{120}$Homer L. Dodge Department of Physics and Astronomy, University of Oklahoma, Norman OK; United States of America.\\
% ...
$^{150}$Department of Physics, Technion, Israel Institute of Technology, Haifa; Israel.\\
$^{151}$Raymond and Beverly Sackler School of Physics and Astronomy, Tel Aviv University, Tel Aviv; Israel.\\
% ...
$^{a}$ Also Affiliated with an institute covered by a cooperation agreement with CERN.\\
$^{b}$ Also at An-Najah National University, Nablus; Palestine.\\
$^{c}$ Also at Borough of Manhattan Community College, City University of New York, New York NY; United States of America.\\
$^{d}$ Also at Center for High Energy Physics, Peking University; China.\\
$^{e}$ Also at Center for Interdisciplinary Research and Innovation (CIRI-AUTH), Thessaloniki; Greece.\\
$^{f}$ Also at Centro Studi e Ricerche Enrico Fermi; Italy.\\
% ...
$^{m}$ Also at Department of Physics, King's College London, London; United Kingdom.\\
% ...

\end{flushleft}


\end{document}